\documentclass{elsarticle}

\usepackage[english]{babel}
\usepackage[braket, qm]{qcircuit}
\usepackage{graphicx}
\usepackage{amsmath}
\usepackage{amssymb}
\usepackage{lineno}
\usepackage{multirow}
\usepackage{booktabs}
\usepackage{tabularx}
\usepackage{framed}
\usepackage{array}
\usepackage{subcaption}
\usepackage{paralist}
\usepackage{makecell}
\usepackage{url}
\makeatletter
\def\ps@pprintTitle{%
 \let\@oddhead\@empty
 \let\@evenhead\@empty
 \def\@oddfoot{}%
 \let\@evenfoot\@oddfoot}
\makeatother
\usepackage{enumitem} 

\newlist{tabitem}{itemize}{1}
\setlist[tabitem]{label=\tiny$\bullet$, leftmargin=*, nosep, before=\vspace{-0.6\baselineskip}, after=\vspace{-1\baselineskip}}

\begin{document}

\begin{frontmatter}

\title{A Longitudinal Analysis of the CEC Single-Objective Competitions (2010–2024) and Implications for Variational Quantum Optimization}


\address[aff1]{Department of Computer Science, Faculty of Electrical Engineering and Computer Science, VSB - Technical University of Ostrava, Ostrava, Czech Republic}
\address[aff2]{IT4Innovations National Supercomputing Center, VSB - Technical University of Ostrava, 708 00 Ostrava, Czech Republic}
\address[aff3]{Department of Mathematics, TUM School of Computation, Information and Technology, Technical University of Munich, Garching bei M\"{u}nchen, Germany}
\address[aff4]{Department of Informatics and Statistics, Marine Research Institute, Klaipeda University, Lithuania}
\address[aff5]{Dipartimento di Ingegneria e Scienze dell'Informazione e Matematica, Universit\`{a} dell'Aquila, Via Vetoio, I-67010 Coppito, L'Aquila, Italy}
\address[aff6]{Electronics and Communication Sciences Unit, Indian Statistical Institute, Kolkata 700108, India}

\author[aff1,aff2,aff4]{Vojt\v{e}ch Nov\'{a}k\corref{cor1}}
\ead{vojtech.novak.st1@vsb.cz}
\author[aff3]{Tom\'{a}\v{s} Bezd\v{e}k}
\author[aff2,aff4]{Ivan Zelinka}
\author[aff6]{Swagatam Das}
\author[aff5]{Martin Beseda}

\cortext[cor1]{Corresponding author}

\begin{abstract}
This paper provides a historical analysis of the IEEE CEC Single Objective Optimization competition results (2010–2024). We analyze how benchmark functions shaped winning algorithms, identifying the 2014 introduction of dense rotation matrices as a key performance filter. This design choice introduced parameter non-separability, reduced effectiveness of coordinate-dependent methods (PSO, GA), and established the dominance of Differential Evolution variants capable of preserving the rotational invariance of their difference vectors, specifically L-SHADE. Post-2020 analysis reveals a shift towards high complexity hybrid optimizers that combine different mechanisms (e.g., Eigenvector Crossover, Societal Sharing, Reinforcement Learning) to maximize ranking stability. We conclude by identifying structural similarities between these modern benchmarks and Variational Quantum Algorithm landscapes, suggesting that evolved CEC solvers possess the specific adaptive capabilities required for quantum control.
\end{abstract}


\end{frontmatter}


\section{Introduction} \label{sec:introduction}

Global optimization of continuous functions is fundamental to computational intelligence and is essential for solving complex problems in engineering and science. Since their inception, competitions organized within the IEEE Congress on Evolutionary Computation (CEC) have established themselves as the de facto standard for benchmarking stochastic optimization algorithms \cite{Molina2018}. These annual events provide a rigorous standardized framework, defined by their respective problem definition reports \cite{tang2010benchmark, das2010problem, liang2013cec2013, liang2013cec2014, wu2017problem, mohamed2019cec2020, kumar2022cec2022}. This allows researchers to evaluate the performance of novel algorithms against state-of-the-art methods under controlled experimental conditions. The resulting datasets and rankings, which we collated from the official repositories maintained by the competition organizers \cite{suganthan2025github}, have significantly influenced the trajectory of evolutionary computation driving the transition from static canonical algorithms to highly adaptive and complex hybrid systems.

While annual reports summarize specific competition results, broader longitudinal analyses are rare. Previous studies by Molina et al. \cite{improvement} and Skvorc et al. \cite{vskvorc2019cec} have critically examined whether newer winners statistically outperform their predecessors, often finding that progress is sometimes the result of overfitting to a specific year's suite rather than general algorithmic improvement. Existing surveys typically focus on specific benchmark suites or provide a snapshot of performance for a single year, often comparing a proposed algorithm only against its immediate predecessors. However, these studies are limited to specific timeframes and provide fragmented views of results.

This paper complements those statistical surveys by analyzing the design evolution of the winning algorithms from 2010 to 2024. Rather than simply tabulating rankings, we trace the lineage of dominant mechanisms, such as Success History Adaptation \cite{tanabe2013success} and Linear Population Size Reduction \cite{Tanabe2014}, to understand how solvers adapted to specific changes in benchmark topology. For example, we identify how the introduction of dense rotation matrices in the CEC 2014 hindered the performance of coordinate-dependent methods such as Particle Swarm Optimization (PSO) \cite{kennedy1995pso} and Genetic Algorithms (GA) \cite{goldberg1989genetic}. This shift forced the community toward Differential Evolution \cite{das2010differential} variants that leverage rotationally invariant mutation \cite{sutton2007differential, tanabe2014improving, price2005differential}, where the addition of population reduction proved decisive in outperforming restart based strategies like Covariance Matrix Adaptation Evolution Strategy (CMA-ES).

The analysis identifies three distinct evolutionary eras that characterize the progress of the field during this period. The first phase spanning from 2010 to 2013 represents a period of algorithmic specialization, where diverse paradigms competed and Canonical Genetic Algorithms claimed victory on Real-World Problems \cite{Elsayed2011_GAMPC}. The second phase, beginning in 2014, marks the dominance of adaptive Differential Evolution with the rise of the L-SHADE family \cite{Tanabe2014, awad2016lshade, brest2016ilshade}, establishing the superiority of adaptation over static heuristics. By 2017, this evolved into a period of hybridization as benchmark functions became increasingly non-separable and multimodal through the use of composition functions. This led to the proliferation of societal algorithms that structurally hybridized distinct search logics to navigate composite landscapes such as AGSK \cite{mohamed2020agsk} and IMODE \cite{sallam2020imode}. The third and most recent phase from 2021 to 2024 reflects a period of intense complexity and refinement characterized by the adoption of Eigenvector Crossover to mimic covariance learning \cite{ea4eign2022, alic2020jde100e} and Success Rate based adaptation to sharpen exploitation \cite{stanovov2024lsrde}. We examine the post 2020 phase and its increasing complexity of algorithms, noting the tendency to combine complexity with necessary innovation \cite{Biedrzycki2024}.

We extend our analysis of CEC results into a forward-looking proposal to apply these evolved optimizers to Variational Quantum Algorithms (VQAs) \cite{cerezo2020variational, mcclean2016theory, kandala2017hardware, illesova2025transformation, beseda2024state}. VQAs leverage hybrid quantum-classical systems to enhance machine learning models with quantum features \cite{Arrasmith2021}, solve large-scale combinatorial optimization problems, and simulate entangled quantum systems. However, their parameter landscapes feature extreme noise, entangled variable interactions, and barren plateaus \cite{larocca2025barren}, challenges akin to post-2014 CEC benchmarks' non-separability and multimodality.

Researchers have already explored gradient-free evolutionary methods to address these issues, such as basic DE to avoid local minima \cite{Failde2023} and mitigate barren plateaus \cite{Arrasmith2021}, as well as comparisons of standard DE and CMA-ES variants on VQA performance \cite{BonetMonroig2023}. While these foundational applications demonstrate the viability of evolutionary approaches, they primarily rely on canonical or basic implementations, leaving room for enhancement through more adaptive mechanisms. CEC-dominant methods, like L-SHADE variants, are particularly suited to extend this work due to their rotational invariance for correlated parameters due to quantum entanglement. The next biggest challenge is finite sampling noise of the quantum state that can be approached with, for example, population averaging in CMA-ES \cite{astete2015evolution, hellwig2016evolution, novak2025reliable}.

\section{Methodology}
\label{sec:methodology}

This study employs a systematic analysis of the winning algorithms from the CEC Single Objective Optimization competitions from 2010 through 2024. While the raw ranking data was collated from the official repositories maintained by the organizers \cite{suganthan2025github}, our analytical framework moves beyond simple tabulation to categorize the evolutionary mechanisms that enabled these algorithms to survive increasing benchmark complexity. The primary data sources for this review are the technical reports and algorithmic descriptions of the top-performing entries (Ranks 1--3) spanning fifteen years of competition. Specific algorithmic details, operator definitions, and parameter adaptation strategies were extracted directly from the individual algorithm papers summarized in the tables in Section \ref{sec:results}.

To accurately map these evolutionary trajectories, we ground our classification in the canonical definitions of the major algorithm families. We analyze foundational frameworks established by Holland for Genetic Algorithms \cite{Holland1975}, Storn and Price for Differential Evolution \cite{Storn1997}, and Hansen and Ostermeier for Evolution Strategies \cite{Hansen2001}. This theoretical baseline allows us to identify the specific structural limitations that led to the decline of coordinate-dependent methods like PSO \cite{Shami2022} and the parallel rise of rotationally invariant strategies \cite{sutton2007differential}.

Our methodology is based on the foundational literature in the field. Das and Suganthan established the baseline taxonomy for Differential Evolution in their surveys \cite{das2010differential, das2016recent}, yet their analysis predates the widespread adoption of the L-SHADE family. Similarly, Molina et al. provided a rigorous analysis of bio-inspired algorithms over a decade of competitions \cite{Molina2018} and critically examined whether newer winners statistically outperform their predecessors \cite{piotrowski2023choice}. However, these studies do not cover the post-2020 phase of structural hybridization or the cooperative decomposition trends noted by Mahdavi et al. \cite{Mahdavi2015}. This review fills that gap by integrating these perspectives with a modern critique of complexity, applying the component-wise sensitivity analysis proposed by Biedrzycki \cite{Biedrzycki2024} and the arguments of Eiben and Jelasity \cite{eiben2002critical} to distinguish between genuine algorithmic innovation and redundant complexity inflation. Algorithms were subsequently classified into structural families to visualize the dominance trends discussed in the following sections.

\section{Results: CEC Results by Optimization Mechanisms (2010–2024)}
\label{sec:results}

\begin{figure}[htbp]
    \centering
    \includegraphics[width=1\linewidth]{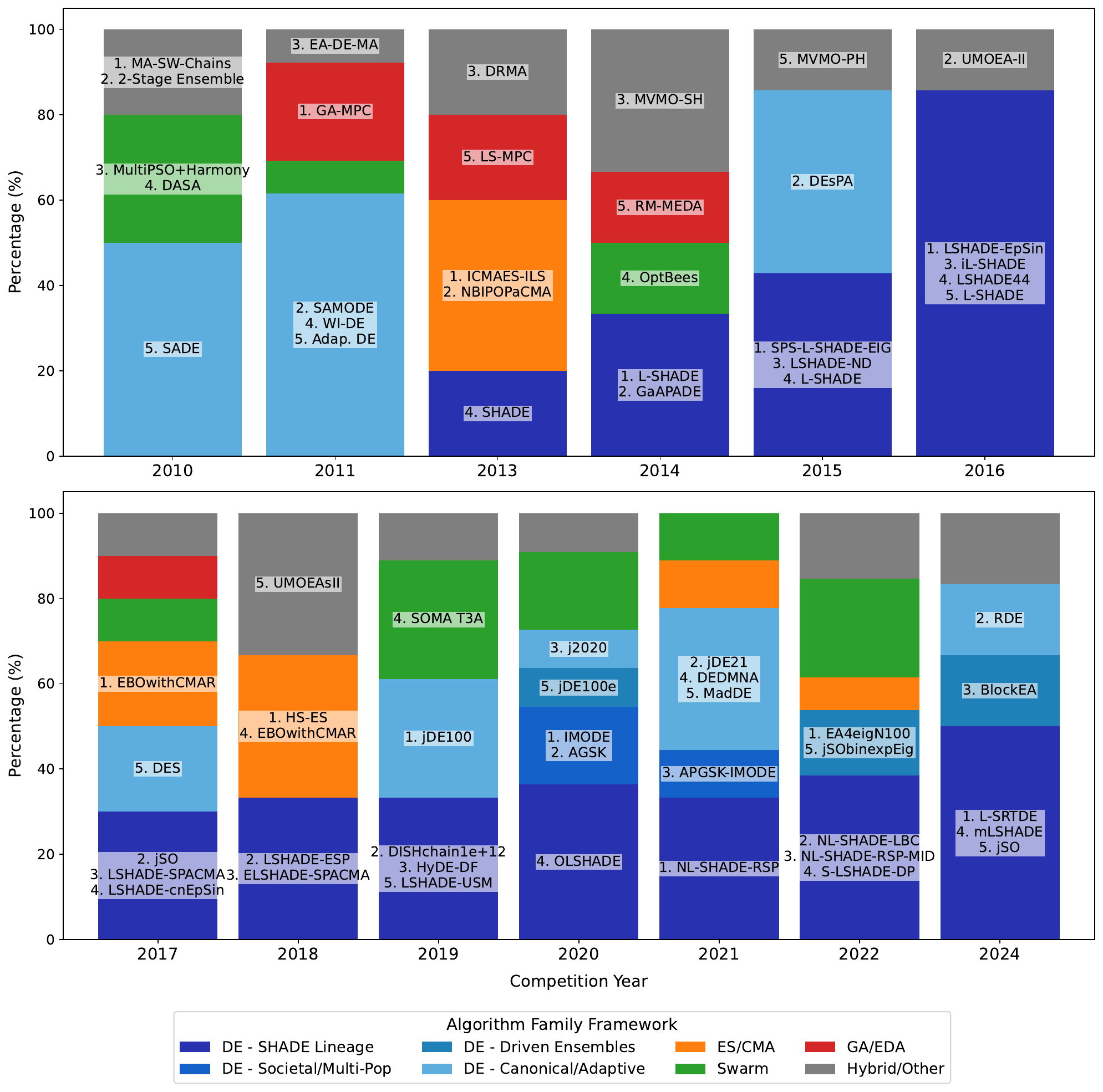}
    \caption{Longitudinal evolution of algorithm families in CEC Competitions (2010–2024). Stacked bars illustrate the percentage distribution of all participating entries by category, with text annotations identifying the top-five-ranked algorithms for each year. To address the structural hybridization characteristic of post-2020 solvers, algorithmic classification follows a strict "Primary Search Engine" protocol. Solvers are categorized based on the specific evolutionary mechanism that consumes the majority ($>50\%$) of the maximum function evaluation budget. To accurately capture evolutionary trajectories, the DE family is subdivided into Canonical/Adaptive, SHADE Lineage, Societal/Multi-Population, and Driven Ensembles. Under this protocol, multi-component architectures are classified by their dominant foundational logic; for instance, ensembles that rely on DE as their core exploration framework but periodically trigger Covariance Matrix Adaptation to navigate rotated valleys (e.g., EA4eigN100 ) are classified as DE-Driven Ensembles rather than generic hybrids. This granular taxonomy confirms a structural shift from early algorithmic diversity to the sustained dominance of highly specialized DE architectures.}
    \label{fig:bar}
\end{figure}

The historical trajectory of the CEC competition benchmarks reflects a clear transition from experimental diversity to rigorous standardization. The increasing complexity of the benchmark functions directly dictated which algorithmic mechanisms survived and which became obsolete. We aggregated all results from 2010 to 2024 in Fig. \ref{fig:bar} where we show the algorithm family with color and the top 5 places for each year. This section is divided into three distinct phases based on optimizer mechanisms, with 2014 being the most significant year that established SHADE variants as the most potent solvers for non-separable functions.

\subsection{Phase 1: Coordinate-Dependent Search and Domain Specialization (2010--2013)}

Prior to the establishment of modern CEC benchmarks, algorithms such as JADE \cite{JADE} and CoDE \cite{CoDE} defined the state-of-the-art. JADE introduced parameter adaptation based on evolutionary history, while CoDE validated the combination of multiple trial vector strategies. These methods established the performance baseline for subsequent DE development.

The 2010 Large-Scale Global Optimization suite \cite{tang2010benchmark} marked the beginning of this phase by addressing the curse of dimensionality. By scaling problems to 1000 dimensions, it emphasized variable interaction over topological ruggedness. As shown in Tab. \ref{tab:cec2010_2011_2013}, this favored Memetic Algorithms and Swarm Intelligence variants capable of decomposing high-dimensional spaces.

The CEC 2011 competition \cite{das2010problem} deviated significantly from standard benchmarking by focusing exclusively on Real-World Problems. Unlike synthetic functions defined by abstract mathematical properties, this suite comprised 22 constrained engineering tasks, ranging from chemical engineering control (Tersoff Potential \cite{lindsay2010optimized}) to power system scheduling (Economic Load Dispatch \cite{farag1995economic}) and space trajectory optimization (Cassini 2 \cite{addis2011global}). These problems introduced practical complexities such as mixed-integer variables and non-linear physical constraints. Consequently, this unique testbed favored algorithms capable of handling discrete and constrained search spaces, explaining the success of Genetic Algorithms like GA-MPC \cite{Elsayed2011_GAMPC} over purely continuous optimizers.

In 2013 \cite{liang2013cec2013}, the competition divided into specific tracks. While the Niching track focused on multimodal diversity, the Real-Parameter Single Objective track sought global optimization efficiency. The latter was dominated by CMA-ES hybrids. The winner, ICMAES-ILS \cite{Liao2013}, combined IPOP-CMA-ES with Iterative Local Search to balance exploration and exploitation. The runner-up, NBIPOPaCMA \cite{Loshchilov2013}, improved standard CMA-ES by incorporating worst-solution updates and maintaining dual populations. DRMA \cite{Lacroix2013} took third place by applying CMA-ES within dynamically resizing hypercubes.

Despite this dominance of hybrid methods, the fourth-ranked algorithm proved most significant for future developments. SHADE \cite{SHADE2013}, a pure DE variant building on JADE, demonstrated competitive performance without complex hybridization. This marked a turning point; as discussed in the following section, the introduction of rotational matrices in 2014 \cite{liang2013cec2014} would see SHADE evolve into L-SHADE, signaling the end of CMA-ES supremacy on these benchmarks.

\begin{table}[htbp]
\centering
\footnotesize
\setlength{\tabcolsep}{3pt}
\caption{CEC 2010 Large Scale Global Optimization Challenge, CEC 2011 Competition on
Testing Evolutionary Algorithms on Real World Optimization Problems and CEC 2013 – Rankings (N MO means Niching
Methods for Multimodal Optimization, R-P means Real-Parameter Single Objective)}
\label{tab:cec2010_2011_2013}
\begin{tabular}{@{}c l l l l@{}}
\toprule
Rank & CEC 2010 (LSGO)                      & CEC 2011                          & CEC 2013 (N M-O)             & CEC 2013 (R-P SO) \\
\midrule
1  & MA-SW-Chains \cite{Molina2010}       & GA-MPC \cite{Elsayed2011_GAMPC}           & NEA2 \cite{Preuss2013}         & ICMAES-ILS \cite{Liao2013} \\
2  & 2-Stage Ensemble \cite{Wang2010}     & SAMODE \cite{Elsayed2011_SAMODE}          & dADE/nrand/1 \cite{Epitropakis2013} & NBIPOPaCMA \cite{Loshchilov2013} \\
3  & MultiPSO+Harmony \cite{LaTorre2010}  & EA-DE-MA \cite{Singh2011_EADEMA}          & CMA-ES                         & DRMA \cite{Lacroix2013} \\
4  & DASA \cite{DASA}                     & WI-DE \cite{Haider2011_WIDE}              & N-VMO \cite{Barrera2013}       & SHADE \cite{SHADE2013} \\
5  & SADE \cite{self-DE}                  & Adap. DE 171 \cite{Asafuddoula2011_AdapDE}& dADE/nrand/2 \cite{Epitropakis2013} & LS-MPC \cite{GA2013} \\
6  & DECC-D \cite{DECC}                   & DE-$\Lambda$ \cite{Reynoso2011_DELambda}  & DE/nrand/2 \cite{Epitropakis2013_DE} & -- \\
7  & Locust Swarms \cite{Locust}          & ED-DE \cite{ED-DE}                & PNA-NSGAII \cite{Bandyopadhyay2013} & -- \\
8  & SDENS \cite{SEQ}                     & DE-RHC \cite{LaTorre2011_DERHC}           & CrowdingDE \cite{Thomsen2004}  & -- \\
9  & Classic DE (CR=0.9)                  & RGA \cite{Saha2011_RGA}                   & DE/nrand/1 \cite{Epitropakis2013_DE} & -- \\
10 & Classic DE (CR=0.0)                  & Mod-DE-LS \cite{Mandal2011_ModDELS}       & NEA1 \cite{Preuss2013}         & -- \\
11 & --                                   & mSBX-GA \cite{Bandaru2011_mSBX}           & DELG                           & -- \\
12 & --                                   & ENSML-DE \cite{Mallipeddi2011_ENSML}      & DELS-aj                        & -- \\
13 & --                                   & CDASA \cite{Korosec2011_CDASA}            & DECG                           & -- \\
14 & --                                   & --                                        & IPOP-CMA-ES \cite{Auger2005}   & -- \\
15 & --                                   & --                                        & A-NSGAII                       & -- \\
\bottomrule
\end{tabular}
\end{table}

\subsection{Phase 2: Parameter Non-Separability and L-SHADE variants (2014--2019)}

A distinct shift in algorithmic performance occurred between 2014 and 2017, driven by the standardization of benchmark features. The defining characteristic of the CEC 2014 suite was the widespread application of dense rotation matrices to the base functions \cite{liang2013cec2014}. This transformation fundamentally altered the landscape topology by destroying separability. Mathematically, a function $f(\mathbf{x})$ is considered additively separable if it can be represented as the sum of $D$ functions, each depending on a single variable $x_i$ \cite{whitley1996evaluating}

\begin{equation}
    f(\mathbf{x}) = \sum_{i=1}^{D} f_i(x_i).
\end{equation}

In a separable landscape, the global optimum can be located by minimizing each dimension independently. However, the introduction of a dense orthogonal rotation matrix $\mathbf{M}$ creates a rotated coordinate system 
\begin{equation}
\label{eq:rotation}
    \mathbf{y} = \mathbf{M}(\mathbf{x} - \mathbf{o}),
\end{equation}
where $\mathbf{x}$ is the candidate solution vector and $\mathbf{o}$ is the shift vector. In this transformed space, every variable $y_i$ becomes a linear combination of all original parameters $x_j$ \cite{salomon1996reevaluating}. This introduces a strong variable linkage (epistasis), which means that the optimal value for one parameter depends entirely on the values of the others. This led to reduced effectiveness of coordinate-wise search strategies \cite{simpson1994genetic}.

This structural change created a performance barrier for coordinate-dependent algorithms. Canonical methods such as Genetic Algorithms and PSO often rely on operators like single-point crossover or axis-aligned velocity updates. In a rotated landscape, modifying a single parameter $x_i$ effectively alters all coordinates in the principal eigensystem, causing these component-wise strategies to degrade, as their performance is severely hindered due to the destruction of variable linkage, effectively reducing their efficiency to that of a random search \cite{salomon1996reevaluating}. While the CMA-ES is theoretically capable of learning the rotation $\mathbf{M}$ via its covariance matrix, it often struggles under the CEC competition constraints. The adaptation of the covariance matrix imposes a learning cost that scales quadratically with the dimension ($O(D^2)$) \cite{hansen2001completely}. This creates a significant "warm-up" phase where the algorithm consumes a disproportionate amount of the fixed evaluation budget before it can effectively exploit the learned topology.

Differential Evolution emerged as the dominant paradigm in this phase primarily because its difference vector mutation is rotationally invariant \cite{price2005differential}. This invariance is not merely a static property but a dynamic one: as the population converges, the difference vectors naturally align with the principal eigen-directions of the landscape's contour. This allows DE to implicitly approximate the second-order information of the Hessian \cite{das2010differential}, effectively "learning" the rotation—without the $O(D^2)$ learning cost that often causes CMA-ES to struggle under fixed evaluation budgets. While the binomial crossover typically breaks this invariance by swapping parameters along axis-aligned coordinates, the success of the L-SHADE \cite{tanabe2014improving} lineage  stems from its ability to adaptively sample crossover rates ($CR$) near 1.0. This adaptive feedback loop, managed by the Success-History mechanism, ensures that on rotated landscapes, the algorithm prioritizes the rotationally invariant mutation over the coordinate-dependent crossover. When coupled with Linear Population Size Reduction (LPSR), this created a robust framework that outperformed coordinate-dependent predecessors by maintaining search efficiency regardless of the coordinate system.

The most statistically significant trend in our dataset is the dominance of the L-SHADE family as we can see in Tab. \ref{tab:cec_results_lshade} and Fig. \ref{fig:bar}. This dominance was cemented between 2014 and 2016 \cite{liang2015problem} when the L-SHADE algorithm and its immediate derivatives: SPS-L-SHADE-EIG \cite{guo2015sps} and LSHADE-EpSin \cite{awad2016lshade}, achieved three consecutive wins. This success solidified LPSR and Success-History Adaptation as common components for any competitive solver. The dominance continued through 2020 with CAL-SHADE \cite{zamuda2017cal} and jDE100 \cite{brest2019jde100}, with the only notable exception being the victory of HS-ES \cite{zhang2018hybrid} in 2018. This represented a rare instance where a Covariance Matrix Adaptation approach outperformed Differential Evolution by structurally hybridizing with univariate sampling to directly address the extreme multimodality of the newly introduced composition functions.

\begin{table}[btp]
\centering
\footnotesize
\setlength{\tabcolsep}{3pt}
\caption{CEC 2014, CEC 2015 and CEC 2016 – Main Track Rankings}
\label{tab:cec_results_lshade}
\begin{tabular}{@{}c l l l@{}}
\toprule
Rank & CEC 2014 & CEC 2015 & CEC 2016 \\
\midrule
1 & L-SHADE \cite{tanabe2014improving} & SPS-L-SHADE-EIG \cite{guo2015sps} & LSHADE-EpSin \cite{awad2016lshade} \\
2 & GaAPADE \cite{wu2014gaapade} & DEsPA \cite{tanabe2015despa} & UMOEA-II \cite{elsayed2016umoea} \\
3 & MVMO-SH \cite{erlich2014evaluating} & LSHADE-ND \cite{polacek2015lshade} & iL-SHADE \cite{brest2016ilshade} \\
4 & OptBees \cite{maia2014optbees} & L-SHADE \cite{tanabe2014improving} & LSHADE44 \cite{polakova2016lshade44} \\
5 & RM-MEDA \cite{zhang2014rmmeda} & MVMO-PH \cite{rueda2015mvmo} & L-SHADE \cite{tanabe2014improving} \\
6 & UMOEAs \cite{elsayed2014umoeas} & DE\_b6e6rl \cite{tvrdik2015de} & ECL-L-SHADE \cite{li2016ecl} \\
7 & & SP-UCI \cite{awad2015spuci} & HGO-LSHADE \cite{bujok2016hgo} \\
\bottomrule
\end{tabular}
\end{table}

Results from the 2019 100-Digit Challenge \cite{Price2019} highlighted limitations in standard DE performance regarding ill-conditioned problems, such as the Inverse Hilbert Matrix (F2). They require extreme numerical precision (10+ digits) rather than a good enough basin finding. This challenge revealed that the dominance of L-SHADE in this phase had a blind spot: conditioning. The winners of this challenge had to adopt specific restarts or high-precision types (long double), which standard CEC winners rarely do. This highlights a limitation of these algorithms as general purpose solvers.

\begin{table}[tbp]
\centering
\scriptsize
\setlength{\tabcolsep}{2pt}
\caption{Algorithm Rankings Part 1: CEC 2017 -- 2019}
\label{tab:rankings_part1}
\begin{tabularx}{\textwidth}{@{}c l l l@{}}
\toprule
\textbf{Rank} & \textbf{CEC 2017} & \textbf{CEC 2018} & \textbf{2019 (100-D)} \\
\midrule
1 & EBOwithCMAR~\cite{kumar2017improving} & HS-ES~\cite{zhang2018hybrid} & jDE100~\cite{brest2019jde100} \\
2 & jSO~\cite{brest2017single} & LSHADE-ESP~\cite{stanovov2018lshade} & DISHchain1e+12~\cite{zamuda2019function} \\
3 & LSHADE-SPACMA~\cite{hadi2017lshade} & ELSHADE\_SPACMA~\cite{hadi2021enhanced} & HyDE-DF~\cite{lezama2019hybrid} \\
4 & LSHADE-cnEpSin~\cite{awad2017ensemble} & EBOwithCMAR~\cite{kumar2017improving} & SOMA T3A~\cite{diep2019soma} \\
5 & DES~\cite{awad2017cec} & UMOEAsII~\cite{sallam2018improved} & LSHADE-USM~\cite{molina2019applying} \\
6 & MM-OEDA~\cite{awad2017cec} & MVMO-PH~\cite{rueda2018hybrid} & SOMA Pareto~\cite{thanh2019pareto} \\
7 & IDEbestNsize~\cite{awad2017cec} & -- & rCIPDE~\cite{zhang2019restart} \\
8 & PPSO~\cite{awad2017cec} & -- & Co-Op~\cite{zhang2019cooperative} \\
9 & DYYPO~\cite{awad2017cec} & -- & DISH~\cite{viktorin2019dish} \\
10 & RB-IPOP-CMA-ES~\cite{biedrzycki2017version} & -- & jDE~\cite{brest2019jde100} \\
11 & -- & -- & mL-SHADE~\cite{yeh2019modified} \\
12 & -- & -- & GADE~\cite{epstein2019gade} \\
13 & -- & -- & CMEAL~\cite{bujok2019cooperative} \\
14 & -- & -- & HTPC~\cite{xu2019hybrid} \\
15 & -- & -- & UMDE-MS~\cite{fu2019univariate} \\
16 & -- & -- & DLABC~\cite{lu2019novel} \\
17 & -- & -- & MiLSHADE-LSP~\cite{molina2019applying} \\
18 & -- & -- & ESP-SOMA~\cite{kadavy2019ensemble} \\
\bottomrule
\end{tabularx}
\end{table}

\subsection{Phase 3: Structural Hybridization and Complexity (2020--2024)}

The evolution of benchmarks culminated in the CEC 2017 benchmark suite which established the standard for single-objective optimization through 2024. This suite solidified the use of Hybrid and Composition functions to test the plasticity of algorithms. Even more challenging are the Composition functions which merge multiple basic functions into a single semi-continuous landscape using bias values and weighted aggregation. To solve these, an algorithm cannot rely on a static strategy and must instead demonstrate strong global exploration to identify the correct basin of attraction followed by aggressive local exploitation to refine the solution.

\begin{table}[htbp]
\centering
\scriptsize
\setlength{\tabcolsep}{2pt}
\caption{Algorithm Rankings Part 2: CEC 2020 -- CEC 2024}
\label{tab:rankings_part2}
\begin{tabularx}{\textwidth}{@{}c l l l l@{}}
\toprule
\textbf{Rank} & \textbf{CEC 2020} & \textbf{CEC 2021} & \textbf{CEC 2022} & \textbf{CEC 2024} \\
\midrule
1 & IMODE~\cite{sallam2020imode} & NL-SHADE-RSP~\cite{stanovov2021nlshade} & EA4eigN100~\cite{ea4eign2022} & L-SRTDE~\cite{stanovov2024lsrde} \\
2 & AGSK~\cite{mohamed2020agsk} & jDE21~\cite{alic2021jde21} & NL-SHADE-LBC~\cite{nlshade2022lbc} & RDE~\cite{tao2024rde} \\
3 & j2020~\cite{alic2020j2020} & APGSK\_IMODE~\cite{gad2021apgsk} & NL-SHADE-RSP-MID~\cite{nlshade2022rsp} & BlockEA~\cite{qi2024blockea} \\
4 & OLSHADE~\cite{stanovov2020olshade} & DEDMNA~\cite{kumar2021dedmna} & S\_LSHADE\_DP~\cite{slshade2022dp} & mLSHADE~\cite{chauhan2024mlshade} \\
5 & jDE100e~\cite{alic2020jde100e} & MadDE~\cite{biswas2021madde} & jSObinexpEig~\cite{jso2022bin} & jSO~\cite{bujok2024jso} \\
6 & RASP-SHADE~\cite{salgotra2020rasp} & MLS-LSHADE~\cite{biswas2021mls} & MTT\_SHADE~\cite{mtt2022shade} & iEACOP~\cite{tangherloni2024ieacop} \\
7 & mpmL-SHADE~\cite{biswas2020mpml} & LSHADE~\cite{salgotra2021lshade} & IUMOEAII~\cite{iumoe2022ii} & -- \\
8 & SOMA\_CL~\cite{kadavy2020soma} & SOMA-CLP~\cite{kadavy2021soma} & IMPML-SHADE~\cite{impml2022shade} & -- \\
9 & MP-EEH~\cite{bolufe2020mp} & RB\_IPOP\_CMAES~\cite{rajabi2021rb} & NLSOMACLP~\cite{nlsoma2022clp} & -- \\
10 & DISH-XX~\cite{viktorin2020dish} & -- & ZOCMAES~\cite{zoc2022maes} & -- \\
11 & CSsin~\cite{salgotra2020cssin} & -- & OMCSOMA~\cite{omc2022soma} & -- \\
12 & -- & -- & Co-PPSO~\cite{coppso2022} & -- \\
13 & -- & -- & SPHI1\_Ensemble~\cite{sphi2022ensemble} & -- \\
\bottomrule
\end{tabularx}
\end{table}

The most recent competition data (2020--2024) indicates a transition from singular adaptive frameworks to highly complex hybridized algorithms. While the L-SHADE family relied on adapting control parameters within a fixed search logic, modern winners often function as multi strategy architectures that integrate distinct search mechanisms to maximize robustness across diverse landscape types. Societal Hybridization is exemplified by algorithms such as the Improved Multi-Operator Differential Evolution (IMODE) \cite{sallam2020imode}, which ranked first in CEC 2020, and the Adaptive Gaining-Sharing Knowledge Based Algorithm (AGSK) \cite{mohamed2020agsk}. These algorithms utilize mechanisms for exchanging information between different sub-populations to improve overall convergence. For example, in AGSK the population is partitioned into junior (exploration) and senior (exploitation) phases. 

NL-SHADE-RSP (CEC 2021 winner) argues that linear reduction is essentially a greedy resource allocation. They introduced a non-linear decay that preserves diversity longer in the middle stages of the search before forcing convergence. While traditional LPSR functioned primarily to force convergence, here we see focus on managing that convergence curve more delicately through \textit{Non-Linear Population Size Reduction} (NLPSR) because the functions became too complex for a straight linear collapse.

\textit{Eigenvector Crossover}, utilized by recent top-performers such as EA4eigN100 \cite{ea4eign2022} and L-SRTDE \cite{stanovov2024lsrde}, explicitly incorporates covariance learning. These methods periodically perform Principal Component Analysis (PCA) on the superior archive to extract eigen-coordinates, effectively borrowing the rotational invariance of CMA-ES to guide Differential Evolution mutation in ill-conditioned valleys without incurring the full computational cost of matrix adaptation.

However, this shift toward ensembles has introduced a challenge of \textit{Complexity Inflation}, where algorithms accumulate redundant components to minimize the risk of failure on specific outlier functions. The CEC 2022 winner, EA4eigN100, exemplifies this trend by aggregating four distinct optimization engines (CoBiDE, IDEbd, CMA-ES, and jSO) into a single framework \cite{ea4eign2022}. While this ensures high ranking stability, it raises concerns regarding structural efficiency as concluded in a rigorous ablation study by Biedrzycki \cite{Biedrzycki2024}. We discuss this phenomenon in more detail in Sec. \ref{sec:complexity-modern}.

This specialization is further challenged by Piotrowski \cite{piotrowski2023choice}, who identified a significant negative correlation between performance on modern synthetic benchmarks (CEC 2020) and classical real-world problems (CEC 2011). While over-parameterized hybrids often dominate recent competitions, they frequently underperform on the CEC 2011 set, a domain where less specialized, standard algorithms maintain superior robustness. This trade-off suggests a troubling divergence in evolutionary computation: modern solvers appear increasingly hyper-tuned to the specific topographies of high-budget synthetic suites, potentially sacrificing the generalizability required for messy, real-world engineering constraints.

\section{Discussion}
\label{sec:discussion}

The results of CEC presented in Section \ref{sec:results} reveal a distinct convergence towards DE architectures. To understand this dominance beyond simple ranking statistics, we must dissect the internal mechanisms that enabled this specific family of algorithms to outperform competing paradigms on standardized benchmarks. The sustained superiority of DE-based solvers warrants critical examination in light of the No Free Lunch (NFL) theorems \cite{Wolpert1997}. Since NFL states that no single optimizer offers superior performance across all possible problems, the consistent success of DE in CEC competitions implies that the benchmark suites exhibit specific structural biases, primarily non-separability and ruggedness, that the DE architecture is uniquely capable of exploiting.

We begin by contrasting the theoretical foundations of adaptive DE against model-based paradigms. A key distinction lies in the handling of landscape geometry: while CMA-ES relies on explicit probabilistic modeling via a covariance matrix, Differential Evolution operates as a model-free heuristic. In high-dimensional, rotated landscapes where the number of parameters required to estimate the covariance scales quadratically, DE’s implicit adaptation allows it to approximate search directions without the prohibitive computational overhead of matrix decomposition. This efficiency is critical under the finite evaluation budgets of the CEC competitions.

The following subsections analyze the architectural innovations that define the modern state-of-the-art. We first trace the evolution of the core "engine": Success-History Adaptation coupled with population reduction strategies, which dynamically balance exploration and exploitation. We then examine the community's specific responses to landscape threats, including the adoption of Eigenvector Crossover to manage coordinate rotation and the shift toward structural hybridization to navigate composite landscapes. Finally, we discuss the recent integration of Reinforcement Learning to automate operator selection, identifying these features not just as algorithmic novelties, but as necessary adaptations to the evolving difficulty of the test suites.

\subsection{Exploiting Separability and Real-World Topology (2010--2013)}

Prior to the 2014 standardization of benchmarks, competitive performance was driven by algorithmic diversity rather than the dominance of a single framework. The success of Memetic Algorithms in 2010 and Genetic Algorithms in 2011 challenges the current narrative: it suggests that Differential Evolution is not an inherently superior universal solver. Instead, its recent success may be a result of its architecture being specifically optimized for the 'rotated valley' topologies that have characterized standard benchmarks since 2014.

In the CEC 2010 Large Scale Global Optimization (LSGO) track, the winner was MA-SW-Chains \cite{Molina2010}, a memetic algorithm that hybridized steady-state genetic search with intensive local search chains. Its success was driven by the specific structure of the high-dimensional benchmarks (1000-D). At this scale, the primary difficulty is the ``curse of dimensionality'' rather than variable correlation. MA-SW-Chains succeeded by exploiting the partial separability of the functions, applying local search to subsets of variables (chaining) to refine solutions rapidly. The runner-up, a 2-Stage Ensemble by Wang et al. \cite{Wang2010}, utilized a similar logic by partitioning the population into exploring and exploiting groups. However, both strategies lack rotational invariance. When the CEC 2014 suite introduced dense rotation matrices to all functions, the strategy of optimizing variables in subsets became mathematically invalid, leading to the rapid decline of this memetic lineage in standard tracks.

The CEC 2011 competition stands as a unique anomaly in the dataset because it focused exclusively on Real-World Problems rather than synthetic functions. The winner, GA-MPC (Genetic Algorithm with Multi-Parent Crossover) \cite{Elsayed2011_GAMPC}, succeeded where Differential Evolution variants failed. A deeper look at the problem definitions reveals why. Problems like the Lennard-Jones Potential minimization create semi-convex structures that differ significantly from the random rotation of synthetic benchmarks. The multi-parent crossover of GA-MPC allowed it to mix features from multiple high-quality parents, effectively navigating these discrete-like combinatorial structures. The second-place algorithm, SAMODE \cite{Elsayed2011_SAMODE}, attempted to adapt DE to this domain but was outperformed by the pure GA approach on discrete-natured problems. This suggests that while DE/SHADE is superior for the abstract, non-separable landscapes of modern benchmarking, Canonical Genetic Algorithms remain highly competitive for physics-based engineering tasks.

The CEC 2013 competition marked the transition toward multimodal optimization. The winner, NEA2 (Niching Evolutionary Algorithm 2) \cite{Preuss2013}, utilized a nearest-better clustering mechanism to identify and maintain multiple basins of attraction simultaneously. This contrasted with the second-place dADE/nrand/1 \cite{Epitropakis2013}, which adapted Differential Evolution for niching by restricting mating to spatial neighborhoods. While dADE demonstrated that DE could be adapted for multimodality, NEA2's explicit topological clustering proved more robust for identifying the global optimum in highly multimodal landscapes without getting trapped in local optima.

The 2013 Real-Parameter Single Objective track further highlighted the need for dynamic strategy allocation. The winning algorithm, ICMAES-ILS \cite{Liao2013}, implemented a loosely coupled cooperative-competitive architecture. Rather than statically combining operators, the algorithm featured an initial competition phase where a restart-based CMA-ES ran in parallel with an Iterated Local Search algorithm utilizing multiple trajectory search. After this brief competition, the algorithm deployed whichever method yielded superior early results for the remainder of the computational budget. This dynamic algorithm selection allowed the solver to aggressively exploit simpler unimodal features using local search while reserving the computationally expensive matrix adaptation for non-separable functions where local search failed.

\subsection{Mechanism of Success-History and Success-Rate Adaptation}
\label{sec:sha_analysis}

While the change in benchmark functions of 2014 hindered coordinate-dependent algorithms, the mechanism that allowed Differential Evolution to permanently capture the leaderboard was the introduction of \textit{Success-History Adaptation} (SHA). First deployed in the CEC 2014 winner L-SHADE \cite{tanabe2014improving}, this mechanism transformed DE from a static heuristic into an adaptive learning system. Our analysis reveals that while peripheral mechanisms have fluctuated, the core adaptive architecture has remained ubiquitous in winning entries from 2014 through 2024. Its persistence suggests that SHA is not merely a feature but the fundamental engine that powers modern DE dominance.

To understand this dominance, one must look at the simplicity of the underlying operator. Despite the complexity of modern benchmarks, the most successful solvers continue to rely on the classical mutation strategy known as \textit{DE/rand/1}. As defined in \cite{das2010differential}, this fundamental operator generates a mutant vector $v_{i,G} \in \mathbb{R}^D$ for each target vector $ x_{i,G} \in \mathbb{R}^D $ via the linear combination
\begin{equation}
    v_{i,G} = x_{r1,G} + F \cdot (x_{r2,G} - x_{r3,G}),
    \label{eq:mutation}
\end{equation}
where $F \in \mathbb{R}$ is a static scaling factor and the indices $r_1, r_2, r_3 \in \{1, \dots, N\}$ are mutually distinct random integers, also different from $i$, chosen from a population of size $N$. The fundamental innovation of the SHADE architecture is the decoupling of these control parameters from the genome, linking them instead to a memory-based feedback loop. Unlike earlier self-adaptive approaches that encoded parameters directly into the individual, SHADE maintains an external historical memory consisting of archives $M_{CR}$ and $M_{F}$ of size $H$ \cite{tanabe2013success}. This structure records the control configurations that successfully generated improved solutions, effectively allowing the algorithm to learn the optimal step sizes for the current landscape features.

The parameter adaptation mechanism operates through a continuous cycle of sampling and weighted aggregation. For each individual, control parameters are sampled from the history using distributions that favor diversity, utilizing the Cauchy distribution for the scaling factor $F$ to allow for occasional long jumps.

A critical innovation appears in the update phase, where the memory utilizes the weighted Lehmer mean \cite{lehmer1971compounding} ($\operatorname{mean}_{WL}$) to aggregate the set of successful parameters $S_F$. To prioritize parameters that yield significant fitness gains, a weight $w_i$ is assigned to each successful individual $i \in S_F$ based on its normalized fitness improvement $\Delta f_i$, given by
\begin{equation}
    w_i = \frac{\Delta f_i}{\sum_{j \in S_F} \Delta f_j},
    \label{eq:weights}
\end{equation}
based on its normalized fitness improvement $\Delta f_i = |f(x_{i,G}) - f(v_{i,G})|$, where $f: \mathbb{R}^D \to \mathbb{R}$ is the objective function.
These weights are subsequently applied to calculate the weighted Lehmer mean, formulated as
\begin{equation}
    \operatorname{mean}_{WL}(S_F) = \frac{\sum_{i \in S_F} w_i \cdot S_{F,i}^2}{\sum_{i \in S_F} w_i \cdot S_{F,i}}.
    \label{eq:lehmer}
\end{equation}
Finally, this aggregated mean updates the specific memory slot $k$ via a moving average operation,
\begin{equation}
    M_{F,k}^{new} = c \cdot \operatorname{mean}_{WL}(S_F) + (1-c) \cdot M_{F,k},
    \label{eq:mem_update}
\end{equation}
which establishes a temporal feedback loop where successful search behaviors reinforce themselves. If the landscape currently favors exploration, high values of $F$ yield fitness improvements and subsequently populate the archive, biasing future samples toward larger steps.

A critical component that works in tandem with SHA is the linear population size reduction. Introduced in the original L-SHADE, this mechanism deterministically reduces the population size $N \in \mathbb{Z}^+$ relative to the cumulative number of function evaluations $\text{NFE} \in \mathbb{N}_0$ according to the linear decay function:

\begin{equation}N_{G+1} = \text{round} \left( \left( \frac{N_{min} - N_{init}}{\text{NFE}_{max}} \right) \cdot \text{NFE} + N_{init} \right),
\end{equation}

where $G$ denotes the current generation index. This formula ensures that the population for the subsequent generation, $N_{G+1}$, is calculated based on the cumulative computational effort $\text{NFE}$ expended up to that point. This creates a dynamic search topology. In the early phases where $N \approx N_{init}$, the high density of vectors maximizes the exploration capability of the difference vectors $(x_{r2} - x_{r3})$. As the search progresses and $N \to N_{min}$, the population naturally clusters, forcing the difference vectors to shorten and automatically transitioning the algorithm from global exploration to local exploitation \cite{tanabe2015despa}. This synergy explains why L-SHADE variants dominate: SHA adapts the \textit{step scale} $F$, while LPSR adapts the \textit{search density} ($N$).

This paradigm evolved further with the NL-SHADE family, specifically the CEC 2021 winner NL-SHADE-RSP \cite{stanovov2021nlshade} and the top-performing CEC 2022 variants like NL-SHADE-LBC. These algorithms refined the main mechanism by replacing linear reduction with NLPSR\cite{stanovov2021nlshade, zhou2024adaptive}. NLPSR keeps the population larger for a longer duration of the search compared to LPSR (see Eq. \eqref{eq:nlpsr}), delaying the convergence phase to avoid premature stagnation on the increasingly deceptive composition functions found in post-2020 suites via the population update rule $N_{G+1} \in \mathbb{Z}^+$ given by
\begin{equation}
\label{eq:nlpsr}
N_{G+1} = \text{round} \left( (N_{min} - N_{max}) \cdot \left( \frac{NFE}{NFE_{max}} \right)^{1 - \frac{NFE}{NFE_{max}}} + N_{max} \right).
\end{equation}
Furthermore, NL-SHADE-RSP integrated \textit{Rank-based Selective Pressure} (RSP). Unlike uniform selection for mutation indices $r_1, r_2$ in Equation \eqref{eq:mutation}, the RSP biases the selection of the base vector towards superior individuals using a ranking probability distribution \cite{stanovov2018lshade}. This injects a controllable selection pressure that helps guide the adaptive history ($M_F, M_{CR}$) toward high-quality regions of the search space more rapidly than random drift \cite{STANOVOV2019100463}. The dominance of NL-SHADE-RSP and NL-SHADE-LBC confirms that modern gains are achieved not by reinventing the operator, but by fine-tuning the \textit{flow} of genetic information (via population size and selection rank) fed into the SHA mechanism.

The most recent evolution in this lineage, exemplified by the CEC 2024 entry L-SRTDE \cite{stanovov2024lsrde}, marks a shift from ``History-based'' to ``Success Rate-based'' adaptation. While SHA relies on a lagging memory of specific successful values, the L-SRTDE algorithm couples the scaling factor $F$ directly to the current \textit{Success Rate} ($SR$) of the population, a real-time metric of landscape navigability \cite{stanovov2025lsrde}.

Following the empirical configuration in \cite{stanovov2024lsrde}, the mean for sampling $F$ is defined as a hyperbolic tangent function of the success rate $SR \in [0, 1]$:
\begin{equation}
mF = 0.4 + 0.25 \cdot \tanh(5 \cdot SR).
\end{equation}
This creates an immediate state-based response: when the algorithm is successful (high $SR$), $mF$ increases, encouraging aggressive steps (exploration). When success drops (low $SR$), $mF$ decreases, forcing smaller steps (exploitation) \cite{stanovov2024lsrde}. Additionally, the greediness of the mutation strategy ($pb$) is also dynamically coupled to $SR$, increasing selective pressure as the success rate drops \cite{stanovov2024lsrde}. This represents a simplification and sharpening of the original SHA concept: rather than maintaining a complex memory archive, the algorithm uses the population's current success state as a direct proxy for landscape difficulty.

\subsection{Structural Hybridization and Multi-Population DE}

Before the explosion of multi population architectures in 2020, the necessity of structural hybridization was demonstrated by the 2018 victory of the Hybrid Sampling Evolution Strategy \cite{zhang2018hybrid}. As discussed previously, the quadratic learning cost of the covariance matrix creates a severe bottleneck under strict competition evaluation budgets. While standard mitigations like periodic restarts or diagonalized covariance matrices attempt to reduce this mathematical overhead, they frequently cause premature convergence when faced with the highly deceptive composition functions introduced in 2017. To overcome this limitation, the 2018 winning architecture structurally hybridized the multivariate sampling of CMA-ES with a computationally cheap univariate sampling method derived from Estimation of Distribution Algorithms. By alternating between expensive covariance learning to exploit rotated valleys and cheap univariate sampling to maintain search diversity and escape multimodal traps, this approach proved that explicitly combining distinct search logics was the most effective way to deploy matrix adaptation within fixed budgets.

Algorithms such as IMODE \cite{sallam2020imode} and AGSK \cite{mohamed2020agsk} utilize Heterogeneous Multi-Population (HMP) frameworks and Deterministic Policy Learning to address composite optimization problems. The Adaptive Gaining-Sharing Knowledge (AGSK) framework \cite{mohamed2020agsk} implements a Dual-Phase Knowledge Sharing mechanism where the population is partitioned into acquisition and refinement stages. The ``Junior'' phase forces information acquisition from neighbors rather than global optima via
\begin{equation}
    x_{i,j}^{new} = x_{i,j} + k_F (x_{r,j} - x_{i,j}) + (x_{s,j} - x_{r,j}),
\end{equation}
where $x_r$ and $x_s$ represent distinct neighbor vectors and $k_F \in \mathbb{R}^+$ denotes the knowledge factor used to scale the experience gained during the sharing process. The ``Senior'' phase biases updates toward the sub-population's elite ($x_{best}$) and middle-class ($x_{better}$) vectors according to
\begin{equation}
    x_{i,j}^{new} = x_{i,j} + k_F (x_{best,j} - x_{i,j}) + (x_{better,j} - x_{best,j}).
\end{equation}

IMODE \cite{sallam2020imode} implements this heterogeneity as a Heterogeneous Modular Ensemble, dividing the population into independent islands that employ distinct operator strategies, such as rotation-invariant mutations versus diversity-preserving archival updates. This structure enables dynamic reallocation of computational resources to the sub-population exhibiting the highest localized success rate.

In the IMODE framework, the optimizer is modeled as a Markov Decision Process where the Q-table $Q(s, a)$ stores the expected long-term utility of taking action $a$ in state $s$. The state space $\mathcal{S}$ characterizes the evolutionary status, quantified by metrics such as population diversity and the remaining $\text{NFE}$, from which a state $s \in \mathcal{S}$ is sampled. Correspondingly, the action space $\mathcal{A}$ comprises the available mutation and crossover strategies, where an action $a \in \mathcal{A}$ denotes the selection of a specific operator. The reward signal $r_{t+1} \in \mathbb{R}$ quantifies the immediate benefit based on the fitness improvement rate relative to the computational cost incurred. Policy updates follow the Bellman equation

\begin{equation}
    Q(s_t, a_t) \leftarrow Q(s_t, a_t) + \alpha [r_{t+1} + \gamma \max_{a} Q(s_{t+1}, a) - Q(s_t, a_t)],
\end{equation}

where $\alpha \in [0, 1]$ is the learning rate governing the integration of new experience and $\gamma \in [0, 1]$ is the discount factor determining the importance of future rewards. This mechanism allows the agent to learn temporal policies that prioritize high-variance exploration in early search phases and transition to conservative local search as the budget nears depletion.

\subsection{The Bias of Rotational Invariance}

The marked decline of coordinate-dependent algorithms (such as Genetic Algorithms and PSO) after 2014 correlates with the introduction of aggressive coordinate rotation in benchmark construction (Eq. \ref{eq:rotation}). This transformation destroys the separability of the underlying base functions, introducing high correlation (linkage) between variables.

We can formally demonstrate why this transformation favors Differential Evolution over other paradigms. Consider a standard crossover operator (e.g., discrete recombination) used in Genetic Algorithms. It produces a child $\mathbf{x}'$ by selecting the $j$-th component from either parent $\mathbf{u}$ or $\mathbf{v}$. In the rotated landscape, the fitness depends on the vector $\mathbf{y} = \mathbf{M}\mathbf{x}$. Because $\mathbf{M}$ is dense, modifying a single coordinate $x_j$ independently alters all coordinates in the principal axis system $\mathbf{y}$. Unless the parents are aligned along the coordinate axes, this operation destroys the linkage information required to navigate the valley.

Conversely, Differential Evolution relies on a vector difference mutation. Let the mutant vector be generated as $\mathbf{v} = \mathbf{x}_{r1} + F \cdot (\mathbf{x}_{r2} - \mathbf{x}_{r3})$. To see how this behaves in the rotated landscape, we apply the transformation to the mutant

\begin{equation}
    \mathbf{y}_{v} = \mathbf{M}(\mathbf{x}_{r1} - \mathbf{o}) + F \cdot (\mathbf{M}(\mathbf{x}_{r2} - \mathbf{o}) - \mathbf{M}(\mathbf{x}_{r3} - \mathbf{o})),
\end{equation}
where $\mathbf{o} \in \mathbb{R}^D$ is the shift vector defining the location of the global optimum in the search space. Due to the linearity of matrix multiplication, this expands to
\begin{equation}
    \mathbf{y}_{v} = \mathbf{y}_{r1} + F \cdot (\mathbf{y}_{r2} - \mathbf{y}_{r3}).
\end{equation}

This equality demonstrates that the mutation operation in the rotated coordinate system $\mathbf{y}$ is mathematically identical to the operation in the original space $\mathbf{x}$. Geometrically, as the population converges into a narrow, diagonal valley, the difference vectors $(\mathbf{x}_{r2} - \mathbf{x}_{r3})$ automatically align with the principal eigen-directions of that valley. Consequently, the DE mutation operator possesses a property of rotational invariance, meaning its search distribution is independent of the coordinate system orientation \cite{price2005differential, das2010differential}.

However, this claim of implicit rotational invariance requires rigorous treatment because it strictly applies only to the mutation operator. Modern Differential Evolution variants typically employ binomial crossover after mutation to mix coordinates between the target vector $\mathbf{x}$ and the mutant vector $\mathbf{v}$ to generate a trial vector $\mathbf{u}$. This operation is mathematically defined for each dimension $j$ as

\begin{equation}
    u_j = \begin{cases} 
        v_j, & \text{if } \text{rand} \leq CR \text{ or } j = j_{rand} \\ 
        x_j, & \text{otherwise.} 
    \end{cases}
\end{equation}

Because this discrete recombination swaps parameters along the original coordinate axes, it inherently breaks the rotational invariance established by the difference vectors. Geometrically, binomial crossover constrains the trial vector $\mathbf{u}$ to the vertices of an axis-aligned hyper-rectangle defined by $\mathbf{x}$ and $\mathbf{v}$. In a rotated, non-separable landscape, even if $\mathbf{x}$ and $\mathbf{v}$ reside within a narrow, low-cost valley, their bounding Cartesian hyper-rectangle invariably intersects the steep, high-cost walls of the objective function. Consequently, mixing coordinates independently destroys the variable linkage required to navigate the valley, effectively yielding trial vectors with severely degraded fitness. Pure rotational invariance is only preserved when the crossover probability $CR$ approaches one.

Top performing solvers navigate this structural limitation through distinct adaptation strategies. Algorithms such as L-SHADE rely on implicit adaptation through success history memory \cite{tanabe2014improving}. When binomial crossover destroys fitness on rotated landscapes, only the trial vectors generated with high crossover probabilities survive, which naturally forces the memory to learn to sample crossover rates near one. Other variants take explicit control of this mechanism to ensure convergence. For instance, NL-SHADE-RSP linearly increases its crossover probability to exactly one by the end of the search to guarantee rotational invariance during the convergence phase \cite{stanovov2021nlshade}. The recent L-SRTDE algorithm maintains success rate based adaptation but repairs the crossover rate to prevent complete stagnation on separable components \cite{stanovov2024lsrde}. 

Alternatively, algorithms like EA4eigN100 address the problem by periodically applying principal component analysis (PCA) to the superior solutions \cite{ea4eign2022}. This extracts the eigenvectors of the rotated valley to form an orthogonal basis. Projecting the target and mutant vectors into this eigenbasis aligns the crossover hyper-rectangle with the principal axes of the landscape contour. This allows the algorithm to perform crossover in a temporarily transformed coordinate system, explicitly preserving linkage and restoring rotational invariance without completely abandoning the crossover operator \cite{ea4eign2022}.

The efficiency of these modern methods relies on strictly managing the cost of learning the search space structure. For algorithms integrating covariance learning, building a covariance matrix and extracting its eigenvectors is fundamentally an $\mathcal{O}(D^3)$ operation. If applied at every generation, this cubic scaling would rapidly exhaust the finite computational budget before convergence. While EA4eigN100 incorporates a full CMA-ES engine alongside its DE components, and L-SRTDE, jDE100e, SPS-L-SHADE-EIG, relies on a structurally lighter architecture using only occasional PCA-guided crossover, both algorithms ensure efficiency by triggering these expensive rotations strictly on demand \cite{ea4eign2022, alic2020jde100e}. The coordinate transformation is typically gated behind stagnation counters or specific interval thresholds, and is computed using a small archive of successful solutions rather than the entire population. By restricting the eigendecomposition to these specific conditions, these solvers maintain the linear $\mathcal{O}(D)$ scalability characteristic of standard Differential Evolution for most of the search, deploying the computationally intensive operations only when navigating severely ill-conditioned valleys.

\subsection{Complexity and Robustness in Modern Solvers}
\label{sec:complexity-modern}
The post-2020 phase (CEC 2021--2024) is characterized by a shift from monolithic algorithms to often \textit{highly complex hybrids}. Unlike earlier dominant methods such as jDE100 \cite{Price2019}, which relied on a single coherent search logic, modern winners function as "algorithm portfolios" that aggregate multiple optimization strategies to minimize the risk of convergence failure. This trend is driven by the increasing heterogeneity of the benchmark suites, where no single operator can effectively navigate the diverse mixture of unimodal, multimodal, and composition functions.

The CEC 2021 winner, NL-SHADE-RSP \cite{stanovov2021nlshade}, initiated this trend by hybridizing stochastic evolution with deterministic heuristics. It introduced rank-based selective pressure, which decouples survival from raw fitness values to maintain diversity, and coupled it with a \textit{Midpoint Target} strategy that exploits the geometric center of the best-so-far solutions. This hybridization was necessary to handle the "Composition" functions, where the landscape consists of multiple overlapping basins that trap standard greedy approaches. Similarly, the third-ranked APGSK-IMODE \cite{gad2021apgsk} merged Differential Evolution with the societal Gaining-Sharing Knowledge (GSK) framework, effectively running two distinct search behaviors, knowledge acquisition (exploration) and knowledge sharing (exploitation) in parallel.

This algorithm hybrid approach culminated in the CEC 2022 winner, EA4eigN100 \cite{ea4eign2022}, which represents the peak of algorithmic complexity in our dataset. The algorithm is an ensemble of four distinct optimizers: \textit{CoBiDE} (for coordinate adaptation), \textit{IDEbd} (for diversity), \textit{jSO} (for exploitation), and \textit{CMA-ES} (for covariance learning). The inclusion of CMA-ES specifically addresses the rotationally invariant valley structures described in second phase. By periodically extracting eigenvectors from the superior archive, the algorithm borrows the rotational invariance of CMA-ES to guide the mutation operators of the DE components.

However, this aggregation strategy introduces significant structural redundancy. A rigorous ablation study by Biedrzycki \cite{Biedrzycki2024} revealed that a simplified version of EA4eigN100, with the CMA-ES and CoBiDE components \textit{completely disabled}, achieved a higher ranking score than the official winning submission. By stacking multiple engines, these algorithms statistically smooth out performance variance across the 10-30 benchmark functions, ensuring a high average rank even if individual components are inefficient or redundant. This study also revealed that many winning algorithms contain hardcoded constants in their source code that are not listed in the accompanying paper. When Biedrzycki performed a sensitivity analysis, he found that these hidden parameters often had a higher impact on performance than the novel mechanisms described in the paper.

The most recent winner, L-SRTDE (CEC 2024) \cite{stanovov2024lsrde}, signals a potential refinement of this paradigm. While still highly complex, it shifts the adaptation mechanism from fitness magnitude to \textit{Success Rate (SR)}. Instead of measuring \textit{how much} a parameter set improves a solution (which can be misleading in ``steep'' basins), L-SRTDE measures the \textit{frequency} of success. This provides a more robust feedback signal for operator selection in the noisy, multimodal landscapes of the 2024 suite, suggesting that future innovations may lie in better signal processing of search history rather than simply adding more engines to the ensemble.

\subsection{Analysis of composition functions and hybridization post-2017}

To understand the structural shift from monolithic algorithms to the complex hybrid ensembles that dominate the current era of evolutionary computation, it is necessary to look beyond overall competition rankings and evaluate algorithmic performance at the per-function-class level. The introduction of the CEC 2017 benchmark suite marked a distinct turning point by heavily emphasizing highly rugged, non-separable composition functions. By tracking the relative performance of the top Differential Evolution, Evolution Strategy (ES/CMA), and Swarm families across specific landscape typologies and evaluation criteria from 2018 to 2024, we can explicitly quantify how these topological challenges acted as strict architectural filters, ultimately forcing the field toward hybridization.

To substantiate the claim that composition functions specifically drove the architectural shift toward hybridization, we conducted a targeted comparative analysis using the CEC 2018 50-D competition data (Table \ref{tab:function_class_2018}). Rather than an exhaustive meta-analysis of all functions across all years, this representative case study quantitatively isolates the performance of top distinct families (ES/CMA vs. DE vs. Hybrids) across the four primary landscape typologies. The data confirms that while DE frameworks excel on Unimodal landscapes and ES/CMA architectures dominate Hybrid/Multimodal functions, the rugged, overlapping nature of Composition functions ($F_{21}-F_{30}$) neutralized these individual advantages. This localized rank convergence explicitly justifies the evolutionary drive toward hybrid ensembles in subsequent years.

\begin{table}[htbp]
\centering
\scriptsize
\setlength{\tabcolsep}{3pt}
\caption{Average Rank per Function Class (CEC 2018, 50-Dimensions). Data demonstrates the performance shift across landscape complexities, justifying the evolutionary drive toward hybridization on Composition functions.}
\label{tab:function_class_2018}
\begin{tabular}{@{} l c c c c c @{}}
\toprule
\textbf{Function} & \makecell{\textbf{HS-ES}\\(ES/CMA)} & \makecell{\textbf{LSHADE-}\\\textbf{RSP}\\(DE)} & \makecell{\textbf{ELSHADE-}\\\textbf{SPACMA}\\(DE)} & \makecell{\textbf{UMOEAsII}\\(Hybrid)} & \makecell{\textbf{MVMO-}\\\textbf{PH}\\(Hybrid)} \\
\midrule
\textbf{Unimodal} ($F_1, F_3$) & 4.00 & \textbf{1.00} & \textbf{1.00} & 5.00 & \textbf{1.00} \\
\textbf{Multimodal} ($F_4-F_{10}$) & \textbf{1.57} & 3.14 & 3.14 & 2.29 & 4.71 \\
\textbf{Hybrid} ($F_{11}-F_{20}$) & \textbf{1.80} & 2.10 & 2.60 & 3.90 & 4.50 \\
\textbf{Comp.} ($F_{21}-F_{30}$) & 2.90 & \textbf{2.50} & 2.70 & 3.00 & 3.90 \\
\midrule
\textbf{Overall Rank} & \textbf{2.29} & 2.36 & 2.57 & 3.39 & 4.07 \\
\bottomrule
\end{tabular}
\end{table}

To further investigate performance boundaries, we analyzed the unique CEC 2019 ``100-Digit Challenge'' (Table \ref{tab:100_digit_2019}), which temporarily removed strict computational budget constraints to test pure accuracy limits. The data reveals that while classic benchmark topologies ($F_1-F_7, F_{10}$) are effectively ``solved'' by modern adaptive algorithms, highly conditioned landscapes like the Happy Cat function ($F_9$) act as harsh architectural filters. Notably, while DE-derived methods (jDE100, DISHchain1e+12) took the top spots, the SOMA T3A framework tied for third, proving that highly adaptive migrating topologies, when incorporating DE-like crossover and greedy selection behaviors, can remain highly competitive in unbound budget scenarios.

\begin{table}[htbp]
\centering
\scriptsize
\setlength{\tabcolsep}{3pt}
\caption{Average Correct Digits Found (CEC 2019, 100-Digit Challenge). The table highlights how the top algorithms easily solved $F_1-F_7, F_{10}$, leaving $F_8$ (Expanded Schaffer) and $F_9$ (Happy Cat) as the ultimate architectural filters. SOMA T3A emerges as the only non-DE algorithm in the top tier.}
\label{tab:100_digit_2019}
\begin{tabular}{@{} l c c c c c @{}}
\toprule
\textbf{Function Group} & \makecell{\textbf{jDE100}\\(DE-R1)} & \makecell{\textbf{DISHchain}\\\textbf{1e+12}\\(DE-R2)} & \makecell{\textbf{HyDE-DF}\\(DE-R3)} & \makecell{\textbf{SOMA T3A}\\(Swarm-R3)} & \makecell{\textbf{ESHADE-}\\\textbf{USM}\\(DE-R5)} \\
\midrule
\textbf{Solved} ($F_1\text{--}F_7, F_{10}$) & 10.00 & 10.00 & 10.00 & 10.00 & 10.00 \\
\textbf{Extr. Persist.} ($F_8$) & 10.00 & 10.00 & 10.00 & 10.00 & 2.00 \\
\textbf{Cond. Trench} ($F_9$) & 10.00 & 7.12 & 3.00 & 3.00 & 3.52 \\
\midrule
\textbf{Total (Max 100)} & \textbf{100.00} & 97.12 & 93.00 & 93.00 & 85.52 \\
\bottomrule
\end{tabular}
\end{table}

The 2021 CEC competition introduced parameterized benchmarking, allowing for the isolation of specific algorithmic vulnerabilities by toggling bias, shift, and rotation transformations. Analysis of these isolated transformations reveals a critical structural phenomenon in modern solver design: the ``center-bias'' illusion (Table \ref{tab:cec2021_transformations}). In non-shifted landscapes where the global optimum resides at the origin, heavily hybridized architectures such as APGSK-IMODE demonstrated overwhelming dominance. However, this success was artifactual, driven by center-biased operators rather than genuine landscape adaptability. Once the shift transformation was activated, displacing the optimum and enforcing a true test of rotational invariance, the leaderboard inverted entirely. Center-biased hybrids collapsed in the rankings, while refined SHADE lineages (specifically NL-SHADE-RSP) reclaimed the top position by demonstrating mathematically rigorous rotational invariance. Notably, pure covariance matrix frameworks (RB-IPOP-CMA-ES) failed to remain competitive across all configurations, cementing the necessity of DE-driven ensembles for modern non-separable landscapes.

\begin{table}[htbp]
\centering
\scriptsize
\setlength{\tabcolsep}{3pt}
\caption{Rank Inversion in CEC 2021 Parameterized Benchmarking. The table illustrates the ``Center-Bias'' effect. Center-biased hybrids (APGSK-IMODE) dominate when the optimum is at the origin (Non-Shifted). However, when the optimum is shifted and rotated, true rotationally invariant DE lineages (NL-SHADE-RSP) reclaim dominance, while pure CMA-ES (RB-IPOP-CMA-ES) fails consistently.}
\label{tab:cec2021_transformations}
\begin{tabular}{@{} l l c c @{}}
\toprule
\textbf{Algorithm} & \textbf{Family} & \makecell{\textbf{Non-Shifted Rank}\\(Opt. at Origin)} & \makecell{\textbf{Rotated/Shifted Rank}\\(True Invariance)} \\
\midrule
\textbf{APGSK-IMODE} & DE (Multi-Pop) & \textbf{1} & 4 \\
\textbf{MadDE}       & DE (Adaptive)  & 2 & 6 \\
\textbf{LSHADE}      & DE (SHADE)     & 3 & 8 \\
\midrule
\textbf{NL-SHADE-RSP}& DE (SHADE)     & 5 & \textbf{1} \\
\textbf{jDE21}       & DE (Adaptive)  & 7 & \textbf{2} \\
\textbf{DEDMNA}      & DE (Adaptive)  & 6 & \textbf{3} \\
\midrule
\textbf{RB-IPOP-CMA} & ES/CMA         & 9 & 9 \\
\bottomrule
\end{tabular}
\end{table}

The architectural evolution hypothesized in Phase 3 is ultimately validated by the design of the CEC 2022 benchmark suite. In response to the center-bias vulnerabilities exposed during the 2021 parameterization tests, the 2022 organizers unilaterally applied ``Shifted and Full Rotated'' transformations across all twelve benchmark functions. This structural decision eliminated coordinate-dependent shortcuts, transforming the entire competition into a strict filter for rotational invariance. Consequently, the 2022 leaderboard was completely dominated by advanced Differential Evolution frameworks and hybrid ensembles. Notably, the winning algorithm (EA4eigN100\_10) and the fifth-place algorithm (jSObinexpEig) explicitly integrated Eigenvector transformations into their crossover strategies to navigate the rotated valleys, while standard Swarm and pure CMA-ES frameworks fell to the bottom quartile. This confirms that for modern, fully rotated, and shifted topological suites, the integration of covariance or eigenvector mechanisms into a foundational DE search engine is no longer optional, but a prerequisite for competitive viability.

The 2024 CEC competition introduced a fundamental paradigm shift in performance evaluation by implementing the U-score (Trial-based dominance), a metric that captures the error value at regular intervals to holistically measure both convergence speed and final accuracy. Returning to the highly complex, rotated topologies of the CEC 2017 suite, the U-score effectively penalized architectures that were accurate but computationally sluggish, as well as those that converged rapidly to local optima. Analysis of the composition functions reveals that modern Differential Evolution lineages have mastered this speed-accuracy trade-off (Table \ref{tab:cec2024_composition}). The winning framework, L-SRDE, secured its rank not by dominating every individual landscape, but through extreme architectural consistency, rarely dropping below third place on any composition function. Conversely, alternative hybrids like EACOP exhibited severe polarization, oscillating violently between first and sixth place depending on the specific landscape. This confirms that the evolutionary trajectory of DE, from the rotational invariance established in Phase 2 to the dynamic, success-rate adaptation of Phase 3, has culminated in frameworks capable of consistently navigating non-separable, multi-basin topologies without sacrificing convergence efficiency.

\begin{table}[htbp]
\centering
\scriptsize
\setlength{\tabcolsep}{6pt}
\caption{U-Score Rankings on CEC 2024 Composition Functions. The U-score evaluates both convergence speed and final accuracy. The data demonstrates that the winning algorithm, L-SRDE, achieved victory through extreme consistency across non-separable topologies, whereas the hybrid EACOP exhibited severe polarization (alternating between 1st and 6th place).}
\label{tab:cec2024_composition}
\begin{tabular}{@{} l c c c c @{}}
\toprule
\textbf{Function} & \makecell{\textbf{L-SRDE}\\(DE-R1)} & \makecell{\textbf{RDE}\\(DE-R2)} & \makecell{\textbf{mLSHADE}\\(DE-R4)} & \makecell{\textbf{EACOP}\\(Hybrid-R6)} \\
\midrule
\textbf{$f_{21}$} & 3 & 4 & 2 & 6 \\
\textbf{$f_{22}$} & 3 & 2 & 5 & \textbf{1} \\
\textbf{$f_{23}$} & 3 & 4 & \textbf{1} & 2 \\
\textbf{$f_{24}$} & 3 & \textbf{1} & 2 & 6 \\
\textbf{$f_{25}$} & 3 & 6 & 5 & \textbf{1} \\
\textbf{$f_{26}$} & 3 & \textbf{1} & 2 & 6 \\
\textbf{$f_{27}$} & 5 & 2 & 6 & \textbf{1} \\
\textbf{$f_{28}$} & 2 & 6 & 5 & \textbf{1} \\
\textbf{$f_{29}$} & \textbf{1} & 3 & 2 & 6 \\
\textbf{$f_{30}$} & 3 & \textbf{1} & 2 & 6 \\
\midrule
\textbf{Behavior} & \textbf{Consistent} & High Var. & Moderate & \textbf{Polarized} \\
\bottomrule
\end{tabular}
\end{table}

In conclusion, the longitudinal, function-specific data from 2017 onward substantiates that the hybridization of search algorithms was not a random trend, but a necessary evolutionary response to benchmark design. The introduction of composition functions exposed the limitations of relying purely on standard difference vectors or isolated covariance matrices. Furthermore, as the CEC organizers actively patched evaluation loopholes, such as penalizing center-biased strategies with mandatory shift operators in 2021 and 2022, and enforcing convergence efficiency via the U-score in 2024, brittle hybrids were systematically filtered out. The empirical evidence hints at the fact that the only architectures capable of surviving this sustained topological pressure are deeply integrated, DE-driven ensembles. These modern frameworks maintain the robust, scalable exploration of the SHADE lineage while surgically deploying matrix-adapted or eigenvector-based operators to traverse non-separable composition valleys, ultimately achieving a highly consistent balance of speed and accuracy.

\section{CEC optimizers for Variational Quantum Algorithms}

In this section, we propose that modern evolutionary and swarm optimizers are structurally well-suited for hybrid classical-quantum computing architectures, such as VQAs \cite{mcclean2016theory, cerezo2020variational, TILLY20221, farhi2014quantum}. In this framework, the classical optimizer acts as a feedback controller for a quantum experiment. The candidate solution vector $\theta$ specifies physical control parameters, typically qubit rotation angles, within a Parameterized Quantum Circuit (PQC) or ansatz \cite{ostaszewski2021structure, benedetti2019parameterized}. This circuit prepares a quantum wave function $|\psi(\theta)\rangle$, which is subsequently measured against a problem-specific Hamiltonian $H$. The optimization objective is to minimize the expectation value of this energy measurement, $\langle\psi(\theta)|H|\psi(\theta)\rangle$, effectively "learning" the optimal quantum state to solve the underlying computational or machine learning problem.

Crucially, this quantum cost function exhibits pathological landscape features that mirror the most challenging classical CEC benchmarks \cite{BonetMonroig2023, Arrasmith2021, Failde2023, illesova2025statistical}. Because the fundamental principles of quantum mechanics dictate that the expectation value must be estimated through a finite number of experimental measurements (shots), the resulting sampling noise follows a binomial distribution. According to the de Moivre–Laplace theorem, a specific case of the Central Limit Theorem, this distribution converges to additive Gaussian noise as the number of shots becomes sufficiently large \cite{lin2021real}. Furthermore, as the expressibility of the PQC increases to model complex data, the optimization landscape frequently flattens into Barren Plateaus, vast regions where gradients vanish exponentially with the number of qubits \cite{cerezo2021cost, sim2019expressibility, bilek2025experimental}. 

This combination of stochastic fitness evaluations, hardware-induced errors, and exponentially flat regions renders standard finite-difference gradient methods highly unstable \cite{lewandowska2025benchmarking}. Consequently, the efficacy of VQAs relies entirely on the classical optimizer's ability to extract a precise learning signal from a noisy, non-convex topography. Advanced derivative-free solvers developed for the CEC domain, capable of maintaining search momentum through flat regions and filtering stochastic noise without relying on localized gradients, are fundamentally equipped to navigate these quantum bottlenecks. This universal requirement underscores the necessity of transferring complex CEC solvers to the quantum domain to train models amid Barren Plateaus and persistent hardware noise, as summarized in Table \ref{tab:vqa_apps}.

\begin{table}[htbp]
\centering
\scriptsize
\setlength{\tabcolsep}{3pt}
\caption{VQA Optimization Under Quantum Noise}
\label{tab:vqa_apps}
\begin{tabularx}{\textwidth}{@{} >{\raggedright\arraybackslash}p{1.8cm} X X X @{}}
\toprule
\textbf{Domain} & \textbf{Methods} & \textbf{Noise Sources} & \textbf{Key Benchmarks} \\
\midrule
Quantum Chemistry & VQE, ADAPT-VQE, SAOOVQE & Gate infidelity, decoherence, shot noise & Chem. accuracy (1.6 mHa) on SC \cite{10.1063/5.0161057, PRXQuantum.2.020310} \\
\addlinespace[3pt]
Optimization & QAOA, RED-QAOA, Recursive & Circuit depth, spectral gaps & Advantage over random; warm-starting \cite{PhysRevA.109.032408} \\
\addlinespace[3pt]
Machine Learning & QCNN, VQC, Hybrid Transfer & Barren plateaus, sampling noise & High effective dim. vs. classical NN \cite{buonaiuto2024effects} \\
\addlinespace[3pt]
Finance & VQE for QUBO, Portfolio Opt. & Landscape roughness, volatility & Arbitrage/portfolio on NISQ \cite{quantum2024finance} \\
\addlinespace[3pt]
Materials & Ham. Simulation, Periodic VQE & Many-body complexity, discretization & Ground state for correlated models \cite{cerezo2021variational} \\
\addlinespace[3pt]
Early Fault Tolerance & EFT-VQA, PEC, pQEC & Physical error rate ($\sim0.1\%$), QEC overhead & 9.27x fidelity gain via pQEC \cite{10.1145/3695053.3731112, PhysRevResearch.6.023118} \\
\bottomrule
\end{tabularx}
\end{table}

\begin{figure}[!t]
    \centering
    \includegraphics[width=1\linewidth]{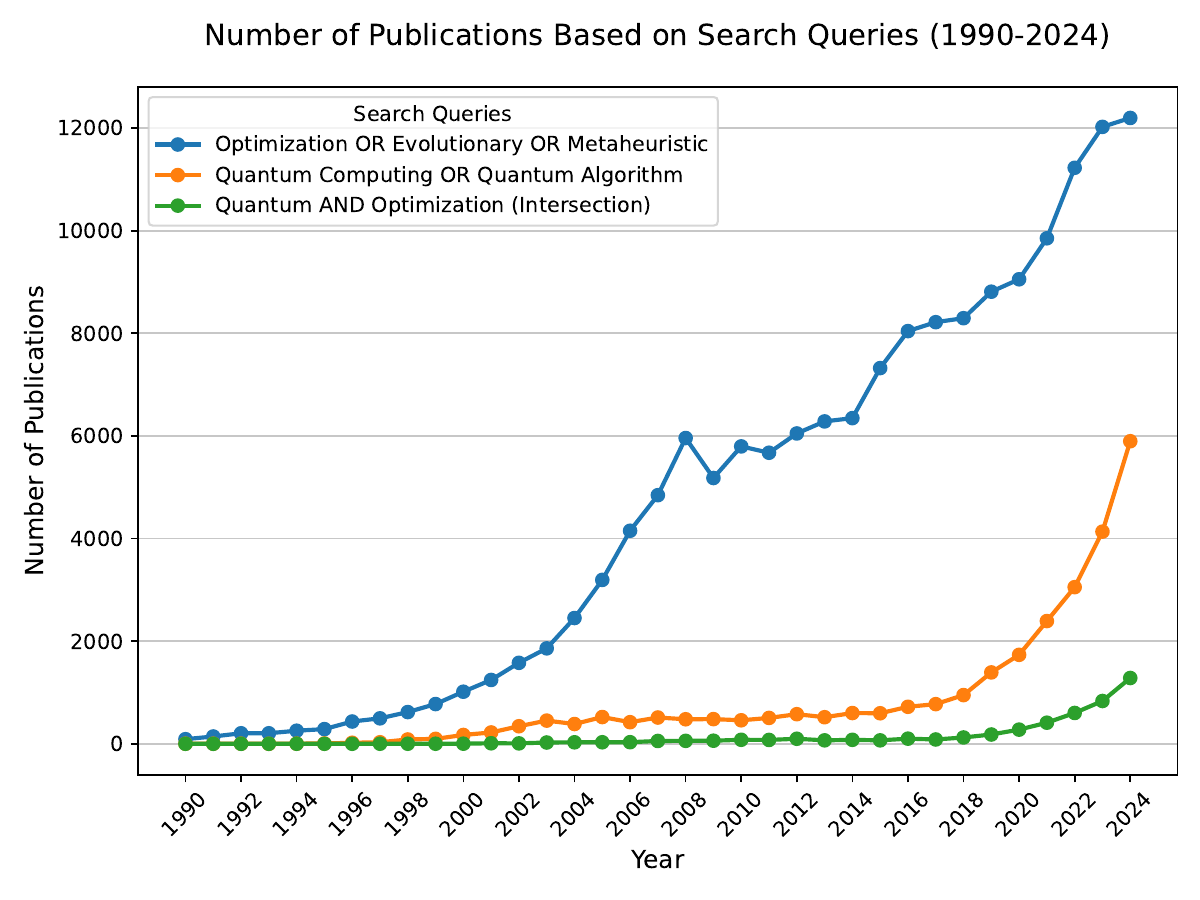}
    \caption{Publication volume across three primary search clusters. The blue line establishes the baseline of the optimization community. The orange line tracks the ``Quantum Revolution'', starting around 2015-2016 when NISQ devices became a major research focus. The green line illustrates the intersection of quantum computing and optimization.}
    \label{fig:scopus}
\end{figure}

This robust optimization capability will determine whether current NISQ hardware can achieve quantum supremacy. The rapid growth and contemporary relevance of this intersection between quantum computing and optimization are evidenced by the bibliometric data shown in Figure~\ref{fig:scopus}, which reveals a significant increase in publication volume. For the last three years (2023–2025), the field exhibits a significant preprint-to-document ratio ($R \approx 0.64$), indicating that a substantial portion of the state-of-the-art is currently circulating via preprint servers prior to formal publication.

\subsection{VQAs and QML optimization landscapes}

To understand the theoretical basis for this cross-domain transferability, one must examine the specific topological features of VQA cost landscapes, which emerge from the interplay between the problem Hamiltonian and the parameterized ansatz \cite{kandala2017hardware, vha}. While simple systems like molecular hydrogen or small transverse Ising models often yield unimodal surfaces due to low entanglement entropy, complexity scales rapidly with system size. Intermediate problems, such as lithium hydride \cite{boy2025energy} or larger spin chains \cite{medina2024variational}, introduce local minima driven by electron correlation or frustration. In high-complexity regimes like the Hubbard model \cite{anselme2022simulating, novak2025optimization} or frustrated 2D Ising systems \cite{uvarov2020variational}, the landscape becomes highly multimodal, containing hundreds of distinct basins arising from competing energy scales and ground state degeneracy. This structural ruggedness mirrors the composite function benchmarks of CEC 2017--2022, where global search capabilities are a prerequisite for survival.

\begin{figure}[!t]
    \centering
    \includegraphics[width=1\linewidth]{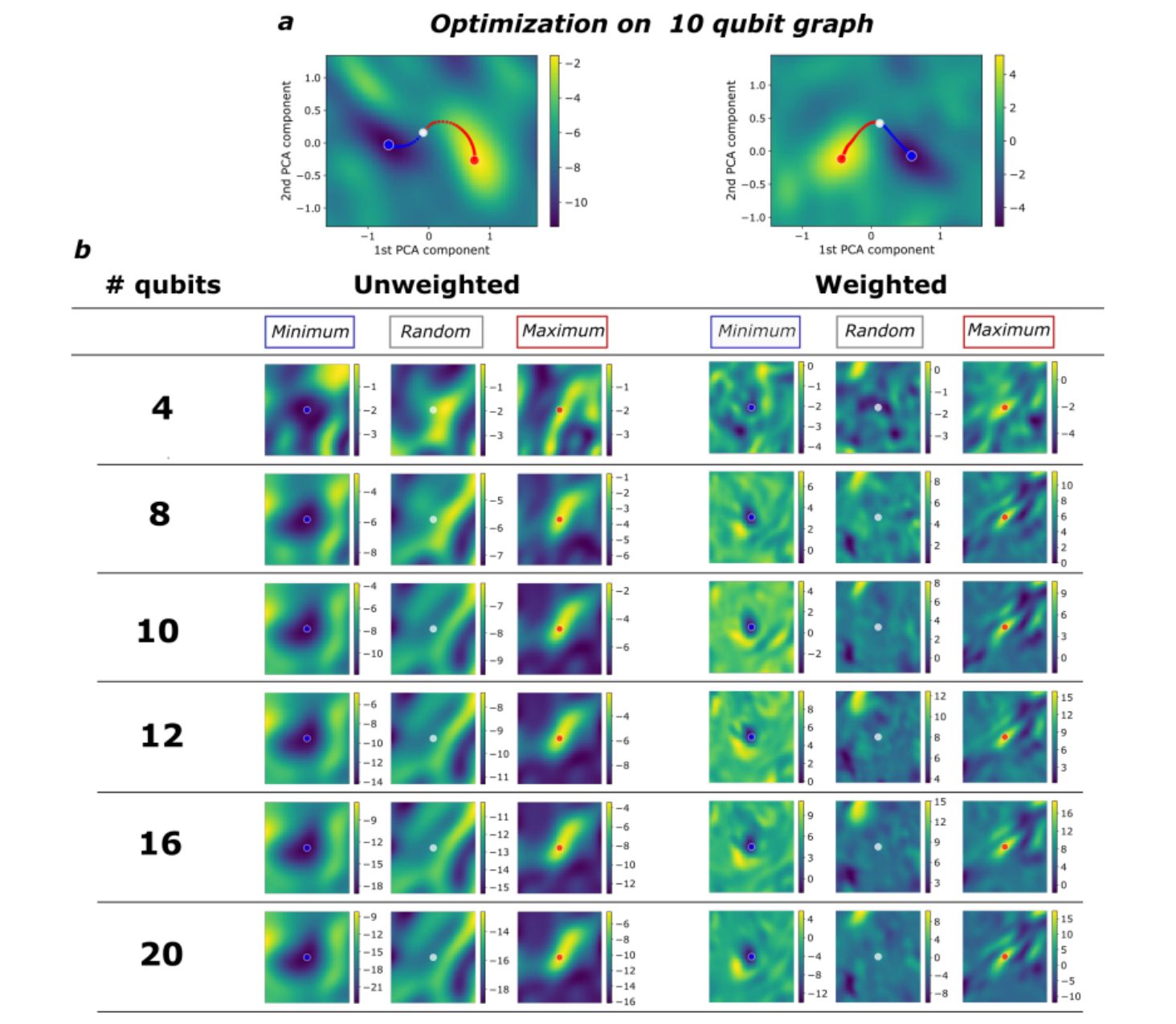}
    \caption{Parameter concentration in QAOA for Max-Cut on 3-regular graphs with unweighted ($w_{ij} = 1$) and weighted ($w_{ij} \in \{\pm 1, \pm 2, \pm 3\}$) edges using circuit depth $p = 4$. (a) PCA visualization of optimization trajectories identifying local minimum (blue), maximum (red), and random initialization (white) points on a 10-qubit graph instance. Panel (b) reveals that the local minima (left columns) possess non-separable, rotated valley structures that persist across increasing problem dimensions ($N=4$ to $20$). Source: \cite{rudolph2021orqvizvisualizin}}
    \label{fig:qaoa-landscape}
\end{figure}

Empirical visualization of QAOA \cite{farhi2014quantum, bezdek2025classical, trovato2025preliminary} cost landscapes confirms the structural similarities with CEC benchmarks. As shown in Fig. \ref{fig:qaoa-landscape} (b), the basins of attraction (see 'Minimum' columns) frequently exhibit diagonal, ellipsoidal topologies. In the taxonomy of CEC, this represents a non-separable, ill-conditioned landscape where variable linkage prevents successful optimization via coordinate-wise methods. This diagonal orientation causes the use of rotationally invariant operators, such as the difference vector mutation in Differential Evolution or the covariance learning in CMA-ES, which can adapt to the principal eigen-directions of the valley.

The navigability of this topology is further dictated by the choice of ansatz. Symmetry-preserving architectures, such as UCCSD \cite{grimsley2019trotterized, ciaramelletti2025detecting} for fermions, constrain the search to physically relevant paths and effectively suppress multimodality. In contrast, hardware-efficient or generic ansatze often suffer from overparameterization \cite{illesova2025qmetric} that creates redundant symmetries and elongated valleys. This structural mismatch can lead to artificial saddle points or Barren Plateaus where gradients vanish exponentially. Consequently, while adaptivity can prune irrelevant parameters to simplify the landscape, generic fixed-topology circuits often require robust global search strategies to distinguish the true ground state from sub-optimal high-energy traps, a challenge analogous to the ``deceptive'' basins found in CEC composition functions.

\begin{table}[htbp]
\centering
\scriptsize 
\setlength{\tabcolsep}{3pt}
\caption{CEC Benchmark Features vs. Quantum landscapes}
\label{tab:cec_quantum_map}
\begin{tabularx}{\textwidth}{@{} >{\bfseries}l X X >{\raggedright\arraybackslash}p{3cm} @{}}
\toprule
\textbf{CEC Feature} & \textbf{Quantum Analog} & \textbf{Physical Origin} & \textbf{Algorithmic Need} \\
\midrule
Multimodal & Rugged landscapes, local traps & Frustrated interactions (Hubbard, Spin Glass) & Global search (DE) to escape local basins \\
\addlinespace[3pt]
Non-Sep. & Parameter Correlation (Entanglement) & Non-commuting gates; $\theta$ dependencies & Rotation-invariant opt. (CMA-ES, DE) \\
\addlinespace[3pt]
Ill-Cond. & Barren Plateaus, Narrow Gorges & Deep ansatz; Lie algebra redundancies & Hessian/Covariance-based navigation \\
\addlinespace[3pt]
Noisy & Stochastic Shot Noise & Finite sampling & Noise-tolerant search \\
\bottomrule
\end{tabularx}
\end{table}

VQAs and Quantum Machine Learning \cite{kumar2021multi, novak2025predicting, illesova2025importance, illesova2025complementarity} optimization landscapes are characterized by high parameter correlation (entanglement), rugged local minima, and statistical measurement noise. Arrasmith et al.~\cite{Arrasmith2021} rigorously demonstrated that standard gradient-free optimizers (e.g., Nelder-Mead, COBYLA) are insufficient for these tasks; their analysis proved that cost function differences are exponentially suppressed in Barren Plateaus. This creates a necessity for global search mechanisms capable of navigating flat landscapes through advanced exploration.

Empirical benchmarks confirm that the specific hybridization strategies dominating the CEC competitions combining DE with CMA-ES are the superior solution for these quantum landscapes. Bonet-Monroig et al.~\cite{BonetMonroig2023} identified that while traditional methods struggle, CMA-ES excels at handling the noisy, non-convex terrain of quantum chemistry problems when properly tuned. Complementing this, Failde et al.~\cite{Failde2023} provided evidence using the Variational Quantum Eigensolver (VQE) that DE is essential for preventing premature convergence; in their experiments on 1D Ising models, DE maintained a 100\% success rate where local optimizers (SLSQP, L-BFGS-B) failed significantly as system size increased. These findings suggest that the ``ensemble'' architectures of recent CEC winners, which blend global DE exploration with local covariance refinement, are not just algorithmic curiosities but are chemically pure analogs to the solvers required for the next generation of quantum computing.

\subsection{Non-separability and quantum entanglement}

Beyond multimodality, quantum objective functions are characterized by intrinsic non-separability. Unlike standard CEC benchmarks where parameter linkage is artificially applied via dense rotation matrices, quantum non-separability arises directly from the sequence of highly entangling, non-commuting unitary gates. 

In continuous optimization, a function $E(\boldsymbol{\theta})$ is additively separable if and only if all mixed partial derivatives are zero
\begin{equation}
\frac{\partial^2 E}{\partial \theta_i \partial \theta_j} = 0.
\end{equation}
If this condition holds, parameters can be optimized independently without the need for complex linkage learning. In VQAs, the objective function is the expectation value of a Hamiltonian $H$, defined as $E(\boldsymbol{\theta}) = \langle \psi_0 | U^\dagger(\boldsymbol{\theta}) H U(\boldsymbol{\theta}) | \psi_0 \rangle$. Consider a minimal two-parameter circuit where $U(\theta_1, \theta_2) = U_2(\theta_2) U_{ent} U_1(\theta_1)$, with $U_i(\theta_i) = e^{-i \theta_i P_i / 2}$ representing local Pauli rotations on distinct qubits, and $U_{ent}$ representing an intermediate entangling gate.

The structural linkage between $\theta_1$ and $\theta_2$ is governed by the nested commutator of their generators \cite{mcclean2018barren}. The mixed partial derivative, evaluated near the origin, is proportional to
\begin{equation}
\frac{\partial^2 E}{\partial \theta_1 \partial \theta_2} \propto \langle \psi_0 | \left[ P_1, U_{ent}^\dagger [P_2, H] U_{ent} \right] | \psi_0 \rangle.
\end{equation}
If the circuit lacks entanglement ($U_{ent} = I$), the local operators commute ($[P_1, P_2] = 0$), the mixed derivative vanishes, and the parameters remain strictly separable. However, when $U_{ent}$ is an entangling gate (e.g., a CNOT), it propagates the operator $P_2$ across multiple qubits. The conjugated operator $U_{ent}^\dagger P_2 U_{ent}$ becomes a multi-qubit Pauli string that no longer commutes with $P_1$. Consequently, the mixed partial derivative becomes non-zero ($\frac{\partial^2 E}{\partial \theta_1 \partial \theta_2} \neq 0$), mathematically guaranteeing parameter linkage.

This physical entanglement manifests mathematically as strong cross-correlations between parameters; the optimal setting for a deep-layer parameter $\theta_k$ is functionally dependent on the state prepared by all preceding layers $\theta_{1 \dots k-1}$. In quantum machine learning, this highly correlated geometry of the quantum state space $\mathbb{CP}^{N-1}$ \cite{Bengtsson2017} is formally captured by the Quantum Fisher Information Matrix (QFIM), representing the Fubini-Study metric tensor \cite{Meyer2021}. It is critical to distinguish this state-space geometry from the topography of the cost landscape itself. Unlike in classical statistical learning, where the Fisher Information Matrix approximates the Hessian of the loss, the quantum Fisher Information does not directly measure the curvature of the objective function. Instead, it defines the steepest descent direction with respect to the underlying quantum information geometry.

To navigate these highly entangled parameter spaces, gradient-based methods such as the Quantum Natural Gradient (QNG) \cite{Stokes2020} precondition descent steps using the QFIM. While computing the exact, dense QFIM on NISQ hardware is computationally demanding, block-diagonal approximations effectively reduce this overhead to a single quantum evaluation per parametrized layer. Nevertheless, even with these efficient metric approximations, local first-order methods remain inherently susceptible to the exponential vanishing of gradients (barren plateaus) and severe statistical shot noise typical of deep or noisy ansätze. Consequently, modern derivative-free CEC optimizers offer a highly robust alternative \cite{wang2021noise}. By computing the empirical covariance matrix of successful trial vectors, algorithms employing CMA or Eigenvector Crossover explicitly approximate the inverse Hessian of the entangled cost landscape, rather than the geometry of the state space. This allows the optimizers to dynamically extract the principal eigen-directions of the parameter space and rotate their search operators to match the entanglement-induced topography, restoring rotational invariance without relying on explicit, noise-corrupted gradient or metric tensor calculations.

\subsection{Optimization Limitations: Barren Plateaus and Hardware Constraints}

While modern CEC benchmarks artificially induce parameter non-separability using dense orthogonal rotation matrices, VQA landscapes differ fundamentally; their non-separability arises intrinsically from the sequence of non-commuting unitary gates within the PQC \cite{fontana2022non}. Furthermore, VQA optimization faces a critical bottleneck absent in classical continuous suites: gradient concentration scaling, known as the Barren Plateau (BP) phenomenon \cite{larocca2025barren, mcclean2018barren}. As problem dimensionality increases, highly expressive ansatzes explore the exponentially large Hilbert space in an unbiased manner, causing the loss function and its partial derivatives to concentrate exponentially around their mean \cite{holmes2022connecting}. For CEC-evolved optimizers relying on difference vectors or covariance matrix updates, this exponential flatness dictates that resolving a true descent direction requires an exponentially scaling number of measurement shots to overcome statistical sampling noise \cite{larocca2025barren}. Consequently, an evolutionary optimizer's behavior is strictly bounded by ansatz expressibility, establishing a topological barrier that classical algorithms do not encounter in standard CEC competitions \cite{larocca2025barren}.

Beyond statistical shot noise, physical hardware noise dictates the optimization landscape by inducing deterministic Barren Plateaus \cite{wang2021noise}. When circuits are subjected to noise channels with the maximally mixed state as a fixed point, the quantum state exponentially decays as circuit depth increases \cite{wang2021noise}. This noise-induced concentration bounds the loss function deviation to an exponentially small regime for all parameter values, independent of the noiseless landscape topology \cite{larocca2025barren}. Optimizers are thus forced to navigate features suppressed by physical decoherence. 

Optimization is further constrained by discrete architectural hyperparameters, including circuit depth and physical qubit connectivity. Hardware-efficient ansatzes utilize specific entangling gate topologies to minimize compilation overhead on restrictive near-term devices \cite{kandala2017hardware, gupta2022how}. While restricting these architectures to shallow depths can prevent probabilistic Barren Plateaus, it introduces severe reachability deficits where the constrained parameter space excludes the target solution \cite{akshay2020reachability, larocca2023theory}. Furthermore, under-parameterized shallow circuits exhibit exponentially many spurious local minima, preventing optimizers from converging even when gradient signals remain non-vanishing \cite{bittel2021training}.

\subsection{Proof-of-Concept: Navigating Underparameterized VQA Topologies}

\begin{figure}[htbp]
    \centering
\scalebox{0.6}{
\Qcircuit @C=0.2em @R=0.2em @!R { \\
	 	\nghost{{q}_{0} :  } & \lstick{{q}_{0} :  } & \gate{\mathrm{R_Y}\,(\mathrm{{\ensuremath{\theta}}[0]})} & \ctrl{1} & \gate{\mathrm{R_Y}\,(\mathrm{{\ensuremath{\theta}}[10]})} & \qw & \qw & \qw & \qw & \qw & \qw & \qw & \qw & \qw & \qw\\
	 	\nghost{{q}_{1} :  } & \lstick{{q}_{1} :  } & \gate{\mathrm{R_Y}\,(\mathrm{{\ensuremath{\theta}}[1]})} & \targ & \ctrl{1} & \gate{\mathrm{R_Y}\,(\mathrm{{\ensuremath{\theta}}[11]})} & \qw & \qw & \qw & \qw & \qw & \qw & \qw & \qw & \qw\\
	 	\nghost{{q}_{2} :  } & \lstick{{q}_{2} :  } & \gate{\mathrm{R_Y}\,(\mathrm{{\ensuremath{\theta}}[2]})} & \qw & \targ & \ctrl{1} & \gate{\mathrm{R_Y}\,(\mathrm{{\ensuremath{\theta}}[12]})} & \qw & \qw & \qw & \qw & \qw & \qw & \qw & \qw\\
	 	\nghost{{q}_{3} :  } & \lstick{{q}_{3} :  } & \gate{\mathrm{R_Y}\,(\mathrm{{\ensuremath{\theta}}[3]})} & \qw & \qw & \targ & \ctrl{1} & \gate{\mathrm{R_Y}\,(\mathrm{{\ensuremath{\theta}}[13]})} & \qw & \qw & \qw & \qw & \qw & \qw & \qw\\
	 	\nghost{{q}_{4} :  } & \lstick{{q}_{4} :  } & \gate{\mathrm{R_Y}\,(\mathrm{{\ensuremath{\theta}}[4]})} & \qw & \qw & \qw & \targ & \ctrl{1} & \gate{\mathrm{R_Y}\,(\mathrm{{\ensuremath{\theta}}[14]})} & \qw & \qw & \qw & \qw & \qw & \qw\\
	 	\nghost{{q}_{5} :  } & \lstick{{q}_{5} :  } & \gate{\mathrm{R_Y}\,(\mathrm{{\ensuremath{\theta}}[5]})} & \qw & \qw & \qw & \qw & \targ & \ctrl{1} & \gate{\mathrm{R_Y}\,(\mathrm{{\ensuremath{\theta}}[15]})} & \qw & \qw & \qw & \qw & \qw\\
	 	\nghost{{q}_{6} :  } & \lstick{{q}_{6} :  } & \gate{\mathrm{R_Y}\,(\mathrm{{\ensuremath{\theta}}[6]})} & \qw & \qw & \qw & \qw & \qw & \targ & \ctrl{1} & \gate{\mathrm{R_Y}\,(\mathrm{{\ensuremath{\theta}}[16]})} & \qw & \qw & \qw & \qw\\
	 	\nghost{{q}_{7} :  } & \lstick{{q}_{7} :  } & \gate{\mathrm{R_Y}\,(\mathrm{{\ensuremath{\theta}}[7]})} & \qw & \qw & \qw & \qw & \qw & \qw & \targ & \ctrl{1} & \gate{\mathrm{R_Y}\,(\mathrm{{\ensuremath{\theta}}[17]})} & \qw & \qw & \qw\\
	 	\nghost{{q}_{8} :  } & \lstick{{q}_{8} :  } & \gate{\mathrm{R_Y}\,(\mathrm{{\ensuremath{\theta}}[8]})} & \qw & \qw & \qw & \qw & \qw & \qw & \qw & \targ & \ctrl{1} & \gate{\mathrm{R_Y}\,(\mathrm{{\ensuremath{\theta}}[18]})} & \qw & \qw\\
	 	\nghost{{q}_{9} :  } & \lstick{{q}_{9} :  } & \gate{\mathrm{R_Y}\,(\mathrm{{\ensuremath{\theta}}[9]})} & \qw & \qw & \qw & \qw & \qw & \qw & \qw & \qw & \targ & \gate{\mathrm{R_Y}\,(\mathrm{{\ensuremath{\theta}}[19]})} & \qw & \qw\\
\\ }}
\caption{Hardware-efficient ansatz with $L=1$ entangling layer for $N=10$ qubits. The circuit utilizes $R_y$ rotation gates and a linear CNOT staircase to produce a 20-dimensional parameter space.}
    \label{fig:ansatz_circuit}
\end{figure}
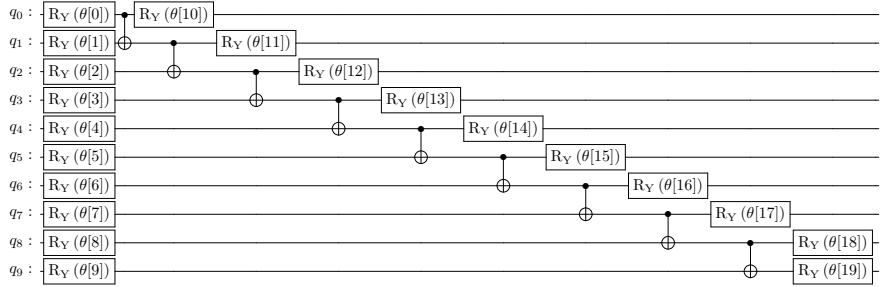

To explicitly evaluate the transferability of CEC competition optimization strategies to quantum landscapes, we present a proof-of-concept classical simulation of a 10-qubit VQE. While this scale does not induce the exponential gradient vanishing characteristic of true Barren Plateaus, it effectively models the spatial periodicity and spurious local minima inherent to underparameterized quantum models. The objective function minimizes the ground state energy of a 1D ferromagnetic Ising model without an external magnetic field, defined by the Hamiltonian
\begin{equation}
 H = -\sum_{i=1}^{N-1} Z_i Z_{i+1},
\end{equation}
with a global minimum of $E = -9.0$. The landscape is mapped using a hardware-efficient ansatz, shown in Fig. \ref{fig:ansatz_circuit}, restricted to $L=1$ entangling layer to strictly enforce an underparameterized regime. Each block consists of independent single-qubit rotations, $R_y(\theta)$, followed by a staircase of linear CNOT gates matching standard adjacent qubit connectivity. This $L=1$ restriction minimizes circuit depth to mitigate gate infidelity and decoherence on near-term hardware. However, this underparameterization, combined with the $2\pi$ periodicity of the $R_y(\theta)$ gates, induces a highly multimodal landscape. This configuration serves to evaluate if metaheuristic optimizers can navigate the resulting local minima and isolate the global ground state despite the restricted parameter density required for noise-resilient execution. This shallow architecture yields a 20-dimensional parameter space.

The optimization suites are evaluated under two discrete regimes: an exact statevector simulation and a noisy simulation. The noisy regime injects additive Gaussian noise proportional to the statistical variance of finite sampling, modeled as $\mathcal{N}(0, 1/\sqrt{N_{shots}})$ with $N_{shots} = 1024$. This methodology isolates the algorithmic robustness of coordinate-dependent updates and covariance matrix adaptations (e.g., L-SHADE, CMA-ES) when navigating non-separable, flat landscapes corrupted by shot-induced stochasticity.

\begin{table}[ht]
\centering
\caption{Results (Mean Best Energy $\pm$ Std and Average Runtime over 20 runs) on the 10-Qubit $L=1$ Ising VQA Landscape over 100,000 FEs. Global Ground State is $E = -9.0$.}
\label{tab:vqa_results}
\resizebox{\textwidth}{!}{
\begin{tabular}{l c c c c}
\toprule
\textbf{Optimizer} & \multicolumn{2}{c}{\textbf{Exact Simulation}} & \multicolumn{2}{c}{\textbf{Noisy (1024 Shots)}} \\
\cmidrule(lr){2-3} \cmidrule(lr){4-5}
& \textbf{Mean Best Energy $\pm$ Std} & \textbf{Time (s)} & \textbf{Mean Best Energy $\pm$ Std} & \textbf{Time (s)} \\
\midrule
L-BFGS-B     & -8.0759 $\pm$ $5.83 \times 10^{-1}$ & \phantom{0}0.20 & -2.8452 $\pm$ $2.03 \times 10^{0}$  & \phantom{0}0.14 \\
SPSA                & -8.2879 $\pm$ $1.00 \times 10^{0}$  & 24.42           & -7.9177 $\pm$ $8.68 \times 10^{-1}$ & 24.65           \\
CMA-ES              & -7.8571 $\pm$ $8.35 \times 10^{-1}$ & \phantom{0}1.60 & -8.3065 $\pm$ $7.74 \times 10^{-1}$ & \phantom{0}6.00 \\
SciPy DE (best1bin) & -9.0000 $\pm$ $2.24 \times 10^{-7}$ & 26.41           & -6.2366 $\pm$ $6.81 \times 10^{-1}$ & 26.47           \\
GA-MPC (2011)       & -8.7432 $\pm$ $1.35 \times 10^{-1}$ & 25.91           & -8.8049 $\pm$ $1.38 \times 10^{-1}$ & 25.93           \\
LSHADE-RSP (2018)   & -8.9994 $\pm$ $1.49 \times 10^{-3}$ & 16.71           & -9.0229 $\pm$ $6.55 \times 10^{-2}$ & 16.48           \\
HS-ES (2018)        & -9.0000 $\pm$ $4.12 \times 10^{-11}$& 16.93           & -9.1263 $\pm$ $7.29 \times 10^{-3}$ & 19.73           \\
jSO (2017)          & -8.9999 $\pm$ $1.61 \times 10^{-4}$ & 25.54           & -9.0687 $\pm$ $3.58 \times 10^{-2}$ & 25.55           \\
iL-SHADE            & -9.0000 $\pm$ $1.12 \times 10^{-15}$& 25.83           & -9.1174 $\pm$ $7.02 \times 10^{-3}$ & 25.61           \\
IMODE (2020)        & -9.0000 $\pm$ $2.56 \times 10^{-8}$ & 15.80           & -9.0937 $\pm$ $8.80 \times 10^{-3}$ & 15.85           \\
NL-SHADE-LBC (2022) & -8.9905 $\pm$ $1.93 \times 10^{-2}$ & 16.24           & -9.0745 $\pm$ $3.69 \times 10^{-2}$ & 16.18           \\
L-SRTDE (2024)      & -9.0000 $\pm$ $3.13 \times 10^{-14}$& 27.60           & -9.1183 $\pm$ $8.91 \times 10^{-3}$ & 27.44           \\
\bottomrule
\end{tabular}}
\end{table}

\begin{figure}
    \centering
    \includegraphics[width=1\linewidth]{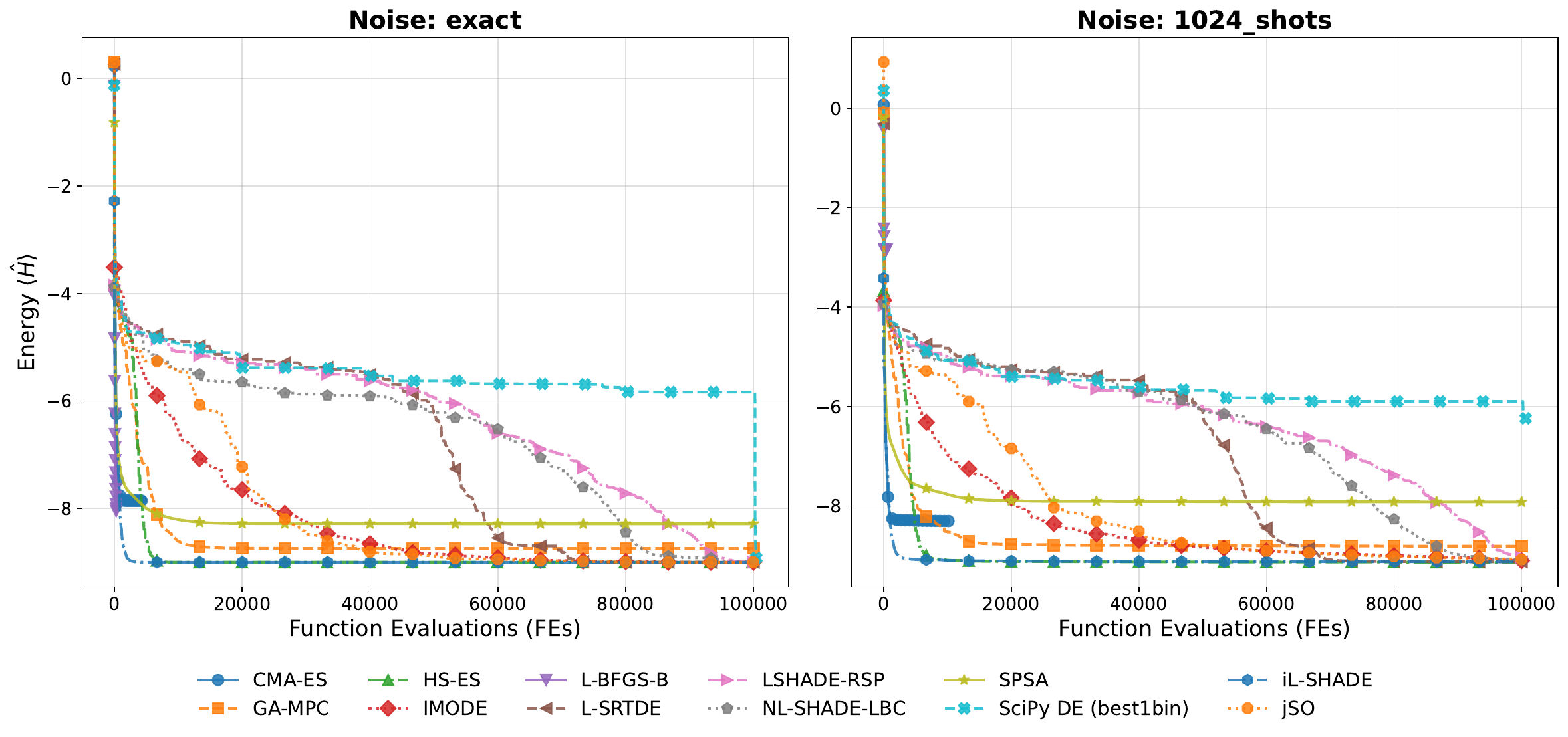}
    \caption{Optimizer convergence on the 10-qubit $L=1$ Ising VQA landscape. (left) Under exact statevector simulation, several advanced metaheuristics successfully navigate the underparameterized landscape to reach the global minimum. (right) Under noisy conditions, modern CEC variants demonstrate superior robustness, maintaining convergence despite stochastic fluctuations in the energy expectation values.}
    \label{fig:vqa_convergence_comparison}
\end{figure}

The empirical data in Table \ref{tab:vqa_results} validates the topological complexity of the underparameterized regime. In the exact simulation, quasi-Newton methods utilizing finite-difference gradients (L-BFGS-B) and stochastic gradient approximations (SPSA) prematurely converge to sub-optimal local traps ($E \approx -8.0$). This confirms the presence of severe reachability deficits inherent to the $L=1$ hardware-efficient ansatz. Conversely, spatial diversity maintenance allows all advanced CEC Differential Evolution variants and CMA-ES to successfully navigate the local minima and locate the global ground state ($E = -9.0$).

The finite-shot noisy regime introduces severe stochastic perturbations that destroy local gradient fidelity. L-BFGS-B completely fails ($E \approx 0.57$), as statistical variance corrupts its finite-difference calculations. SPSA and CMA-ES demonstrate partial robustness but fail to fully minimize the energy. Basic differential evolution strategies (SciPy DE) also collapse under the noise profile ($E \approx -6.57$).

Crucially, modern CEC variants (LSHADE-RSP, HS-ES, jSO, iL-SHADE, IMODE, NL-SHADE-LBC, and L-SRTDE) exhibit total robustness to shot noise. As demonstrated by the exact statevector simulations (Table \ref{tab:vqa_results}), these algorithms definitively isolate the true global basin ($E = -9.0$). In the noisy regime, the final expectation values frequently dip below the theoretical $-9.0$ physical baseline. This is an expected statistical artifact of reaching the noise floor, driven by the additive nature of the $\mathcal{N}(0, 1/\sqrt{1024})$ variance. Because the solvers evaluate fitness purely via these noisy finite-shot estimations, aggressive selection pressures naturally converge on vectors experiencing favorable stochastic fluctuations within the correct global basin. If higher measurement precision were required, an asymptotic increase in $N_{shots}$ during the final exploitation phase would smoothly converge these values to the exact physical ground state.

\begin{figure}[htbp]
    \centering
    \includegraphics[width=\textwidth]{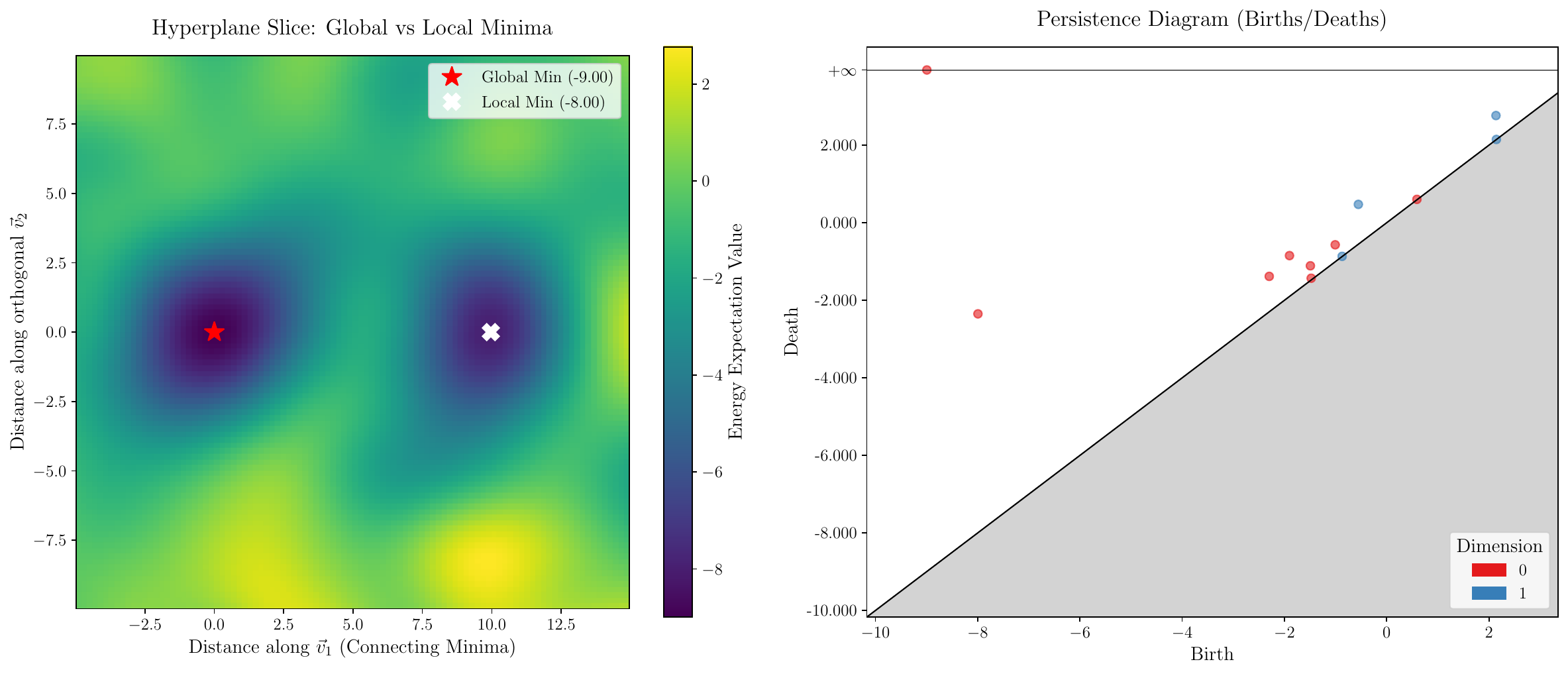}
    \caption{Topological analysis of the $L=1$ Ising VQA landscape. (Left) A 2D hyperplane slice explicitly constructed to intersect the global minimum ($E = -9.0$) and the primary sub-optimal local trap ($E = -8.0$). (Right) The corresponding persistence diagram tracking the sublevel set filtration. The $H_0$ generators (red dots) quantify the birth and death of distinct energy basins.}
    \label{fig:persistence}
\end{figure}

To quantify the topological barriers that cause these local solvers to stagnate, we conducted a persistent homology analysis of the VQA manifold using the GUDHI \cite{gudhi:urm} package. Figure \ref{fig:persistence} illustrates a 2D hyperplane slice constructed using an orthonormal basis that explicitly connects the global minimum discovered by iL-SHADE and the sub-optimal local minimum where L-BFGS-B converges. The accompanying persistence diagram tracks the topological evolution of the landscape's sublevel sets. The $H_0$ features, representing connected components, correspond to the birth of local energy basins. The diagram explicitly confirms the existence of a deep topological trap at $E = -8.0$. 

Crucially, the death coordinate of this local minimum in the persistence diagram, which represents the saddle point where the sub-optimal basin merges into the global basin within this projected plane, occurs at an exceptionally high energy level ($E \approx -2.0$). This reveals a massive $6.0$ energy-unit topological barrier separating the two minima. Because this 2D hyperplane provides a constrained cross-section of the 20-dimensional parameter space, we explicitly note that this 6.0 energy-unit barrier represents an upper bound on the true saddle point energy; there may theoretically exist lower-energy contour paths connecting the minima in the remaining 18 orthogonal dimensions. However, this restricted landscape effectively illustrates why local gradient-exploiting methods (such as L-BFGS-B) fail: lacking non-local search mechanisms, they are unable to evaluate higher-dimensional escape trajectories and become immediately trapped by localized geometric barriers. In contrast, the spatial diversity, large initial populations, and crossover operators of the top-tier CEC variants allow them to sample across or around this barrier during the initial exploratory phase, entirely bypassing the sub-optimal manifold to isolate the true ground state.

The failure of local gradient-based methods in this regime underscores the necessity of the non-local search mechanisms inherent to CEC metaheuristics, which successfully navigate these large-scale geometric traps. However, it is important to note that the $L=1$ hardware-efficient ansatz employed here represents a relatively low-entanglement regime. Consequently, the specific mechanisms by which success-history adaptation and distance-based mutation handle the extreme non-separability and high-dimensional correlations characteristic of more complex, deeply entangled quantum states remain to be fully quantified. Future empirical work must therefore evaluate these evolved CEC solvers against more expressive, highly entangled circuits and realistic hardware noise models. Furthermore, benchmarking these strategies against quantum-aware second-order methods, such as the Quantum Natural Gradient, will be essential to determine their scalability limits when transitioning from underparameterized multimodal landscapes to the regime of true noise-induced and expressivity-driven Barren Plateaus.

\section{Conclusions}
\label{sec:conclusions}

This review of the CEC Single Objective Optimization Competition (2010–2024) reveals a distinct evolutionary trajectory, transitioning from the algorithmic diversity of the early decade, exemplified by the victory of GA-MPC on the CEC 2011 Real-World Problem suite, to the dominance of adaptive Differential Evolution. Our analysis identifies the introduction of aggressive coordinate rotation in the post-2013 benchmark suites as the primary cause that limited the effectiveness of coordinate-dependent algorithms like PSO, GA, Memetic, ACO and others. In this environment, DE thrived due to its implicit rotational adaptation and the standardization of Success-History Adaptation paired with Linear Population Size Reduction. This mechanism, pivotal in the dominance of the L-SHADE family (2014–2018), eventually evolved into Non-Linear Population Size Reduction in later NL-SHADE variants, an adaptation designed to prevent premature convergence by maintaining population diversity longer during the search of increasingly rugged multimodal landscapes.

The post-2020 phase (2020–2024) marks a transition from simple parameter adaptation to Structural Hybridization and Heterogeneous Population-based Optimization. To address the extreme heterogeneity of modern composite benchmarks, winning architectures such as IMODE and AGSK introduced heterogeneous population structures, where individuals are partitioned into distinct sub-groups, such as junior and senior phases, to execute specialized exploration and exploitation logics simultaneously. This paradigm allows solvers to maintain a diverse portfolio of strategies within a single framework, maximizing robustness across conflicting landscape topologies. While this trend initially favored massive ensemble architectures like EA4eigN100, which aggregated multiple engines to ensure ranking stability, the most recent 2024 winner, L-SRTDE, signals a critical refinement of this complexity. By coupling operator selection to Success Rate rather than raw fitness magnitude, L-SRTDE demonstrates that superior performance now relies on high-fidelity signal processing of the search history, effectively filtering the noise of deceptive landscapes better than its predecessors.

We propose the integration of swarm and evolutionary heuristics as the standard control mechanism for the classical-quantum High-Performance Computing paradigm. The operation of Noisy Intermediate-Scale Quantum (NISQ) devices generates objective landscapes hindered by stochastic shot noise, multimodality, non-separability, and Barren Plateaus where gradient-based optimizers become ineffective and unstable. While local methods fail to navigate these features, the composite global-local hybrid strategies dominant in recent CEC competitions demonstrate the specific robustness required for this domain. Ultimately, the attainment of Quantum Advantage relies on this algorithmic transfer; high-precision outcomes in critical applications, from ground-state estimation in chemistry to the modeling of highly entangled systems, are strictly reliant upon the optimizer's ability to locate global optima within noisy quantum cost functions.

\section*{Acknowledgments}
This project has received funding from the Research Council of Lithuania (LMTLT), agreement No. P-ITP-24-9. This research was also supported by research grants SGS No. SP2026/063 of VSB-Technical University of Ostrava, Czech Republic. This work was supported by the Ministry of Education, Youth and Sports of the Czech Republic through the e-INFRA CZ (ID:90254 ). Martin Beseda is supported by Italian Government (Ministero dell'Università e della Ricerca, PRIN 2022 PNRR) -- cod.P2022SELA7: ''RECHARGE: monitoRing, tEsting, and CHaracterization of performAnce Regressions`` -- Decreto Direttoriale n. 1205 del 28/7/2023.

\section*{Data availability}

The source code and datasets required to reproduce the findings of this study are available across two primary repositories. Data for Figure 1 and Figure 2, as well as the optimizers for the 10-qubit Ising experiment, are available at \\ \url{https://github.com/VojtechNovak/CEC_analysis}. All CEC results and benchmark functions used in this study are available at \url{https://github.com/P-N-Suganthan}.


\appendix
\section{Examples of benchmark functions}

In this section we display some of the more complex widely used benchmark functions with corresponding formulas how to build them. These formulations are drawn from the CEC 2017 benchmark suite \cite{wu2017problem}, which remains a primary baseline for evaluating algorithmic robustness despite the emergence of newer competition standards such as CEC 2020 and CEC 2022. While the specific definitions for IEEE Congress on Evolutionary Computation competitions evolve annually to address landscape biases, the CEC 2017 suite persists in contemporary research (2024--2025) due to the high difficulty of its Hybrid and Composition functions.

\textbf{Expanded Schaffer's Function (F3)} is based on the definition
\begin{equation}
g(x,y) = 0.5 + \frac{\sin^2\left(\sqrt{x^2+y^2}\right) - 0.5}{\left(1+0.001(x^2+y^2)\right)^2},
\end{equation}
which is applied pairwise as
\begin{equation}
f_3(\mathbf{x}) = g(x_1,x_2) + g(x_2,x_3) + \ldots + g(x_{D-1},x_D) + g(x_D,x_1),
\end{equation}
yielding the final transformed function
\begin{equation}
F_3(\mathbf{x}) = f_3\left(\mathbf{M}\left(\frac{0.5(\mathbf{x}-\mathbf{o}_3)}{100}\right)\right) + F_3^*.
\end{equation}

For \textbf{Rastrigin's Function (F4)}, the base function is given by
\begin{equation}
f_4(\mathbf{x}) = \sum_{i=1}^{D} \left(x_i^2 - 10\cos(2\pi x_i) + 10\right),
\end{equation}
formulating the transformed version as
\begin{equation}
F_4(\mathbf{x}) = f_4\left(\mathbf{M}\left(\frac{5.12(\mathbf{x}-\mathbf{o}_4)}{100}\right)\right) + F_4^*.
\end{equation}

The \textbf{Levy Function (F5)} is defined as
\begin{equation}
f_5(\mathbf{x}) = \sin^2(\pi w_1) + \sum_{i=1}^{D-1} (w_i - 1)^2 [1 + 10\sin^2(\pi w_i - 1)] + (w_D - 1)^2 [1 + \sin^2(2\pi w_D)],
\end{equation}
where $w_i = 1 + \frac{x_i - 1}{4}$, resulting in the transformed version
\begin{equation}
F_5(\mathbf{x}) = f_5\left(\mathbf{M}\left(\frac{5.12(\mathbf{x}-\mathbf{o}_5)}{100}\right)\right) + F_5^*.
\end{equation}

In these equations, $\mathbf{M}$ represents the rotation matrix, $\mathbf{o}_i$ the shift vector for function $i$, and $F_i^*$ the optimal value for function $i$.

\begin{figure}[htpb]
    \centering
    \begin{subfigure}{0.31\textwidth}
        \includegraphics[width=\linewidth]{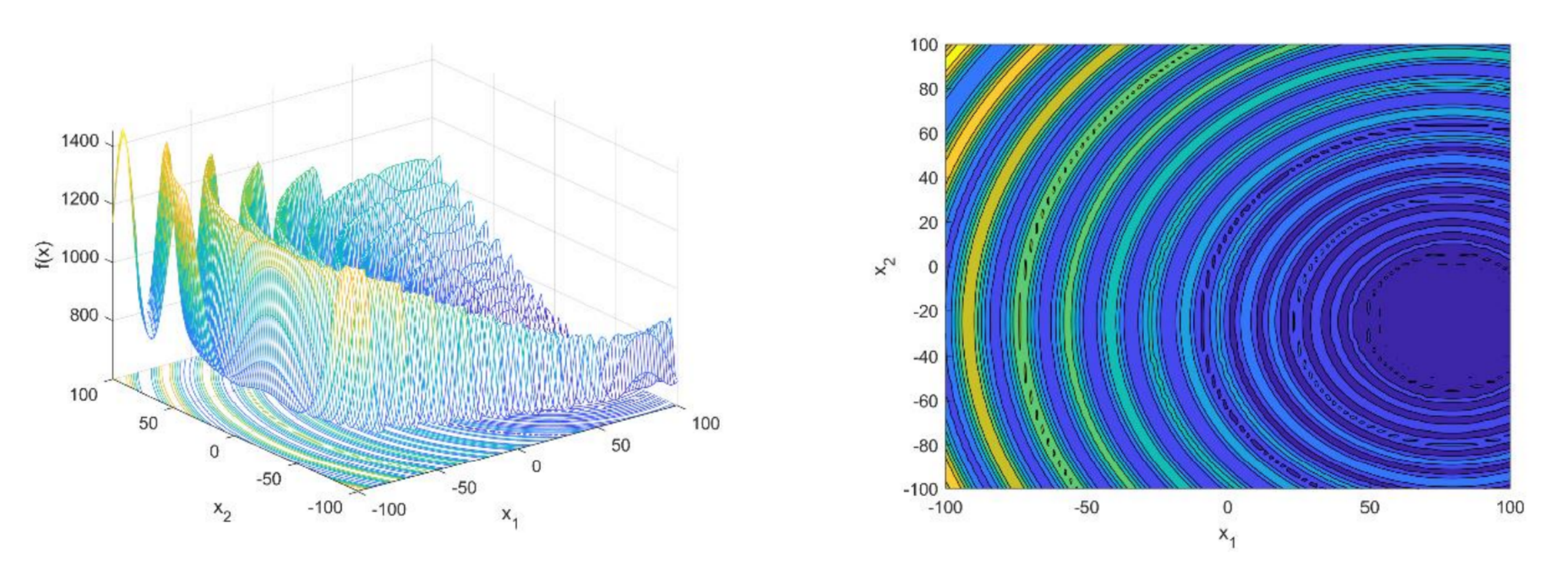}
        \caption{Schaffer's F3}
    \end{subfigure}\hfill
    \begin{subfigure}{0.31\textwidth}
        \includegraphics[width=\linewidth]{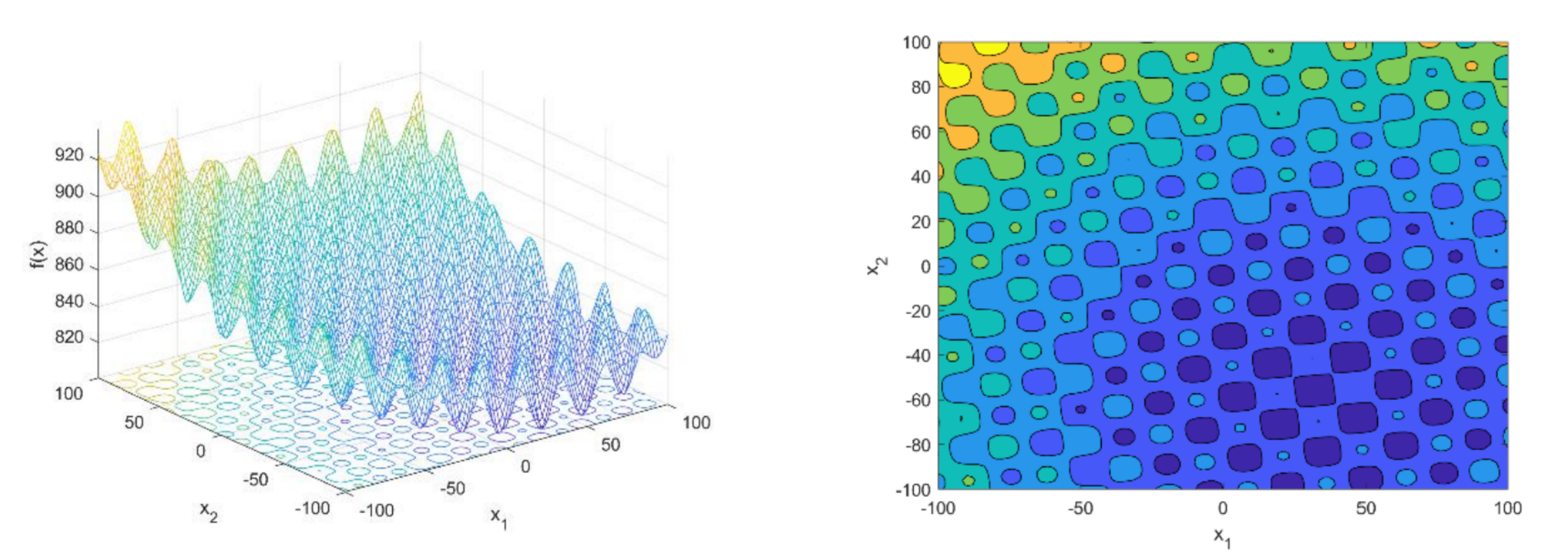}
        \caption{Rastrigin's F4}
    \end{subfigure}\hfill
    \begin{subfigure}{0.31\textwidth}
        \includegraphics[width=\linewidth]{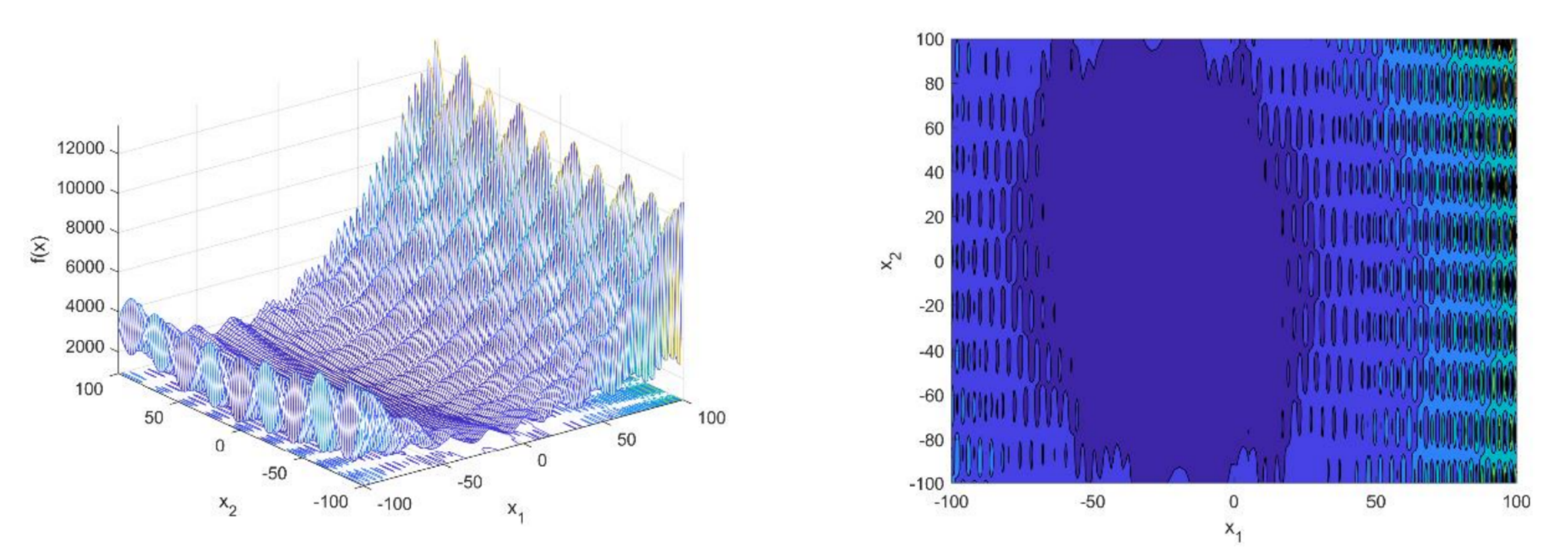}
        \caption{Levy F5}
    \end{subfigure}
    
    \vspace{1em} 
    
    \begin{subfigure}{0.8\textwidth}
        \includegraphics[width=\linewidth]{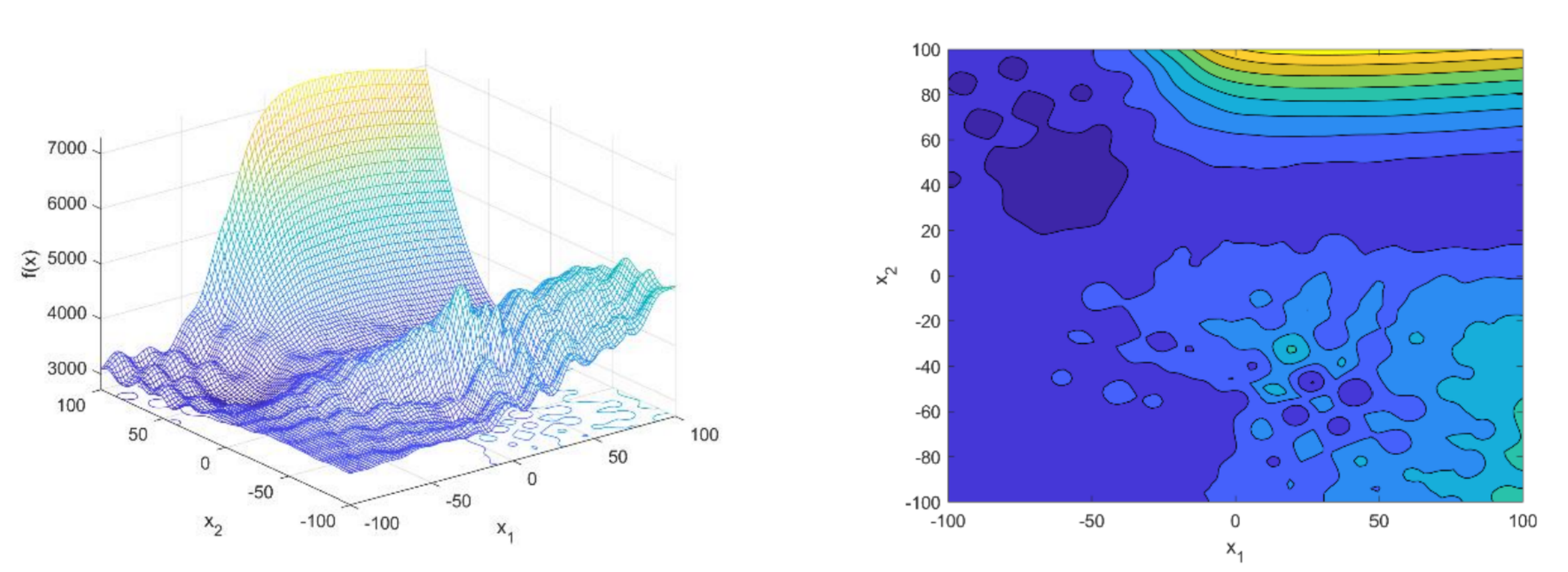}
        \caption{Composition Function (N=6) landscape}
    \end{subfigure}
    
    \caption{Visualizations of the benchmark functions used in this study. The top row shows base landscapes, while the bottom shows the high-complexity composite landscape.}
    \label{fig:all_benchmarks}
\end{figure}

The composition function (Fig.~\ref{fig:all_benchmarks} (d)) presents particular difficulty due to its hybrid construction mechanism
\begin{equation}
F(\mathbf{x}) = \sum_{i=1}^N \left\{ \omega_i \cdot \left[\lambda_i g_i(\mathbf{x} - \mathbf{o}_i) + \text{bias}_i\right]\right\} + F^*.
\label{eq:composition}
\end{equation}

Here, each component function $g_i$ is transformed such that the shift $\mathbf{o}_i$ defines new optimum positions, the scaling $\lambda_i$ controls function height ($10^{-26}$ to $10$), the bias determines the global optimum location (0-500), and the coverage $\sigma_i$ governs the influence region (10-60). The weight calculation follows
\begin{equation}
\omega_i = \frac{1}{\sigma_i\sqrt{2\pi}}\exp\left(-\frac{\|\mathbf{x} - \mathbf{o}_i\|^2}{2\sigma_i^2}\right),
\end{equation}
creating a landscape where different regions exhibit characteristics of different base functions, with transitions controlled by the $\sigma$ parameters. The extreme range of $\lambda$ values ($10^{-26}$ to $10$) creates dramatic variations in sensitivity across the search space.

\scriptsize
\bibliographystyle{elsarticle-num}
\bibliography{CECbib}

@inproceedings{brest2017single,
  title={Single objective real-parameter optimization: Algorithm jSO},
  author={Brest, Janez and Mau{\v{c}}ec, Mirjam Sepesy and Bo{\v{s}}kovi{\'c}, Borko},
  booktitle={2017 IEEE Congress on Evolutionary Computation (CEC)},
  pages={1311--1318},
  year={2017},
  organization={IEEE}
}

@techreport{awad2017cec,
  title={CEC 2017 Special Session on Single Objective Numerical Optimization: Single Bound Constrained Real-Parameter Numerical Optimization (Competition Results)},
  author={Awad, Noor H. and Ali, Mostafa Z. and Liang, J. J. and Qu, B. Y. and Suganthan, Ponnuthurai N.},
  institution={Nanyang Technological University and Zhengzhou University},
  year={2017},
  note={Presented at the IEEE Congress on Evolutionary Computation (CEC)}
}

@inproceedings{hadi2017lshade,
  title={{LSHADE} with semi-parameter adaptation hybrid with {CMA-ES} for solving {CEC} 2017 benchmark problems},
  author={Hadi, Anas A. and Mohamed, Ali W. and Jambi, Kamal M.},
  booktitle={2017 IEEE Congress on Evolutionary Computation (CEC)},
  pages={145--152},
  year={2017},
  organization={IEEE}
}

@inproceedings{awad2017ensemble,
  title={Ensemble sinusoidal differential covariance matrix adaptation with Euclidean neighborhood for solving {CEC2017} benchmark problems},
  author={Awad, Noor H. and Ali, Mostafa Z. and Suganthan, Ponnuthurai N. and Reynolds, Robert G.},
  booktitle={2017 IEEE Congress on Evolutionary Computation (CEC)},
  pages={2055--2062},
  year={2017},
  organization={IEEE}
}

@inproceedings{biedrzycki2017version,
  title={A version of {IPOP-CMA-ES} algorithm with midpoint for {CEC} 2017 single objective bound constrained problems},
  author={Biedrzycki, Rafa{\l} and Arabas, Jaros{\l}aw},
  booktitle={2017 IEEE Congress on Evolutionary Computation (CEC)},
  pages={2063--2070},
  year={2017},
  organization={IEEE}
}

@article{vha,
  title={Numerical Optimization Strategies for the Variational Hamiltonian Ansatz in Noisy Quantum Environments},
  author={Ill{\'e}sov{\'a}, Silvie and Nov{\'a}k, Vojt{\v{e}}ch and Bezd{\v{e}}k, Tom{\'a}{\v{s}} and Beseda, Martin and Possel, Clemens},
  journal={arXiv preprint arXiv:2505.22398},
  year={2025},
  eprint={2505.22398},
  doi={2505.22398},
  archivePrefix={arXiv},
  primaryClass={quant-ph},
}

@article{novak2025reliable,
  title={Reliable Optimization Under Noise in Quantum Variational Algorithms},
  author={Nov{\'a}k, Vojt{\v{e}}ch and Ill{\'e}sov{\'a}, Silvie and Bezd{\v{e}}k, Tom{\'a}{\v{s}} and Zelinka, Ivan and Beseda, Martin},
  journal={arXiv preprint arXiv:2511.08289},
  year={2025},
  doi={10.48550/arXiv.2511.08289},
}

@article{novak2025optimization,
  title={Optimization Strategies for Variational Quantum Algorithms in Noisy Landscapes},
  author={Nov{\'a}k, Vojt{\v{e}}ch and Zelinka, Ivan and Sn{\'a}{\v{s}}el, V{\'a}clav},
  journal={arXiv preprint arXiv:2506.01715},
  year={2025},
  doi={10.48550/arXiv.2506.01715},
}

@article{novak2025predicting,
  title={Predicting Post-Surgical Complications with Quantum Neural Networks: A Clinical Study on Anastomotic Leak},
  author={Nov{\'a}k, Vojt{\v{e}}ch and Zelinka, Ivan and P{\v{r}}ibylov{\'a}, Lenka and Mart{\'i}nek, Lubom{\'i}r and Ben{\v{c}}urik, V{\'a}clav},
  journal={arXiv preprint arXiv:2506.01708},
  year={2025},
  doi={10.48550/arXiv.2506.01708},
}

@article{illesova2025statistical,
  title={Statistical Benchmarking of Optimization Methods for Variational Quantum Eigensolver under Quantum Noise},
  author={Ill{\'e}sov{\'a}, Silvie and Bezd{\v{e}}k, Tom{\'a}{\v{s}} and Nov{\'a}k, Vojt{\v{e}}ch and Senjean, Bruno and Beseda, Martin},
  journal={arXiv preprint arXiv:2510.08727},
  year={2025},
  doi={10.48550/arXiv.2510.08727},
}

@article{illesova2025complementarity,
  title={On the Complementarity of Classical Convolution and Quantum Neural Networks in Image Classification},
  author={Ill{\'e}sov{\'a}, Silvie and Obeng, Emmanuel and Bezd{\v{e}}k, Tom{\'a}{\v{s}} and Nov{\'a}k, Vojt{\v{e}}ch and Beseda, Martin},
  journal={Preprints},
  year={2025},
  doi={10.20944/preprints202512.1348.v1},
  note={Preprint}
}

@article{bezdek2025classical,
  title={Classical Optimization Strategies for Variational Quantum Algorithms: A Systematic Study of Noise Effects and Parameter Efficiency},
  author={Bezd{\v{e}}k, Tom{\'a}{\v{s}} and Yuan, Haomu and Nov{\'a}k, Vojt{\v{e}}ch and Ill{\'e}sov{\'a}, Silvie and Beseda, Martin},
  journal={arXiv preprint arXiv:2511.09314},
  year={2025},
  doi={10.48550/arXiv.2511.09314},
}

@inproceedings{gupta2022how,
  title={How Quantum Computing-Friendly Multispectral Data can be?},
  author={Gupta, Manish K. and Beseda, Martin and Gawron, Piotr},
  booktitle={IGARSS 2022-2022 IEEE International Geoscience and Remote Sensing Symposium},
  pages={4153--4156},
  year={2022},
  organization={IEEE},
  doi={10.1109/IGARSS46834.2022.9883676},
}

@article{ciaramelletti2025detecting,
  title={Detecting Quasidegenerate Ground States in Topological Models via the Variational Quantum Eigensolver},
  author={Ciaramelletti, C. and Beseda, Martin and Consiglio, M. and Lepori, L. and Apollaro, T. J. G.},
  journal={Physical Review A},
  volume={111},
  number={2},
  pages={022437},
  year={2025},
  publisher={APS},
  doi={10.1103/PhysRevA.111.022437},
}

@inproceedings{illesova2025qmetric,
  title={QMetric: Benchmarking Quantum Neural Networks Across Circuits, Features, and Training Dimensions},
  author={Ill{\'e}sov{\'a}, Silvie and Rybotycki, Tomasz and Beseda, Martin},
  booktitle={QualITA 2025: The Fourth Conference on System and Service Quality},
  series={CEUR Workshop Proceedings},
  volume={4080},
  year={2025},
  publisher={CEUR-WS.org},
}

@article{mcclean2018barren,
  title={Barren plateaus in quantum neural network training landscapes},
  author={McClean, Jarrod R and Boixo, Sergio and Smelyanskiy, Vadim N and Babbush, Ryan and Neven, Hartmut},
  journal={Nature Communications},
  volume={9},
  number={1},
  pages={1--6},
  year={2018},
  publisher={Nature Publishing Group}
}

@article{holmes2022connecting,
  title={Connecting ansatz expressibility to gradient magnitudes and barren plateaus},
  author={Holmes, Zo{\"e} and Sharma, Kunal and Cerezo, M and Coles, Patrick J},
  journal={PRX Quantum},
  volume={3},
  number={1},
  pages={010313},
  year={2022},
  publisher={APS}
}

@article{wang2021noise,
  title={Noise-induced barren plateaus in variational quantum algorithms},
  author={Wang, Samson and Fontana, Enrico and Cerezo, M and Sharma, Kunal and Sone, Akira and Cincio, Lukasz and Coles, Patrick J},
  journal={Nature Communications},
  volume={12},
  number={1},
  pages={1--11},
  year={2021},
  publisher={Nature Publishing Group}
}

@article{akshay2020reachability,
  title={Reachability deficits in quantum approximate optimization},
  author={Akshay, V and Philathong, H and Morales, ME and Biamonte, JD},
  journal={Physical Review Letters},
  volume={124},
  number={9},
  pages={090504},
  year={2020},
  publisher={APS}
}

@article{fontana2022non,
  title={Non-trivial symmetries in quantum landscapes and their resilience to quantum noise},
  author={Fontana, Enrico and Cerezo, M and Arrasmith, Andrew and Rungger, Ivan and Coles, Patrick J},
  journal={Quantum},
  volume={6},
  pages={804},
  year={2022},
  publisher={Verein zur F{\"o}rderung des Open Access Publizierens in den Quantenwissenschaften}
}

@article{larocca2023theory,
  title={Theory of overparametrization in quantum neural networks},
  author={Larocca, Mart{\'\i}n and Ju, Nathan and Garc{\'\i}a-Mart{\'\i}n, Diego and Coles, Patrick J and Cerezo, M},
  journal={Nature Computational Science},
  volume={3},
  number={6},
  pages={542--551},
  year={2023},
  publisher={Nature Publishing Group}
}

@article{bittel2021training,
  title={Training variational quantum algorithms is NP-hard},
  author={Bittel, Lennart and Kliesch, Martin},
  journal={Physical Review Letters},
  volume={127},
  number={12},
  pages={120502},
  year={2021},
  publisher={APS}
}

@article{Stokes2020,
  title = {Quantum Natural Gradient},
  author = {Stokes, James and Izaac, Josh and Killoran, Nathan and Giuseppe, Carleo},
  journal = {Quantum},
  volume = {4},
  pages = {269},
  year = {2020},
  doi = {10.22331/q-2020-05-25-269},
  note = {Defines the use of the Quantum Fisher Information Matrix as a metric tensor to navigate entangled parameter spaces[cite: 482].}
}

@book{gudhi:urm
, title        = "{GUDHI} User and Reference Manual"
, author      = "{The GUDHI Project}"
, publisher     = "{GUDHI Editorial Board}"
, year         = 2015
, url =    "http://gudhi.gforge.inria.fr/doc/latest/"
}

@inproceedings{sutton2007differential,
  title={Differential evolution and non-separability: Using selective pressure to focus search},
  author={Sutton, Andrew M and Lunacek, Monte and Whitley, L Darrell},
  booktitle={Proceedings of the 9th annual conference on Genetic and evolutionary computation},
  pages={1428--1435},
  year={2007}
}

@article{zhou2024adaptive,
  title={An adaptive archive differential evolution with non-linear population size reduction and selective pressure},
  author={Zhou, Benben and Huang, Ying},
  journal={Information Sciences},
  volume={682},
  pages={121273},
  year={2024},
  publisher={Elsevier}
}

@book{Bengtsson2017,
  title={Geometry of Quantum States: An Introduction to Quantum Entanglement},
  author={Bengtsson, Ingemar and {\.Z}yczkowski, Karol},
  year={2017},
  publisher={Cambridge University Press},
  doi={10.1017/9781139505468},
  note={The standard reference for $\mathbb{CP}^{N-1}$ and the geometry of Hilbert space.}
}

@article{Meyer2021,
  title = {Fisher Information in Quantum Machine Learning Models},
  author = {Meyer, Johannes Jakob},
  journal = {Quantum},
  volume = {5},
  pages = {539},
  year = {2021},
  doi = {10.22331/q-2021-09-09-539}
}

@inproceedings{trovato2025preliminary,
author = {Trovato, Antonio and Beseda, Martin and Di Nucci, Dario},
title = {A Preliminary Investigation on the Usage of Quantum Approximate Optimization Algorithms for Test Case Selection},
year = {2025},
isbn = {9798400718328},
publisher = {Association for Computing Machinery},
address = {New York, NY, USA},
url = {https://doi.org/10.1145/3727967.3756821},
doi = {10.1145/3727967.3756821},
booktitle = {Proceedings of the 2025 29th International Conference on Evaluation and Assessment in Software Engineering Companion},
pages = {56–60},
numpages = {5},
series = {EASE Companion '25}
}

@article{illesova2025importance,
author = {Ill{\'e}sov{\'a}, Silvie and Rybotycki, Tomasz and Gawron, Piotr and Beseda, Martin},
title = {On the importance of fundamental properties in quantum-classical machine learning models},
journal = {International Journal of Parallel, Emergent and Distributed Systems},
volume = {0},
number = {0},
pages = {1--28},
year = {2026},
publisher = {Taylor \& Francis},
doi = {10.1080/17445760.2026.2626759},
URL = {https://doi.org/10.1080/17445760.2026.2626759},
eprint = {https://doi.org/10.1080/17445760.2026.2626759}
}

@article{lewandowska2025benchmarking,
  title={Benchmarking Gate-Based Quantum Devices via Certification of Qubit von Neumann Measurements},
  author={Lewandowska, Paulina and Beseda, Martin},
  journal={arXiv preprint arXiv:2506.03514},
  year={2025},
  doi={10.48550/arXiv.2506.03514},
}

@article{bilek2025experimental,
  author   = {B{\'\i}lek, Adam and Hlisnikovsk{\'y}, Jan and Bezd{\v{e}}k, Tom{\'a}{\v{s}} and Kukulski, Ryszard and Lewandowska, Paulina},
  title    = {Experimental study of multiple-shot unitary channels discrimination using the IBM Q computers},
  journal  = {Scientific Reports},
  year     = {2026},
  month    = {Feb},
  day      = {08},
  volume   = {16},
  number   = {1},
  pages    = {6142},
  issn     = {2045-2322},
  doi      = {10.1038/s41598-025-31665-z},
  url      = {https://doi.org/10.1038/s41598-025-31665-z}
}

@article{beseda2024state,
  title={State-Averaged Orbital-Optimized VQE: A quantum algorithm for the democratic description of ground and excited electronic states},
  author={Beseda, Martin and Ill{\'e}sov{\'a}, Silvie and Yalouz, Saad and Senjean, Bruno},
  journal={Journal of Open Source Software},
  volume={9},
  number={101},
  pages={6036},
  year={2024},
  doi = {10.21105/joss.06036}, 
  url = {https://doi.org/10.21105/joss.06036},
  publisher = {The Open Journal}
}

@article{illesova2025transformation,
  title={Transformation-free generation of a quasi-diabatic representation from the state-average orbital-optimized variational quantum eigensolver},
  author={Ill{\'e}sov{\'a}, Silvie and Beseda, Martin and Yalouz, Saad and Lasorne, Benjamin and Senjean, Bruno},
  journal={Journal of Chemical Theory and Computation},
  publisher={ACS Publications},
  year={2025}
}

@article{grimsley2019trotterized,
  title={Is the trotterized uccsd ansatz chemically well-defined?},
  author={Grimsley, Harper R and Claudino, Daniel and Economou, Sophia E and Barnes, Edwin and Mayhall, Nicholas J},
  journal={Journal of chemical theory and computation},
  volume={16},
  number={1},
  pages={1--6},
  year={2019},
  publisher={ACS Publications}
}

@article{Price2019,
  title={The 2019 100-digit challenge on real-parameter, single objective optimization: analysis of results},
  author={Price, K and Awad, NH and Ali, MZ and Suganthan, P},
  journal={Dept. Computer Science, University of Freiburg, Freiburg, Germany, School of Computer Information Systems, Jordan University of Science and Technology, Jordan, and School of EEE, Nanyang Technol. Univ., Singapore, Rep},
  volume={2019},
  year={2019}
}

@article{kandala2017hardware,
  title={Hardware-efficient variational quantum eigensolver for small molecules and quantum magnets},
  author={Kandala, Abhinav and Mezzacapo, Antonio and Temme, Kristan and Takita, Maika and Brink, Markus and Chow, Jerry M and Gambetta, Jay M},
  journal={nature},
  volume={549},
  number={7671},
  pages={242--246},
  year={2017},
  publisher={Nature Publishing Group}
}

@article{benedetti2019parameterized,
  title={Parameterized quantum circuits as machine learning models},
  author={Benedetti, Marcello and Lloyd, Erika and Sack, Stefan and Fiorentini, Mattia},
  journal={Quantum Science and Technology},
  volume={4},
  number={4},
  pages={043001},
  year={2019},
  publisher={IOP Publishing}
}

@article{ostaszewski2021structure,
  title={Structure optimization for parameterized quantum circuits},
  author={Ostaszewski, Mateusz and Grant, Edward and Benedetti, Marcello},
  journal={Quantum},
  volume={5},
  pages={391},
  year={2021},
  publisher={Verein zur F{\"o}rderung des Open Access Publizierens in den Quantenwissenschaften}
}

@article{sim2019expressibility,
  title={Expressibility and entangling capability of parameterized quantum circuits for hybrid quantum-classical algorithms},
  author={Sim, Sukin and Johnson, Peter D and Aspuru-Guzik, Al{\'a}n},
  journal={Advanced Quantum Technologies},
  volume={2},
  number={12},
  pages={1900070},
  year={2019},
  publisher={Wiley Online Library}
}

@article{cerezo2021cost,
  title={Cost function dependent barren plateaus in shallow parameterized quantum circuits},
  author={Cerezo, Marco and Sone, Akira and Volkoff, Tyler and Cincio, Lukasz and Coles, Patrick J},
  journal={Nature Communications},
  volume={12},
  number={1},
  pages={1791},
  year={2021},
  publisher={Nature Publishing Group}
}

@article{cerezo2021variational,
  author = {Cerezo, M. and Arrasmith, Andrew and Babbush, Ryan and Benjamin, Simon C. and Endo, Suguru and Fujii, Keisuke and McClean, Jarrod R. and Mitarai, Kosuke and Yuan, Xiao and Cincio, Lukasz and Coles, Patrick J.},
  title = {{Variational quantum algorithms}},
  journal = {Nature Reviews Physics},
  volume = {3},
  number = {9},
  pages = {625--644},
  year = {2021},
  month = {Sep},
  issn = {2522-5820},
  doi = {10.1038/s42254-021-00348-9},
}

@article{buonaiuto2024effects,
 author = {Buonaiuto, G. and Gargiulo, F. and De Pietro, G. and Esposito, M. and Pota, M.},
 doi = {10.1007/s42484-024-00144-5},
 journal = {Quantum Machine Intelligence},
 pages = {12},
 title = {The effects of quantum hardware properties on the performances of variational quantum learning algorithms},
 volume = {6},
 year = {2024}
}

@inproceedings{10.1145/3695053.3731112,
author = {Dangwal, Siddharth and Vittal, Suhas and Seifert, Lennart Maximilian and Chong, Frederic T. and Ravi, Gokul Subramanian},
title = {Variational Quantum Algorithms in the era of Early Fault Tolerance},
year = {2025},
isbn = {9798400712616},
publisher = {Association for Computing Machinery},
address = {New York, NY, USA},
doi = {10.1145/3695053.3731112},
booktitle = {Proceedings of the 52nd Annual International Symposium on Computer Architecture},
pages = {1417–1431},
numpages = {15},
location = {
},
series = {ISCA '25}
}

@article{PhysRevResearch.6.023118,
  title = {Modeling the performance of early fault-tolerant quantum algorithms},
  author = {Liang, Qiyao and Zhou, Yiqing and Dalal, Archismita and Johnson, Peter},
  journal = {Phys. Rev. Res.},
  volume = {6},
  issue = {2},
  pages = {023118},
  numpages = {16},
  year = {2024},
  month = {May},
  publisher = {American Physical Society},
  doi = {10.1103/PhysRevResearch.6.023118}
}

@ARTICLE{quantum2024finance,
  author={Egger, Daniel J. and Gambella, Claudio and Marecek, Jakub and McFaddin, Scott and Mevissen, Martin and Raymond, Rudy and Simonetto, Andrea and Woerner, Stefan and Yndurain, Elena},
  journal={IEEE Transactions on Quantum Engineering},
  title={Quantum Computing for Finance: State-of-the-Art and Future Prospects},
  year={2020},
  volume={1},
  number={},
  pages={1-24},
  doi={10.1109/TQE.2020.3030314}}

@article{PhysRevA.109.032408,
  title = {Challenges of variational quantum optimization with measurement shot noise},
  author = {Scriva, Giuseppe and Astrakhantsev, Nikita and Pilati, Sebastiano and Mazzola, Guglielmo},
  journal = {Phys. Rev. A},
  volume = {109},
  issue = {3},
  pages = {032408},
  numpages = {14},
  year = {2024},
  month = {Mar},
  publisher = {American Physical Society},
  doi = {10.1103/PhysRevA.109.032408},
}

@inproceedings{kumar2021multi,
  title={Multi model implementation on general medicine prediction with quantum neural networks},
  author={Kumar, Swarn Avinash and Kumar, Abhishek and Dutt, Vishal and Agrawal, Rashmi},
  booktitle={2021 Third International Conference on Intelligent Communication Technologies and Virtual Mobile Networks (ICICV)},
  pages={1391--1395},
  year={2021},
  organization={IEEE}
}

@article{tang2010benchmark,
  title={Benchmark functions for the CEC’2010 special session and competition on large-scale global optimization},
  author={Tang, Ke and Li, Xiaodong and Suganthan, Ponnuthurai Nagaratnam and Yang, Zhenyu and Weise, Thomas},
  journal={Nature inspired computation and applications laboratory, USTC, China},
  volume={24},
  pages={1--18},
  year={2007}
}

@article{PRXQuantum.2.020310,
  title = {Qubit-ADAPT-VQE: An Adaptive Algorithm for Constructing Hardware-Efficient Ans\"atze on a Quantum Processor},
  author = {Tang, Ho Lun and Shkolnikov, V.O. and Barron, George S. and Grimsley, Harper R. and Mayhall, Nicholas J. and Barnes, Edwin and Economou, Sophia E.},
  journal = {PRX Quantum},
  volume = {2},
  issue = {2},
  pages = {020310},
  numpages = {16},
  year = {2021},
  month = {Apr},
  publisher = {American Physical Society},
  doi = {10.1103/PRXQuantum.2.020310},
}

@article{10.1063/5.0161057,
    author = {Singh, Harshdeep and Majumder, Sonjoy and Mishra, Sabyashachi},
    title = {Benchmarking of different optimizers in the variational quantum algorithms for applications in quantum chemistry},
    journal = {The Journal of Chemical Physics},
    volume = {159},
    number = {4},
    pages = {044117},
    year = {2023},
    month = {07},
    issn = {0021-9606},
    doi = {10.1063/5.0161057},
    eprint = {https://pubs.aip.org/aip/jcp/article-pdf/doi/10.1063/5.0161057/18065624/044117\_1\_5.0161057.pdf},
}

@article{das2010problem,
  title={Problem definitions and evaluation criteria for CEC 2011 competition on testing evolutionary algorithms on real world optimization problems},
  author={Das, Swagatam and Suganthan, Ponnuthurai N},
  journal={Jadavpur University, Nanyang Technological University, Kolkata},
  pages={341--359},
  year={2010}
}

@techreport{liang2013cec2013,
  title={Problem definitions and evaluation criteria for the CEC 2013 special session on real-parameter optimization},
  author={Liang, JJ and Qu, B-Y and Suganthan, Ponnuthurai N and Hern{\'a}ndez-D{\'\i}az, Alfredo G},
  institution={Computational Intelligence Laboratory, Zhengzhou University, China and Nanyang Technological University, Singapore},
  year={2013},
  number={201212}
}

@techreport{liang2013cec2014,
  title={Problem definitions and evaluation criteria for the CEC 2014 special session on real-parameter numerical optimization},
  author={Liang, JJ and Qu, B-Y and Suganthan, Ponnuthurai N},
  institution={Computational Intelligence Laboratory, Zhengzhou University, China and Nanyang Technological University, Singapore},
  year={2013},
  number={635}
}

@article{liang2015problem,
  title={Problem definitions and evaluation criteria for the CEC 2015 competition on learning-based real-parameter single objective optimization},
  author={Liang, JJ and Qu, BY and Suganthan, PN and Chen, Q},
  journal={Technical Report201411A, Computational Intelligence Laboratory, Zhengzhou University, Zhengzhou China and Technical Report, Nanyang Technological University, Singapore},
  volume={29},
  pages={625--640},
  year={2014}
}

@inproceedings{vskvorc2019cec,
  title={CEC real-parameter optimization competitions: Progress from 2013 to 2018},
  author={{\v{S}}kvorc, Urban and Eftimov, Tome and Koro{\v{s}}ec, Peter},
  booktitle={2019 IEEE congress on evolutionary computation (CEC)},
  pages={3126--3133},
  year={2019},
  organization={IEEE}
}

@article{lindsay2010optimized,
  title={Optimized Tersoff and Brenner empirical potential parameters for lattice dynamics and phonon thermal transport in carbon nanotubes and graphene},
  author={Lindsay, LDAB and Broido, DA},
  journal={Physical Review B—Condensed Matter and Materials Physics},
  volume={81},
  number={20},
  pages={205441},
  year={2010},
  publisher={APS}
}

@article{farag1995economic,
  title={Economic load dispatch multiobjective optimization procedures using linear programming techniques},
  author={Farag, Ahmed and Al-Baiyat, Samir and Cheng, TC},
  journal={IEEE Transactions on Power systems},
  volume={10},
  number={2},
  pages={731--738},
  year={1995},
  publisher={IEEE}
}

@article{addis2011global,
  title={A global optimization method for the design of space trajectories},
  author={Addis, Bernardetta and Cassioli, Andrea and Locatelli, Marco and Schoen, Fabio},
  journal={Computational Optimization and Applications},
  volume={48},
  number={3},
  pages={635--652},
  year={2011},
  publisher={Springer}
}

@article{wu2017problem,
  title={Problem definitions and evaluation criteria for the CEC 2017 competition on constrained real-parameter optimization},
  author={Wu, Guohua and Mallipeddi, Rammohan and Suganthan, Ponnuthurai Nagaratnam},
  journal={National University of Defense Technology, Changsha, Hunan, PR China and Kyungpook National University, Daegu, South Korea and Nanyang Technological University, Singapore, Technical Report},
  volume={9},
  pages={2017},
  year={2017},
  publisher={National University of Defense Technology Changsha, China}
}

@article{larocca2025barren,
  title={Barren plateaus in variational quantum computing},
  author={Larocca, Martin and Thanasilp, Supanut and Wang, Samson and Sharma, Kunal and Biamonte, Jacob and Coles, Patrick J and Cincio, Lukasz and McClean, Jarrod R and Holmes, Zo{\"e} and Cerezo, Marco},
  journal={Nature Reviews Physics},
  pages={1--16},
  year={2025},
  publisher={Nature Publishing Group UK London}
}

@techreport{mohamed2019cec2020,
  title={Problem definitions and evaluation criteria for the CEC 2020 special session and competition on single objective bound constrained numerical optimization},
  author={Mohamed, Ali Wagdy and Hadi, Anas A and Mohamed, Ali Khater},
  institution={Nanyang Technological University, Singapore},
  year={2019}
}

@misc{suganthan2025github,
  author = {Suganthan, Ponnuthurai N},
  title = {{CEC} Competition Datasets and Codes},
  year = {2025},
  publisher = {GitHub},
  journal = {GitHub repository},
  howpublished = {\url{https://github.com/P-N-Suganthan}}
}

@inproceedings{hellwig2016evolution,
  title={Evolution under strong noise: A self-adaptive evolution strategy can reach the lower performance bound-the pccmsa-es},
  author={Hellwig, Michael and Beyer, Hans-Georg},
  booktitle={International Conference on Parallel Problem Solving from Nature},
  pages={26--36},
  year={2016},
  organization={Springer}
}

@inproceedings{astete2015evolution,
  title={Evolution strategies with additive noise: A convergence rate lower bound},
  author={Astete-Morales, Sandra and Cauwet, Marie-Liesse and Teytaud, Olivier},
  booktitle={Proceedings of the 2015 ACM Conference on Foundations of Genetic Algorithms XIII},
  pages={76--84},
  year={2015}
}

@article{Arrasmith2021,
  title={Effect of barren plateaus on gradient-free optimization},
  author={Arrasmith, Andrew and Cerezo, Marco and Czarnik, Piotr and Cincio, Lukasz and Coles, Patrick J},
  journal={Quantum},
  volume={5},
  pages={558},
  year={2021},
  publisher={Verein zur F{\"o}rderung des Open Access Publizierens in den Quantenwissenschaften}
}

@inproceedings{tanabe2013success,
  title={Success-history based parameter adaptation for differential evolution},
  author={Tanabe, Ryoji and Fukunaga, Alex},
  booktitle={2013 IEEE congress on evolutionary computation},
  pages={71--78},
  year={2013},
  organization={IEEE}
}

@article{cerezo2020variational,
 author = {Cerezo, M. and Poremba, A. and Cincio, L. and Coles, P. J.},
 doi = {10.22331/q-2020-03-26-248},
 journal = {Quantum},
 pages = {248},
 title = {Variational quantum fidelity estimation},
 volume = {4},
 year = {2020}
}

@misc{farhi2014quantum,
      title={A Quantum Approximate Optimization Algorithm},
      author={Edward Farhi and Jeffrey Goldstone and Sam Gutmann},
      year={2014},
      eprint={1411.4028},
      archivePrefix={arXiv},
      primaryClass={quant-ph},
      url={https://arxiv.org/abs/1411.4028},
}

@article{mcclean2016theory,
 author = {McClean, J. R. and Romero, J. and Babbush, R. and Aspuru-Guzik, A.},
 doi = {10.1088/1367-2630/18/2/023023},
 journal = {New Journal of Physics},
 number = {2},
 pages = {023023},
 title = {The theory of variational hybrid quantum-classical algorithms},
 volume = {18},
 year = {2016}
}

@article{TILLY20221,
title = {The Variational Quantum Eigensolver: A review of methods and best practices},
journal = {Physics Reports},
volume = {986},
pages = {1-128},
year = {2022},
note = {The Variational Quantum Eigensolver: a review of methods and best practices},
issn = {0370-1573},
doi = {https://doi.org/10.1016/j.physrep.2022.08.003},
author = {Jules Tilly and Hongxiang Chen and Shuxiang Cao and Dario Picozzi and Kanav Setia and Ying Li and Edward Grant and Leonard Wossnig and Ivan Rungger and George H. Booth and Jonathan Tennyson}
}

@article{piotrowski2023choice,
  title={Choice of benchmark optimization problems does matter},
  author={Piotrowski, Adam P and Napiorkowski, Jaroslaw J and Piotrowska, Agnieszka E},
  journal={Swarm and Evolutionary Computation},
  volume={83},
  pages={101378},
  year={2023},
  publisher={Elsevier}
}

@article{das2016recent,
  title={Recent advances in differential evolution--an updated survey},
  author={Das, Swagatam and Mullick, Sankha Subhra and Suganthan, Ponnuthurai N},
  journal={Swarm and evolutionary computation},
  volume={27},
  pages={1--30},
  year={2016},
  publisher={Elsevier}
}

@article{das2010differential,
  title={Differential evolution: A survey of the state-of-the-art},
  author={Das, Swagatam and Suganthan, Ponnuthurai Nagaratnam},
  journal={IEEE transactions on evolutionary computation},
  volume={15},
  number={1},
  pages={4--31},
  year={2010},
  publisher={IEEE}
}

@article{BonetMonroig2023,
  title={Performance comparison of optimization methods on variational quantum algorithms},
  author={Bonet-Monroig, Xavier and Wang, Hao and Vermetten, Diederick and Senjean, Bruno and Moussa, Charles and B{\"a}ck, Thomas and Dunjko, Vedran and O'Brien, Thomas E},
  journal={Physical Review A},
  volume={107},
  number={3},
  pages={032407},
  year={2023},
  publisher={APS}
}

@article{Failde2023,
  title={Using differential evolution to avoid local minima in variational quantum algorithms},
  author={Fa{\'i}lde, Daniel and Viqueira, Jos{\'e} Daniel and Juane, Mariamo Mussa and G{\'o}mez, Andr{\'e}s},
  journal={arXiv preprint arXiv:2303.12186},
  year={2023}
}

@inproceedings{improvement,
  title={Analysis among winners of different IEEE CEC competitions on real-parameters optimization: Is there always improvement?},
  author={Molina, Daniel and Moreno-Garc{\'\i}a, Francisco and Herrera, Francisco},
  booktitle={2017 IEEE congress on evolutionary computation (CEC)},
  pages={805--812},
  year={2017},
  organization={IEEE}
}

@article{Wolpert1997,
  author  = {Wolpert, D.H. and Macready, W.G.},
  title   = {No free lunch theorems for optimization},
  journal = {IEEE Transactions on Evolutionary Computation},
  year    = {1997},
  volume  = {1},
  number  = {1},
  pages   = {67--82},
  doi     = {10.1109/4235.585893}
}

@article{Biedrzycki2024,
  title={Revisiting CEC 2022 ranking: A new ranking method and influence of parameter tuning},
  author={Biedrzycki, Rafa{\l}},
  journal={Swarm and Evolutionary Computation},
  volume={89},
  pages={101623},
  year={2024},
  publisher={Elsevier}
}

@inproceedings{Auger2005,
  title={A restart CMA evolution strategy with increasing population size},
  author={Auger, Anne and Hansen, Nikolaus},
  booktitle={2005 IEEE Congress on Evolutionary Computation},
  volume={2},
  pages={1769--1776},
  year={2005},
  organization={IEEE}
}

@article{Molina2018,
  title={An insight into bio-inspired and evolutionary algorithms for global optimization: Review, analysis, and lessons learnt over a decade of competitions},
  author={Molina, Daniel and LaTorre, Antonio and Herrera, Francisco},
  journal={Cognitive Computation},
  volume={10},
  pages={517--544},
  year={2018},
  publisher={Springer}
}

@inproceedings{kennedy1995pso,
  title={Particle swarm optimization},
  author={Kennedy, James and Eberhart, Russell},
  booktitle={Proceedings of ICNN'95 - International Conference on Neural Networks},
  volume={4},
  pages={1942--1948},
  year={1995},
  organization={IEEE},
  doi={10.1109/ICNN.1995.488968}
}

@book{goldberg1989genetic,
  title={Genetic Algorithms in Search, Optimization, and Machine Learning},
  author={Goldberg, David E.},
  year={1989},
  publisher={Addison-Wesley},
  address={Reading, MA}
}

@inproceedings{Tanabe2014,
  title={Improving the search performance of SHADE using linear population size reduction},
  author={Tanabe, Ryoji and Fukunaga, Alex S},
  booktitle={2014 IEEE Congress on Evolutionary Computation (CEC)},
  pages={1658--1665},
  year={2014},
  organization={IEEE}
}

@inproceedings{DECC,
  title={Cooperative co-evolution with delta grouping for large scale non-separable function optimization},
  author={Omidvar, Mohammad Nabi and Li, Xiaodong and Yao, Xin},
  booktitle={IEEE congress on evolutionary computation},
  pages={1--8},
  year={2010},
  organization={IEEE}
}

@INPROCEEDINGS{DASA,
  author={Korošec, Peter and Tashkova, Katerina and Šilc, Jury},
  booktitle={IEEE Congress on Evolutionary Computation}, 
  title={The differential Ant-Stigmergy Algorithm for large-scale global optimization}, 
  year={2010},
  volume={},
  number={},
  pages={1-8},
  doi={10.1109/CEC.2010.5586201}}

@article{Locust,
  title={Locust Swarms for Large Scale Global Optimization of Nonseparable Problems},
  author={Chen, Stephen},
  journal={Kukkonen, Benchmarking the Classic Differential Evolution Algorithm on Large-Scale Global Optimization},
  year = {2010}
}

@INPROCEEDINGS{SEQ,
  author={Wang, Hui and Wu, Zhijian and Rahnamayan, Shahryar and Jiang, Dazhi},
  booktitle={IEEE Congress on Evolutionary Computation}, 
  title={Sequential DE enhanced by neighborhood search for Large Scale Global Optimization}, 
  year={2010},
  volume={},
  number={},
  pages={1-7},
  doi={10.1109/CEC.2010.5586358}}

@INPROCEEDINGS{self-DE,
  author={Brest, Janez and Bošković, 1Borko and Zamuda, Aleš and Fister, Iztok and Maučec, Mirjam Sepesy},
  booktitle={2012 IEEE Congress on Evolutionary Computation}, 
  title={Self-adaptive differential evolution algorithm with a small and varying population size}, 
  year={2012},
  volume={},
  number={},
  pages={1-8},
  doi={10.1109/CEC.2012.6252909}}

@inproceedings{tanabe2014improving,
  title={Improving the search performance of SHADE using linear population size reduction},
  author={Tanabe, Ryoji and Fukunaga, Alex S},
  booktitle={2014 IEEE Congress on Evolutionary Computation (CEC)},
  pages={1658--1665},
  year={2014},
  organization={IEEE}
}

@inproceedings{wu2014gaapade,
  title={Differential evolution with multi-population based ensemble of mutation strategies},
  author={Wu, Guohua and Mallipeddi, Rammohan and Suganthan, Ponnuthurai N},
  booktitle={2014 IEEE Congress on Evolutionary Computation (CEC)},
  pages={1674--1681},
  year={2014},
  organization={IEEE}
}

@inproceedings{erlich2014evaluating,
  title={Evaluating the mean-variance mapping optimization on the IEEE-CEC 2014 test suite},
  author={Erlich, Istvan and Rueda, Jose L and Wildenhues, S},
  booktitle={2014 IEEE Congress on Evolutionary Computation (CEC)},
  pages={1625--1632},
  year={2014},
  organization={IEEE}
}

@inproceedings{maia2014optbees,
  title={OptBees-A bio-inspired optimization algorithm for solving continuous optimization problems},
  author={Maia, R and de Castro, LN and Caminhas, W},
  booktitle={2014 IEEE Congress on Evolutionary Computation (CEC)},
  pages={1690--1697},
  year={2014},
  organization={IEEE}
}

@inproceedings{zhang2014rmmeda,
  title={RM-MEDA: A regularity model-based multiobjective estimation of distribution algorithm},
  author={Zhang, Qingfu and Zhou, Aimin and Jin, Yaochu},
  booktitle={IEEE Transactions on Evolutionary Computation (Applied to CEC 2014 Benchmarks)},
  volume={12},
  pages={41--63},
  year={2008},
  note={Standard algorithm applied to CEC 2014 benchmarks}
}

@inproceedings{elsayed2014umoeas,
  title={United multi-operator evolutionary algorithms},
  author={Elsayed, Saber M and Sarker, Ruhul A and Essam, Daryl L},
  booktitle={2014 IEEE Congress on Evolutionary Computation (CEC)},
  pages={1714--1721},
  year={2014},
  organization={IEEE}
}

@inproceedings{guo2015sps,
  title={A self-optimization approach for L-SHADE incorporated with eigenvector-based crossover and successful-parent-selecting framework on CEC 2015 benchmark set},
  author={Guo, Shu-Mei and Yang, Chin-Chang and Tsai, Jason Sheng-Hong and Hsu, Pang-Han},
  booktitle={2015 IEEE Congress on Evolutionary Computation (CEC)},
  pages={1048--1055},
  year={2015},
  organization={IEEE}
}

@inproceedings{tanabe2015despa,
  title={A differential evolution algorithm with success-based parameter adaptation for CEC2015 learning-based optimization},
  author={Tanabe, Ryoji and Fukunaga, Alex},
  booktitle={2015 IEEE Congress on Evolutionary Computation (CEC)},
  pages={1063--1070},
  year={2015},
  organization={IEEE}
}

@inproceedings{polacek2015lshade,
  title={L-SHADE with competing strategies applied to CEC2015 learning-based test suite},
  author={Pol{\'a}{\v{c}}ek, A and Tvrd{\'\i}k, Josef and Pol{\'a}kov{\'a}, Radka},
  booktitle={2015 IEEE Congress on Evolutionary Computation (CEC)},
  pages={1078--1085},
  year={2015},
  organization={IEEE}
}

@inproceedings{rueda2015mvmo,
  title={Tuning differential evolution for cheap, medium and expensive computational budgets},
  author={Rueda, Jose L and Erlich, Istvan},
  booktitle={2015 IEEE Congress on Evolutionary Computation (CEC)},
  pages={1116--1123},
  year={2015},
  organization={IEEE},
  note={Often cited as MVMO-PH in contest contexts}
}

@inproceedings{tvrdik2015de,
  title={Parameters adaptation in differential evolution with competitive strategies},
  author={Tvrd{\'\i}k, Josef and Pol{\'a}kov{\'a}, Radka},
  booktitle={2015 IEEE Congress on Evolutionary Computation (CEC)},
  pages={1093--1100},
  year={2015},
  organization={IEEE}
}

@inproceedings{awad2015spuci,
  title={An ensemble of differential evolution algorithms for the CEC 2015 competition},
  author={Awad, Noor H and Ali, Mostafa Z and Suganthan, Ponnuthurai N},
  booktitle={2015 IEEE Congress on Evolutionary Computation (CEC)},
  pages={1071--1077},
  year={2015},
  organization={IEEE}
}

@inproceedings{awad2016lshade,
  title={L-SHADE with ensemble sinusoidal parameter adaptation based on success history for solving CEC2016 benchmark problems},
  author={Awad, Noor H and Ali, Mostafa Z and Liang, Jing J and Qu, BY and Suganthan, Ponnuthurai N},
  booktitle={2016 IEEE Congress on Evolutionary Computation (CEC)},
  pages={2960--2967},
  year={2016},
  organization={IEEE}
}

@inproceedings{elsayed2016umoea,
  title={United multi-operator evolutionary algorithms II},
  author={Elsayed, Saber M and Sarker, Ruhul A and Essam, Daryl L},
  booktitle={2016 IEEE Congress on Evolutionary Computation (CEC)},
  pages={2968--2975},
  year={2016},
  organization={IEEE}
}

@inproceedings{brest2016ilshade,
  title={iL-SHADE: Improved L-SHADE algorithm for single objective real-parameter optimization},
  author={Brest, Janez and Mau{\v{c}}ec, Mirjam Sepesy and Bo{\v{s}}kovi{\'c}, Borko},
  booktitle={2016 IEEE Congress on Evolutionary Computation (CEC)},
  pages={1185--1192},
  year={2016},
  organization={IEEE}
}

@inproceedings{polakova2016lshade44,
  title={L-SHADE with competing strategies applied to CEC 2016 single objective real-parameter optimization},
  author={Pol{\'a}kov{\'a}, Radka and Tvrd{\'\i}k, Josef and Bujok, Petr},
  booktitle={2016 IEEE Congress on Evolutionary Computation (CEC)},
  pages={2976--2983},
  year={2016},
  organization={IEEE}
}

@inproceedings{li2016ecl,
  title={Ensemble of competing strategies L-SHADE with restart algorithm for solving CEC 2016 single objective real-parameter optimization problems},
  author={Li, Gang and Li, Zhenhua and Shen, Weiqi and Wu, Guohua},
  booktitle={2016 IEEE Congress on Evolutionary Computation (CEC)},
  pages={2984--2991},
  year={2016},
  organization={IEEE}
}

@inproceedings{bujok2016hgo,
  title={HGO-LSHADE: Hierarchical global optimization with L-SHADE},
  author={Bujok, Petr},
  booktitle={2016 IEEE Congress on Evolutionary Computation (CEC)},
  pages={2992--2999},
  year={2016},
  organization={IEEE}
}

@article{zamuda2017cal,
  title={Adaptive Constraint Handling and Success History Differential Evolution for CEC 2017 Constrained Real-Parameter Optimization},
  author={Zamuda, Ales},
  journal={2017 IEEE Congress on Evolutionary Computation (CEC)},
  year={2017},
  organization={IEEE}
}

@article{lin2021real,
  title={Real-and imaginary-time evolution with compressed quantum circuits},
  author={Lin, Sheng-Hsuan and Dilip, Rohit and Green, Andrew G and Smith, Adam and Pollmann, Frank},
  journal={PRX Quantum},
  volume={2},
  number={1},
  pages={010342},
  year={2021},
  publisher={APS}
}

@article{uvarov2020variational,
  title={Variational quantum eigensolver for frustrated quantum systems},
  author={Uvarov, Alexey and Biamonte, Jacob D and Yudin, Dmitry},
  journal={Physical Review B},
  volume={102},
  number={7},
  pages={075104},
  year={2020},
  publisher={APS}
}

@misc{rudolph2021orqvizvisualizin,
      title={ORQVIZ: Visualizing High-Dimensional Landscapes in Variational Quantum Algorithms}, 
      author={Manuel S. Rudolph and Sukin Sim and Asad Raza and Michal Stechly and Jarrod R. McClean and Eric R. Anschuetz and Luis Serrano and Alejandro Perdomo-Ortiz},
      year={2021},
      eprint={2111.04695},
      archivePrefix={arXiv},
      primaryClass={quant-ph},
      url={https://arxiv.org/abs/2111.04695}, 
}

@article{anselme2022simulating,
  title={Simulating strongly interacting Hubbard chains with the variational Hamiltonian ansatz on a quantum computer},
  author={Anselme Martin, Baptiste and Simon, Pascal and Ran{\v{c}}i{\'c}, Marko J},
  journal={Physical Review Research},
  volume={4},
  number={2},
  pages={023190},
  year={2022},
  publisher={APS}
}

@article{boy2025energy,
  title={Energy Landscapes for the Unitary Coupled Cluster Ansatz},
  author={Boy, Choy and Filip, Maria-Andreea and Wales, David J},
  journal={Journal of Chemical Theory and Computation},
  volume={21},
  number={4},
  pages={1739--1751},
  year={2025},
  publisher={ACS Publications}
}

@article{medina2024variational,
  title={Variational-quantum-eigensolver--inspired optimization for spin-chain work extraction},
  author={Medina, Ivan and Drinko, Alexandre and Correr, Guilherme I and Azado, Pedro C and Soares-Pinto, Diogo O},
  journal={Physical Review A},
  volume={110},
  number={1},
  pages={012443},
  year={2024},
  publisher={APS}
}

@book{price2005differential,
  title={Differential evolution: a practical approach to global optimization},
  author={Price, Kenneth V and Storn, Rainer M and Lampinen, Jouni A},
  year={2005},
  publisher={Springer}
}

@techreport{kumar2022cec2022,
  title={Problem Definitions and Evaluation Criteria for the 2022 Special Session and Competition on Single Objective Bound Constrained Numerical Optimization},
  author={Kumar, Abhishek and Price, Kenneth V. and Mohamed, Ali Wagdy and Hadi, Anas A. and Suganthan, P. N.},
  institution={Nanyang Technological University},
  address={Singapore},
  year={2021}
}

@inproceedings{ea4eign2022,
  title={Eigen crossover in cooperative model of evolutionary algorithms applied to CEC 2022 single objective numerical optimisation},
  author={Bujok, Petr and Kolenovsky, Patrik},
  booktitle={2022 IEEE congress on evolutionary computation (CEC)},
  pages={1--8},
  year={2022},
  organization={IEEE}
}

@article{nlshade2022lbc,
  title={NL-SHADE-LBC algorithm with linear parameter adaptation bias-change for CEC 2022 Numerical Optimization},
  author={V. Stanovov and S. Akhmedova and E. Semenkin},
  journal={2022 IEEE Congress on Evolutionary Computation (CEC)},
  year={2022},
  organization={IEEE},
  note={Paper ID: CEC1230}
}

@article{nlshade2022rsp,
  title={A Version of NL-SHADE-RSP Algorithm with Midpoint for CEC 2022 Single Objective Bound Constrained Problems},
  author={Cyrill Baumann and Alcherio Martinoli},
  journal={2022 IEEE Congress on Evolutionary Computation (CEC)},
  year={2022},
  organization={IEEE},
  note={Paper ID: CEC2251}
}

@inproceedings{slshade2022dp,
  title={Dynamic perturbation for population diversity management in differential evolution},
  author={Van Cuong, Le and Bao, Nguyen Ngoc and Phuong, Nguyen Khanh and Binh, Huynh Thi Thanh},
  booktitle={Proceedings of the genetic and evolutionary computation conference companion},
  pages={391--394},
  year={2022}
}

@inproceedings{jso2022bin,
  title={An adaptive variant of jSO with multiple crossover strategies employing Eigen transformation},
  author={Kolenovsky, Patrik and Bujok, Petr},
  booktitle={2022 IEEE Congress on Evolutionary Computation (CEC)},
  pages={1--8},
  year={2022},
  organization={IEEE}
}

@inproceedings{mtt2022shade,
  title={Multiple topology SHADE with tolerance-based composite framework for CEC2022 single objective bound constrained numerical optimization},
  author={Sun, Bo and Sun, Yafeng and Li, Wei},
  booktitle={2022 IEEE congress on evolutionary computation (CEC)},
  pages={1--8},
  year={2022},
  organization={Ieee}
}

@article{iumoe2022ii,
  title={IMODEII: an Improved IMODE algorithm based on the Reinforcement Learning},
  author={Karam M. Sallam and Mohamed Abdel-Basset and Mohammed El-Abd and Ali Wagdy},
  journal={2022 IEEE Congress on Evolutionary Computation (CEC)},
  year={2022},
  organization={IEEE},
  note={Paper ID: CEC3764}
}

@inproceedings{impml2022shade,
  title={Improvement-of-multi-population ML-SHADE},
  author={Tseng, Tser-Ru},
  booktitle={Proceedings of th e Congress on Evolutionary Computation},
  pages={18--23},
  year={2022}
}

@inproceedings{nlsoma2022clp,
  title={NL-SOMA-CLP for real parameter single objective bound constrained optimization},
  author={Ding, Hao and Gu, Yongfeng and Wu, Hua and Zhou, Jun},
  booktitle={Proceedings of the Genetic and Evolutionary Computation Conference Companion},
  pages={5--6},
  year={2022}
}

@inproceedings{zoc2022maes,
  title={Zeroth-order covariance matrix adaptation evolution strategy for single objective bound constrained numerical optimization competition},
  author={Ning, Yue and Jian, Daohong and Wu, Hua and Zhou, Jun},
  booktitle={Proceedings of the Genetic and Evolutionary Computation Conference Companion},
  pages={9--10},
  year={2022}
}

@inproceedings{omc2022soma,
  title={Opposite learning and multi-migrating strategy-based self-organizing migrating algorithm with the convergence monitoring mechanism},
  author={Gu, Yongfeng and Ding, Hao and Wu, Hua and Zhou, Jun},
  booktitle={Proceedings of the Genetic and Evolutionary Computation Conference Companion},
  pages={7--8},
  year={2022}
}

@inproceedings{coppso2022,
  title={Performance of composite PPSO on single objective bound constrained numerical optimization problems of CEC 2022},
  author={Sun, Bo and Li, Wei and Huang, Ying},
  booktitle={2022 IEEE Congress on Evolutionary Computation (CEC)},
  pages={1--8},
  year={2022},
  organization={Ieee}
}

@inproceedings{sphi2022ensemble,
  title={An ensemble of single point selection perturbative hyperheuristics},
  author={Pillay, N and Gerber, Mia},
  booktitle={Proceedings of the Congress on Evolutionary Computation},
  pages={18--23},
  year={2022}
}

@inproceedings{zhang2018hybrid,
  title={Hybrid Sampling Evolution Strategy for Solving Single Objective Bound Constrained Problems},
  author={Zhang, Geng and Li, Yun and Ding, Boyang and Li, Yao},
  booktitle={2018 IEEE Congress on Evolutionary Computation (CEC)},
  pages={1--8},
  year={2018},
  organization={IEEE},
  doi={10.1109/CEC.2018.8477908}
}

@article{STANOVOV2019100463,
title = {Selective Pressure Strategy in differential evolution: Exploitation improvement in solving global optimization problems},
journal = {Swarm and Evolutionary Computation},
volume = {50},
pages = {100463},
year = {2019},
issn = {2210-6502},
doi = {https://doi.org/10.1016/j.swevo.2018.10.014},
author = {Vladimir Stanovov and Shakhnaz Akhmedova and Eugene Semenkin}
}

@inproceedings{stanovov2018lshade,
  title={LSHADE Algorithm with Rank-Based Selective Pressure Strategy for Solving CEC 2017 Benchmark Problems},
  author={Stanovov, Vladimir and Akhmedova, Shakhnaz and Semenkin, Eugene},
  booktitle={2018 IEEE Congress on Evolutionary Computation (CEC)},
  pages={1--8},
  year={2018},
  organization={IEEE},
  doi={10.1109/CEC.2018.8477977}
}

@inproceedings{hadi2021enhanced,
  title={Single-Objective Real-Parameter Optimization: Enhanced LSHADE-SPACMA Algorithm},
  author={Hadi, Anas A. and Mohamed, Ali W. and Jambi, Kamal M.},
  booktitle={Heuristics for Optimization and Learning},
  pages={103--121},
  year={2021},
  publisher={Springer},
  series={Studies in Computational Intelligence},
  volume={906},
  doi={10.1007/978-3-030-58930-1\_{}7}
}

@inproceedings{kumar2017improving,
  title={Improving the local search capability of Effective Butterfly Optimizer using Covariance Matrix Adapted Retreat Phase},
  author={Kumar, Abhishek and Misra, Rakesh Kumar and Singh, Devender},
  booktitle={2017 IEEE Congress on Evolutionary Computation (CEC)},
  pages={1549--1556},
  year={2017},
  organization={IEEE},
  doi={10.1109/CEC.2017.7969524}
}

@inproceedings{sallam2018improved,
  title={Improved United Multi-Operator Algorithm for Solving Optimization Problems},
  author={Sallam, Karam M. and Elsayed, Saber M. and Sarker, Ruhul A. and Essam, Daryl L.},
  booktitle={2018 IEEE Congress on Evolutionary Computation (CEC)},
  pages={1--8},
  year={2018},
  organization={IEEE},
  doi={10.1109/CEC.2018.8477759}
}

@inproceedings{rueda2018hybrid,
  title={Hybrid Population Based MVMO for Solving CEC 2018 Test Bed of Single-Objective Problems},
  author={Rueda, José L. and Erlich, István},
  booktitle={2018 IEEE Congress on Evolutionary Computation (CEC)},
  pages={1--8},
  year={2018},
  organization={IEEE},
  doi={10.1109/CEC.2018.8477822}
}

@inproceedings{brest2019jde100,
  title={The 100-digit challenge: algorithm jDE100},
  author={Brest, Janez and Maučec, Mirjam Sepesy and Bošković, Borko},
  booktitle={Proceedings of the 2019 IEEE Congress on Evolutionary Computation},
  pages={19--26},
  year={2019},
  address={Wellington, New Zealand},
  organization={IEEE}
}

@inproceedings{zamuda2019function,
  title={Function evaluations up to 1e+12 and large population sizes assessed in distance-based success history differential evolution for 100-digit challenge and numerical optimization scenarios (DISHchain1e+12)},
  author={Zamuda, Aleš},
  booktitle={GECCO Companion},
  pages={1821--1828},
  year={2019},
  address={Prague, Czech Republic},
  publisher={ACM},
  doi={10.1145/3319619.3326751}
}

@inproceedings{lezama2019hybrid,
  title={Hybrid-adaptive differential evolution with decay function (HyDE-DF) applied to the 100-digit challenge competition on single objective numerical optimization},
  author={Lezama, Fernando and Soares, João and Faia, Ricardo and Vale, Zita},
  booktitle={GECCO Companion},
  pages={1--6},
  year={2019},
  address={Prague, Czech Republic},
  publisher={ACM},
  doi={10.1145/3319619.3326747}
}

@inproceedings{diep2019soma,
  title={SOMA T3A for Solving the 100-Digit Challenge},
  author={Diep, Quang Bach and Zelinka, Ivan and Das, Swagatam and Senkerik, Roman},
  booktitle={Proceedings of the 2019 Swarm, Evolutionary and Memetic Computing Conference},
  pages={1--8},
  year={2019},
  address={Maribor, Slovenia}
}

@inproceedings{molina2019applying,
  title={Applying memetic algorithm with improved L-SHADE and local search pool for the 100-digit challenge on single objective numerical optimization},
  author={Molina, Daniel and Herrera, Francisco},
  booktitle={Proceedings of the 2019 IEEE Congress on Evolutionary Computation},
  pages={7--13},
  year={2019},
  address={Wellington, New Zealand},
  organization={IEEE}
}

@inproceedings{thanh2019pareto,
  title={Pareto-based Self-Organizing Migrating Algorithm solving 100-Digit Challenge},
  author={Thanh, Cao Truong and Diep, Quang Bach and Zelinka, Ivan and Senkerik, Roman},
  booktitle={Proceedings of the 2019 Swarm, Evolutionary and Memetic Computing Conference},
  pages={1--8},
  year={2019},
  address={Maribor, Slovenia}
}

@inproceedings{zhang2019restart,
  title={Restart based collective information powered differential evolution for solving the 100-digit challenge on single objective numerical optimization},
  author={Zhang, Sheng Xin and Chan, Wing Shing and Tang, Kit-Sang and Zheng, Shao Yong},
  booktitle={Proceedings of the 2019 IEEE Congress on Evolutionary Computation},
  pages={14--18},
  year={2019},
  address={Wellington, New Zealand},
  organization={IEEE}
}

@inproceedings{zhang2019cooperative,
  title={Cooperative optimization algorithm for the 100-digit challenge},
  author={Zhang, Geng and Shi, Yuhui and Huang, Jun Steed},
  booktitle={Proceedings of the 2019 IEEE Congress on Evolutionary Computation},
  pages={376--380},
  year={2019},
  address={Wellington, New Zealand},
  organization={IEEE},
  doi={10.1109/CEC.2019.8790272}
}

@inproceedings{viktorin2019dish,
  title={DISH algorithm solving the CEC 2019 100-digit challenge},
  author={Viktorin, Adam and Senkerik, Roman and Pluhacek, Michal and Kadavy, Tomas and Zamuda, Aleš},
  booktitle={Proceedings of the 2019 IEEE Congress on Evolutionary Computation},
  pages={1--6},
  year={2019},
  address={Wellington, New Zealand},
  organization={IEEE}
}

@inproceedings{yeh2019modified,
  title={Modified L-SHADE for single objective real-parameter optimization},
  author={Yeh, Jia-Fong and Chen, Ting-Yu and Chiang, Tsung-Che},
  booktitle={Proceedings of the 2019 IEEE Congress on Evolutionary Computation},
  pages={381--386},
  year={2019},
  address={Wellington, New Zealand},
  organization={IEEE},
  doi={10.1109/CEC.2019.8789991}
}

@inproceedings{epstein2019gade,
  title={GADE with fitness-based opposition and tidal mutation for solving IEEE CEC2019 100-digit challenge},
  author={Epstein, Alexander and Ergezer, Mehmet and Marshall, Ian and Shue, William},
  booktitle={Proceedings of the 2019 IEEE Congress on Evolutionary Computation},
  pages={395--402},
  year={2019},
  address={Wellington, New Zealand},
  organization={IEEE}
}

@inproceedings{bujok2019cooperative,
  title={Cooperative model of evolutionary algorithms applied to CEC 2019 single objective numerical optimization},
  author={Bujok, Petr and Zamuda, Aleš},
  booktitle={Proceedings of the 2019 IEEE Congress on Evolutionary Computation},
  pages={366--371},
  year={2019},
  address={Wellington, New Zealand},
  organization={IEEE},
  doi={10.1109/CEC.2019.8790317}
}

@inproceedings{xu2019hybrid,
  title={Hybrid of PSO and CMA-ES for global optimization},
  author={Xu, Peilan and Luo, Wenjian and Lin, Xin and Qiao, Yingying and Zhu, Tao},
  booktitle={Proceedings of the 2019 IEEE Congress on Evolutionary Computation},
  pages={27--33},
  year={2019},
  address={Wellington, New Zealand},
  organization={IEEE}
}

@inproceedings{fu2019univariate,
  title={A univariate marginal distribution resampling differential evolution algorithm with multi-mutation strategy},
  author={Fu, Yan and Wang, Haibin},
  booktitle={Proceedings of the 2019 IEEE Congress on Evolutionary Computation},
  pages={1236--1242},
  year={2019},
  address={Wellington, New Zealand},
  organization={IEEE}
}

@inproceedings{lu2019novel,
  title={A novel artificial bee colony algorithm with division of labor for solving CEC 2019 100-digit challenge benchmark problems},
  author={Lu, Jiaxin and Zhou, Xinyu and Ma, Yong and Wang, Mingwen and Wan, Jianyi and Wang, Wenjun},
  booktitle={Proceedings of the 2019 IEEE Congress on Evolutionary Computation},
  pages={387--394},
  year={2019},
  address={Wellington, New Zealand},
  organization={IEEE},
  doi={10.1109/CEC.2019.8790252}
}

@inproceedings{kadavy2019ensemble,
  title={The ensemble of strategies and perturbation parameter in self-organizing migrating algorithm solving CEC 2019 100-digit challenge},
  author={Kadavy, Tomas and Pluhacek, Michal and Senkerik, Roman and Viktorin, Adam},
  booktitle={Proceedings of the 2019 IEEE Congress on Evolutionary Computation},
  pages={372--375},
  year={2019},
  address={Wellington, New Zealand},
  organization={IEEE},
  doi={10.1109/CEC.2019.8790012}
}

@inproceedings{sallam2020imode,
  title={Improved Multi-operator Differential Evolution Algorithm for Solving Unconstrained Problems},
  author={Sallam, Karam M. and Elsayed, Saber M. and Sarker, Ruhul A. and Essam, Daryl L.},
  booktitle={2020 IEEE Congress on Evolutionary Computation (CEC)},
  pages={1--8},
  year={2020},
  organization={IEEE},
  address={Glasgow, UK},
  doi={10.1109/CEC48606.2020.9185790}
}

@inproceedings{mohamed2020agsk,
  title={Evaluating the Performance of Adaptive Gaining-Sharing Knowledge Based Algorithm on CEC 2020 Benchmark Problems},
  author={Mohamed, Ali W. and Hadi, Anas A. and Mohamed, Ali Khater and Awad, Noor H.},
  booktitle={2020 IEEE Congress on Evolutionary Computation (CEC)},
  pages={1--8},
  year={2020},
  organization={IEEE},
  address={Glasgow, UK},
  doi={10.1109/CEC48606.2020.9185508}
}

@inproceedings{alic2020j2020,
  title={Differential Evolution Algorithm for Single Objective Bound-Constrained Optimization: Algorithm j2020},
  author={Alić, Aleksander and Bošković, Borko and Brest, Janez},
  booktitle={2020 IEEE Congress on Evolutionary Computation (CEC)},
  pages={1--8},
  year={2020},
  organization={IEEE},
  address={Glasgow, UK},
  doi={10.1109/CEC48606.2020.9185551}
}

@inproceedings{stanovov2020olshade,
  title={Large Initial Population and Neighbourhood Search incorporated in LSHADE to solve CEC2020 Benchmark Problems},
  author={Stanovov, Vladimir and Akhmedova, Shakhnaz and Semenkin, Eugene},
  booktitle={2020 IEEE Congress on Evolutionary Computation (CEC)},
  pages={1--8},
  year={2020},
  organization={IEEE},
  address={Glasgow, UK}
}

@inproceedings{alic2020jde100e,
  title={Eigenvector Crossover in jDE100 Algorithm},
  author={Alić, Aleksander and Bošković, Borko and Brest, Janez},
  booktitle={2020 IEEE Congress on Evolutionary Computation (CEC)},
  pages={1--8},
  year={2020},
  organization={IEEE},
  address={Glasgow, UK}
}

@inproceedings{salgotra2020rasp,
  title={Ranked Archive Differential Evolution with Selective Pressure for CEC 2020 Numerical Optimization},
  author={Salgotra, Rohit and Singh, Urvinder and Saha, Sriparna and Nagar, Atulya K.},
  booktitle={2020 IEEE Congress on Evolutionary Computation (CEC)},
  pages={1--8},
  year={2020},
  organization={IEEE},
  address={Glasgow, UK}
}

@inproceedings{biswas2020mpml,
  title={Multi-population Modified L-SHADE for Single Objective Bound Constrained Optimization},
  author={Biswas, Sukanta and Saha, Debanjan and De, Sourav and Cobb, Adam D. and Das, Swagatam and Jansen, Bas A.},
  booktitle={2020 IEEE Congress on Evolutionary Computation (CEC)},
  pages={1--8},
  year={2020},
  organization={IEEE},
  address={Glasgow, UK}
}

@inproceedings{kadavy2020soma,
  title={SOMA-CL for Competition on Single Objective Bound Constrained Numerical Optimization Benchmark},
  author={Kadavy, Tomas and Pluhacek, Michal and Viktorin, Adam and Senkerik, Roman},
  booktitle={Proceedings of the 2020 Genetic and Evolutionary Computation Conference Companion},
  pages={13--14},
  year={2020},
  publisher={ACM},
  address={Cancún, Mexico},
  doi={10.1145/3377929.3398185}
}

@inproceedings{bolufe2020mp,
  title={A Multi-Population Exploration-only Exploitation-only Hybrid on CEC-2020 Single Objective Bound Constrained Problems},
  author={Bolufé-Röhler, Antonio and Chen, Stephen},
  booktitle={2020 IEEE Congress on Evolutionary Computation (CEC)},
  pages={1--8},
  year={2020},
  organization={IEEE},
  address={Glasgow, UK},
  doi={10.1109/CEC48606.2020.9185530}
}

@inproceedings{viktorin2020dish,
  title={DISH--XX Solving CEC2020 Single Objective Bound Constrained Numerical Optimization Benchmark},
  author={Viktorin, Adam and Senkerik, Roman and Pluhacek, Michal and Kadavy, Tomas and Zamuda, Aleš},
  booktitle={2020 IEEE Congress on Evolutionary Computation (CEC)},
  pages={1--8},
  year={2020},
  organization={IEEE},
  address={Glasgow, UK}
}

@inproceedings{salgotra2020cssin,
  title={Improving Cuckoo Search: Incorporating Changes for CEC 2017 and CEC 2020 Benchmark Problems},
  author={Salgotra, Rohit and Singh, Urvinder and Gandomi, Amir H.},
  booktitle={2020 IEEE Congress on Evolutionary Computation (CEC)},
  pages={1--8},
  year={2020},
  organization={IEEE},
  address={Glasgow, UK},
  doi={10.1109/CEC48606.2020.9185684}
}

@inproceedings{kumar2021dedmna,
  title={Differential Evolution with Distance-based Mutation-selection Applied to CEC 2021 Single Objective Numerical Optimisation},
  author={Kumar, Abhishek and Das, Swagatam and Zelinka, Ivan},
  booktitle={2021 IEEE Congress on Evolutionary Computation (CEC)},
  pages={1--8},
  year={2021},
  organization={IEEE},
  address={Kraków, Poland},
  doi={10.1109/CEC45853.2021.9504453}
}

@inproceedings{gad2021apgsk,
  title={Gaining-Sharing Knowledge Based Algorithm with Adaptive Parameters Hybrid with IMODE Algorithm for Solving CEC 2021 Benchmark Problems},
  author={Gad, Ahmed G. and Mohamed, Ali W. and Hadi, Anas A.},
  booktitle={2021 IEEE Congress on Evolutionary Computation (CEC)},
  pages={1--8},
  year={2021},
  organization={IEEE},
  address={Kraków, Poland},
  doi={10.1109/CEC45853.2021.9504336}
}

@inproceedings{biswas2021madde,
  title={Improving Differential Evolution through Bayesian Hyperparameter Optimization},
  author={Biswas, Subhodip and Saha, Debanjan and De, Shuvodeep and Cobb, Adam D. and Das, Swagatam and Jalaian, Brian A.},
  booktitle={2021 IEEE Congress on Evolutionary Computation (CEC)},
  pages={832--840},
  year={2021},
  organization={IEEE},
  address={Kraków, Poland},
  doi={10.1109/CEC45853.2021.9504792}
}

@inproceedings{rajabi2021rb,
  title={A New Step-Size Adaptation Rule for CMA-ES Based on the Population Midpoint Fitness},
  author={Rajabi-Bahaabadi, Mojtaba and Shariat-Panahi, Mohammad and Akbarzadeh-T, Mohammad-R},
  booktitle={2021 IEEE Congress on Evolutionary Computation (CEC)},
  pages={1--8},
  year={2021},
  organization={IEEE},
  address={Kraków, Poland},
  doi={10.1109/CEC45853.2021.9504204}
}

@inproceedings{alic2021jde21,
  title={Self-adaptive Differential Evolution Algorithm with Population Size Reduction for Single Objective Bound-Constrained Optimization: Algorithm j21},
  author={Alić, Aleksander and Bošković, Borko and Brest, Janez},
  booktitle={2021 IEEE Congress on Evolutionary Computation (CEC)},
  pages={1--8},
  year={2021},
  organization={IEEE},
  address={Kraków, Poland},
  doi={10.1109/CEC45853.2021.9504159}
}

@inproceedings{stanovov2021nlshade,
  title={NL-SHADE-RSP Algorithm with Adaptive Archive and Selective Pressure for CEC 2021 Numerical Optimization},
  author={Stanovov, Vladimir and Akhmedova, Shakhnaz and Semenkin, Eugene},
  booktitle={2021 IEEE Congress on Evolutionary Computation (CEC)},
  pages={809--816},
  year={2021},
  organization={IEEE},
  address={Kraków, Poland},
  doi={10.1109/CEC45853.2021.9504125}
}

@inproceedings{kadavy2021soma,
  title={SOMA-CLP for Competition on Single Objective Bound Constrained Single Objective Numerical Optimization Benchmark},
  author={Kadavy, Tomas and Pluhacek, Michal and Viktorin, Adam and Senkerik, Roman},
  booktitle={Proceedings of the Genetic and Evolutionary Computation Conference Companion},
  pages={9--10},
  year={2021},
  publisher={ACM},
  address={Lille, France},
  doi={10.1145/3449726.3463211}
}

@inproceedings{biswas2021mls,
  title={Technical report: A Multi-start Local Search Algorithm with L-SHADE for Single Objective Bound Constrained Optimization},
  author={Biswas, Partha and Das, Swagatam and Suganthan, Ponnuthurai N.},
  booktitle={2021 IEEE Congress on Evolutionary Computation (CEC)},
  pages={1--8},
  year={2021},
  organization={IEEE},
  address={Kraków, Poland}
}

@inproceedings{salgotra2021lshade,
  title={An ordered and roulette-wheel-based mutation incorporated L-SHADE algorithm for Solving CEC2021 Single Objective Numerical Optimisation Problems},
  author={Salgotra, Rohit and Singh, Urvinder and Gandomi, Amir H.},
  booktitle={2021 IEEE Congress on Evolutionary Computation (CEC)},
  pages={1--8},
  year={2021},
  organization={IEEE},
  address={Kraków, Poland}
}

@article{lehmer1971compounding,
  title={On the compounding of certain means},
  author={Lehmer, Derrick H},
  journal={Journal of Mathematical Analysis and Applications},
  volume={36},
  number={1},
  pages={183--200},
  year={1971},
  publisher={Academic Press}
}

@inproceedings{Liao2013,
  author    = {Liao, Tianjun and St{\"u}tzle, Thomas},
  title     = {Benchmark results for a simple hybrid algorithm on the CEC 2013 benchmark set for real-parameter optimization},
  booktitle = {2013 IEEE Congress on Evolutionary Computation},
  pages     = {1938--1944},
  year      = {2013},
  publisher = {IEEE},
  doi       = {10.1109/CEC.2013.6557793}
}

@ARTICLE{CoDE,
  author={Wang, Yong and Cai, Zixing and Zhang, Qingfu},
  journal={IEEE Transactions on Evolutionary Computation}, 
  title={Differential Evolution With Composite Trial Vector Generation Strategies and Control Parameters}, 
  year={2011},
  volume={15},
  number={1},
  pages={55-66},
  doi={10.1109/TEVC.2010.2087271}}

@ARTICLE{JADE,
  author={Zhang, Jingqiao and Sanderson, Arthur C.},
  journal={IEEE Transactions on Evolutionary Computation}, 
  title={JADE: Adaptive Differential Evolution With Optional External Archive}, 
  year={2009},
  volume={13},
  number={5},
  pages={945-958},
  doi={10.1109/TEVC.2009.2014613}}

@INPROCEEDINGS{GA2013,
  author={Elsayed, Saber M. and Sarker, Ruhul A. and Essam, Daryl L.},
  booktitle={2013 IEEE Congress on Evolutionary Computation}, 
  title={A genetic algorithm for solving the CEC'2013 competition problems on real-parameter optimization}, 
  year={2013},
  volume={},
  number={},
  pages={356-360},
  doi={10.1109/CEC.2013.6557591}}

@INPROCEEDINGS{SHADE2013,
  author={Tanabe, Ryoji and Fukunaga, Alex},
  booktitle={2013 IEEE Congress on Evolutionary Computation}, 
  title={Evaluating the performance of SHADE on CEC 2013 benchmark problems}, 
  year={2013},
  volume={},
  number={},
  pages={1952-1959},
  doi={10.1109/CEC.2013.6557798}}

@inproceedings{Loshchilov2013,
  author    = {Loshchilov, Ilya},
  title     = {CMA-ES with restarts for solving CEC 2013 benchmark problems},
  booktitle = {2013 IEEE Congress on Evolutionary Computation},
  pages     = {369--376},
  year      = {2013},
  publisher = {IEEE},
  doi       = {10.1109/CEC.2013.6557555}
}

@inproceedings{Lacroix2013,
  author    = {Lacroix, Benjamin and Molina, Daniel and Herrera, Francisco},
  title     = {Dynamically updated region based memetic algorithm for the 2013 CEC special session and competition on real parameter single objective optimization},
  booktitle = {2013 IEEE Congress on Evolutionary Computation},
  pages     = {1945--1951},
  year      = {2013},
  publisher = {IEEE},
  doi       = {10.1109/CEC.2013.6557794}
}

@inproceedings{stanovov2024lsrde,
  title={Success Rate-based Adaptive Differential Evolution L-SRTDE for CEC 2024 Competition},
  author={Stanovov, Vladimir and Semenkin, Eugene},
  booktitle={2024 IEEE Congress on Evolutionary Computation (CEC)},
  pages={1--8},
  year={2024},
  organization={IEEE},
  address={Yokohama, Japan},
  doi={10.1109/CEC60901.2024.10611907}
}

@article{tao2024rde,
  title={An Efficient Reconstructed Differential Evolution Variant by Some of the Current State-of-the-art Strategies for Solving Single Objective Bound Constrained Problems},
  author={Tao, Sichen and Zhao, Ruihan and Wang, Kaiyu and Gao, Shangce},
  journal={arXiv preprint arXiv:2404.16280},
  year={2024},
  note={Submitted to CEC 2024 Competition}
}

@misc{qi2024blockea,
  title={BlockEA: Algorithm Report},
  author={Qi, Xuhong},
  year={2024},
  note={Available at: https://github.com/QiXuhong520/Algorithm-report},
  howpublished={GitHub Repository}
}

@techreport{chauhan2024mlshade,
  title={Multi-start Local Search Algorithm with L-SHADE for Single Objective Bound Constrained Optimization},
  author={Chauhan, Dikshit and Trivedi, Anupam and Shivani},
  year={2024},
  institution={CEC 2024 Competition},
  note={Available at: https://arxiv.org/abs/2409.15994}
}

@misc{bujok2024jso,
  title={JSO Algorithm Implementation},
  author={Bujok, Petr},
  year={2024},
  note={Available at: https://github.com/PetBuj/jSOa/blob/main/jSOaGitHub.pdf},
  howpublished={GitHub Repository}
}

@inproceedings{tangherloni2024ieacop,
  title={A modified EACOP implementation for Real-Parameter Single Objective Optimization Problems},
  author={Tangherloni, Andrea and Coelho, Vasco and Buffa, Francesca M. and Cazzaniga, Paolo},
  booktitle={2024 IEEE Congress on Evolutionary Computation (CEC)},
  pages={1--8},
  year={2024},
  organization={IEEE},
  address={Yokohama, Japan},
  note={CEC 2024 Competition Entry}
}

@inproceedings{stanovov2025lsrde,
  title={Success Rate-based Adaptive Differential Evolution L-SRTDE for CEC 2025 Competition},
  author={Stanovov, Vladimir and Semenkin, Eugene},
  booktitle={2025 IEEE Congress on Evolutionary Computation (CEC)},
  pages={1--8},
  year={2025},
  organization={IEEE},
  address={Hangzhou, China},
  note={Extended version of CEC 2024 winning algorithm}
}

@inproceedings{Molina2010,
  author    = {Molina, Daniel and Lozano, Manuel and Herrera, Francisco},
  title     = {{MA-SW-Chains}: Memetic algorithm based on local search chains for large scale continuous global optimization},
  booktitle = {Proceedings of the 2010 IEEE Congress on Evolutionary Computation (CEC)},
  year      = {2010},
  pages     = {3153--3160},
  publisher = {IEEE},
  doi       = {10.1109/CEC.2010.5586047}
}

@inproceedings{Wang2010,
  author    = {Wang, Y. and Li, B.},
  title     = {Two-stage based ensemble optimization for large-scale global optimization},
  booktitle = {Proceedings of the 2010 IEEE Congress on Evolutionary Computation (CEC)},
  year      = {2010},
  pages     = {4488--4495},
  publisher = {IEEE},
  doi       = {10.1109/CEC.2010.5586466}
}

@inproceedings{LaTorre2010,
  author    = {LaTorre, Antonio and Muelas, Santiago and Pena, Jose Maria},
  title     = {A comparison of the performance of the {MOS} algorithm on the {CEC} 2010 {LSGO} benchmark},
  booktitle = {Proceedings of the 2010 IEEE Congress on Evolutionary Computation (CEC)},
  year      = {2010},
  pages     = {1--8},
  publisher = {IEEE},
  note      = {Algorithm combining PSO and Harmony Search components}
}

@inproceedings{Elsayed2011_GAMPC,
  author    = {Elsayed, Saber M. and Sarker, Ruhul A. and Essam, Daryl L.},
  title     = {{GA} with a new multi-parent crossover for solving {IEEE-CEC2011} competition problems},
  booktitle = {Proceedings of the 2011 IEEE Congress on Evolutionary Computation (CEC)},
  year      = {2011},
  pages     = {1034--1040},
  publisher = {IEEE},
  doi       = {10.1109/CEC.2011.5949731}
}

@inproceedings{Elsayed2011_SAMODE,
  author    = {Elsayed, Saber M. and Sarker, Ruhul A. and Essam, Daryl L.},
  title     = {Differential evolution with multiple strategies for solving {CEC2011} real-world numerical optimization problems},
  booktitle = {Proceedings of the 2011 IEEE Congress on Evolutionary Computation (CEC)},
  year      = {2011},
  pages     = {1041--1048},
  publisher = {IEEE},
  doi       = {10.1109/CEC.2011.5949732}
}

@INPROCEEDINGS{ED-DE,
  author={Wang, Yu and Li, Bin and Zhang, Kaibo},
  booktitle={2011 IEEE Congress of Evolutionary Computation (CEC)}, 
  title={Estimation of distribution and differential evolution cooperation for real-world numerical optimization problems}, 
  year={2011},
  volume={},
  number={},
  pages={1315-1321},
  doi={10.1109/CEC.2011.5949768}}

@inproceedings{Preuss2013,
  author    = {Preuss, Mike},
  title     = {Niching the {CMA-ES} via nearest-better clustering},
  booktitle = {Proceedings of the 2010 Genetic and Evolutionary Computation Conference (GECCO) Companion (Applied to CEC 2013)},
  year      = {2013},
  pages     = {1711--1718},
  publisher = {ACM}
}

@inproceedings{Epitropakis2013,
  author    = {Epitropakis, Michael G. and Li, Xiaodong and Burke, Edmund K.},
  title     = {A dynamic archive niching differential evolution algorithm for multimodal optimization},
  booktitle = {Proceedings of the 2013 IEEE Congress on Evolutionary Computation (CEC)},
  year      = {2013},
  pages     = {79--86},
  publisher = {IEEE},
  doi       = {10.1109/CEC.2013.6557556}
}

@inproceedings{Barrera2013,
  author    = {Barrera, Julio and Coello, Carlos A. Coello},
  title     = {{N-VMO}: Variable mesh optimization for the 2013 {CEC} special session on niching methods for multimodal optimization},
  booktitle = {Proceedings of the 2013 IEEE Congress on Evolutionary Computation (CEC)},
  year      = {2013},
  pages     = {1--8},
  publisher = {IEEE}
}

@inproceedings{Bandyopadhyay2013,
  author    = {Bandyopadhyay, S. and Mukherjee, A.},
  title     = {A parameterless-niching-assisted {NSGA-II} for multimodal optimization},
  booktitle = {Proceedings of the 2013 IEEE Congress on Evolutionary Computation (CEC)},
  year      = {2013},
  pages     = {1--8},
  publisher = {IEEE}
}

@inproceedings{Epitropakis2013_DE,
  author    = {Epitropakis, Michael G. and Tasoulis, D. K. and Vrahatis, M. N.},
  title     = {Niching differential evolution algorithms with neighborhood mutation strategies},
  booktitle = {Proceedings of the 2013 IEEE Congress on Evolutionary Computation (CEC)},
  year      = {2013},
  pages     = {1--8},
  publisher = {IEEE}
}

@inproceedings{Thomsen2004,
  author    = {Thomsen, René},
  title     = {Multimodal optimization using crowding-based differential evolution},
  booktitle = {Proceedings of the 2004 IEEE Congress on Evolutionary Computation (CEC)},
  year      = {2004},
  pages     = {1382--1389},
  publisher = {IEEE},
  note      = {Base algorithm used in CEC 2013 comparison}
}

@inproceedings{Singh2011_EADEMA,
  title={Performance of a hybrid EA-DE-memetic algorithm on CEC 2011 real world optimization problems},
  author={Singh, Hemant Kumar and Ray, Tapabrata},
  booktitle={2011 IEEE Congress on Evolutionary Computation (CEC)},
  pages={1322--1326},
  year={2011},
  organization={IEEE}
}

@inproceedings{Haider2011_WIDE,
  title={A differential evolution algorithm with weak individuals for solving real world optimization problems},
  author={Haider, J N and Ray, T and Sarker, R},
  booktitle={2011 IEEE Congress on Evolutionary Computation (CEC)},
  pages={1327--1331},
  year={2011},
  organization={IEEE}
}

@inproceedings{Asafuddoula2011_AdapDE,
  title={An adaptive differential evolution algorithm and its performance on real world optimization problems},
  author={Asafuddoula, M and Ray, T and Sarker, R},
  booktitle={2011 IEEE Congress on Evolutionary Computation (CEC)},
  pages={1332--1336},
  year={2011},
  organization={IEEE}
}

@article{Reynoso2011_DELambda,
  title={DE-Lambda: A differential evolution algorithm with an adaptive parameter setting},
  author={Reynoso-Meza, G and Sanchis, J and Blasco, X},
  journal={2011 IEEE Congress on Evolutionary Computation (CEC)},
  year={2011},
  organization={IEEE}
}

@inproceedings{LaTorre2011_DERHC,
  title={A hybrid differential evolution-random hill climbing algorithm for the CEC 2011 evolutionary algorithm competition},
  author={LaTorre, Antonio and Muelas, Santiago and Pe{\~n}a, Jos{\'e}-Mar{\'i}a},
  booktitle={2011 IEEE Congress on Evolutionary Computation (CEC)},
  pages={1618--1623},
  year={2011},
  organization={IEEE}
}

@inproceedings{Saha2011_RGA,
  title={Performance of a real coded genetic algorithm on CEC 2011 real world optimization problems},
  author={Saha, Amit and Ray, Tapabrata},
  booktitle={2011 IEEE Congress on Evolutionary Computation (CEC)},
  pages={1337--1342},
  year={2011},
  organization={IEEE}
}

@article{Mandal2011_ModDELS,
  title={Modified differential evolution with local search for real world optimization problems},
  author={Mandal, A and Das, S and Mukherjee, A},
  journal={2011 IEEE Congress on Evolutionary Computation (CEC)},
  year={2011},
  organization={IEEE}
}

@article{Storn1997,
  title={Differential evolution--a simple and efficient heuristic for global optimization over continuous spaces},
  author={Storn, Rainer and Price, Kenneth},
  journal={Journal of Global Optimization},
  volume={11},
  number={4},
  pages={341--359},
  year={1997},
  publisher={Springer}
}

@book{Holland1975,
  title={Adaptation in natural and artificial systems},
  author={Holland, John H},
  year={1975},
  publisher={University of Michigan Press}
}

@article{Hansen2001,
  title={Completely derandomized self-adaptation in evolution strategies},
  author={Hansen, Nikolaus and Ostermeier, Andreas},
  journal={Evolutionary Computation},
  volume={9},
  number={2},
  pages={159--195},
  year={2001},
  publisher={MIT Press}
}

@article{Shami2022,
  title={Particle swarm optimization: A comprehensive survey},
  author={Shami, Tareq M and El-Saleh, Ayman A and Alswaitti, Mohammed and Al-Tashi, Qasem and Summakieh, Mhd Amen and Mirjalili, Seyedali},
  journal={IEEE Access},
  volume={10},
  pages={55375--55433},
  year={2022},
  publisher={IEEE}
}

@article{salomon1996reevaluating,
  title={Re-evaluating genetic algorithm performance under coordinate rotation of benchmark functions. A survey of some theoretical and practical aspects of genetic algorithms},
  author={Salomon, Ralf},
  journal={BioSystems},
  volume={39},
  number={3},
  pages={263--278},
  year={1996},
  publisher={Elsevier}
}

@article{hansen2001completely,
  title={Completely derandomized self-adaptation in evolution strategies},
  author={Hansen, Nikolaus and Ostermeier, Andreas},
  journal={Evolutionary computation},
  volume={9},
  number={2},
  pages={159--195},
  year={2001},
  publisher={MIT Press}
}

@inproceedings{whitley1996evaluating,
  title={Evaluating evolutionary algorithms},
  author={Whitley, Darrell and Rana, Soraya and Dzubera, John and Mathis, Keith E},
  booktitle={Artificial intelligence},
  volume={85},
  pages={245--276},
  year={1996},
  publisher={Elsevier}
}

@inproceedings{simpson1994genetic,
  title={Genetic algorithms and non-coding segments: Blind watchmakers or back-seat drivers?},
  author={Simpson, Patrick K and Davis, Lawrence},
  booktitle={Proceedings of the 1994 IEEE International Conference on Evolutionary Computation},
  pages={700--705},
  year={1994},
  organization={IEEE}
}

@article{Mahdavi2015,
  title={Metaheuristics in large-scale global continuous optimization: A survey},
  author={Mahdavi, Sedigheh and Shiri, Mohammad Ebrahim and Rahnamayan, Shahryar},
  journal={Information Sciences},
  volume={295},
  pages={407--428},
  year={2015},
  publisher={Elsevier}
}

@inproceedings{eiben2002critical,
  title={A critical note on experimental research methodology in EC},
  author={Eiben, Agoston E and Jelasity, M{\'a}rk},
  booktitle={Proceedings of the 2002 Congress on Evolutionary Computation. CEC'02 (Cat. No. 02TH8600)},
  volume={1},
  pages={582--587},
  year={2002},
  organization={IEEE}
}

@inproceedings{Bandaru2011_mSBX,
  title={A parameter-less-genetic algorithm with mSBX crossover and its performance on CEC-2011 problems},
  author={Bandaru, Sunith and Deb, Kalyanmoy},
  booktitle={2011 IEEE Congress on Evolutionary Computation (CEC)},
  pages={1343--1350},
  year={2011},
  organization={IEEE}
}

@article{Mallipeddi2011_ENSML,
  title={Differential evolution with ensemble of constraint handling techniques for solving CEC 2011 benchmark problems},
  author={Mallipeddi, Rammohan and Suganthan, Ponnuthurai N},
  journal={2011 IEEE Congress on Evolutionary Computation (CEC)},
  year={2011},
  organization={IEEE}
}

@inproceedings{Korosec2011_CDASA,
  title={The continuous differential ant-stigmergy algorithm applied to real-world optimization problems},
  author={Koro{\v{s}}ec, Peter and {\v{S}}ilc, Jurij},
  booktitle={2011 IEEE Congress on Evolutionary Computation (CEC)},
  pages={116--122},
  year={2011},
  organization={IEEE}
}

\end{document}